%% file: Mainfile_RSE.tex
\documentclass[]{elsarticle}
\usepackage{lineno}
\usepackage{graphicx}
\usepackage[usenames,dvipsnames,svgnames,table]{xcolor}
\usepackage[utf8x]{inputenc}
\usepackage{amsmath}
\usepackage{amssymb}
\usepackage{amsthm}
\usepackage{multirow}
\usepackage{booktabs}
\usepackage{framed}
\usepackage{epstopdf}
\usepackage{url}
\usepackage{ucs}
\usepackage{soul} %To highlight pasts of the paper

\usepackage{color}

\input{notation}

\modulolinenumbers[5]
\journal{Remote Sensing of Environment}

\usepackage{hyperref}
\hypersetup{
    bookmarks=true,         % show bookmarks bar?
    unicode=false,          % non-Latin characters in Acrobat’s bookmarks
    pdftoolbar=true,        % show Acrobat’s toolbar?
    pdfmenubar=true,        % show Acrobat’s menu?
    pdffitwindow=false,     % window fit to page when opened
    pdfstartview={FitH},    % fits the width of the page to the window
    pdftitle={TRY upscaling},    % title
    pdfauthor={Moreno et al.},     % author
    pdfsubject={TRY upscaling},   % subject of the document
    pdfcreator={Moreno et al.},   % creator of the document
    pdfproducer={Moreno et al.}, % producer of the document
    pdfkeywords={keyword1} {key2} {key3}, % list of keywords
    pdfnewwindow=true,      % links in new window
    colorlinks=true,       % false: boxed links; true: colored links
    linkcolor=blue,          % color of internal links (change box color with linkbordercolor)
    citecolor=blue,        % color of links to bibliography
    filecolor=blue,      % color of file links
    urlcolor=blue           % color of external links
}
\biboptions{authoryear}
\graphicspath{../figures/}

\begin{document}

\begin{frontmatter}

\title{A Methodology to Derive Global Maps of Leaf Traits Using Remote Sensing and Climate Data}

%% Group authors per affiliation:
\author{\'Alvaro Moreno-Mart\'inez\fnref{address1}} \ead{alvaro.moreno@ntsg.umt.edu}
\author{Gustau Camps-Valls\fnref{address2}}
\author{Jens Kattge\fnref{address3}}
\author{Nathaniel Robinson\fnref{address1}}
\author{Markus Reichstein\fnref{address3}}
\author{Peter van Bodegom\fnref{address15}}
\author{Koen Kramer\fnref{address4}}
\author{J. Hans C. Cornelissen\fnref{address5}}
\author{Peter Reich\fnref{address6}}
\author{Michael Bahn\fnref{address16}}
\author{{\"U}lo Niinemets\fnref{address8}}
\author{Josep Pe{\~n}uelas\fnref{address10}}
\author{Joseph Craine\fnref{address17}}
\author{Bruno E.L. Cerabolini\fnref{address7}}
\author{Vanessa Minden\fnref{address11}}
\author{Daniel C. Laughlin\fnref{address12}}
\author{Lawren Sack\fnref{address13}}
\author{Brady Allred\fnref{address1}}
\author{Christopher Baraloto\fnref{address14}}
%\author{Matthew O. Jones\fnref{address1}}
\author{Chaeho Byun\fnref{address9}}
\author{Nadejda A. Soudzilovskaia\fnref{address15}}
\author{Steve W. Running\fnref{address1}}
\address[address1]{Numerical Terradynamic Simulation Group (NTSG), College of Forestry and Conservation, University of Montana, USA}
\address[address2]{Image Processing Laboratory (IPL), Universitat de Val\`encia, Val\`encia, Spain}
\address[address3]{Max-Planck-Institute for Biogeochemistry (MPI-BGC), Jena, Germany}
\address[address4]{Wageningen Environmental Research (WUR), Wageningen University, Wageningen, The Netherlands}
\address[address5]{Systems Ecology, Department of Ecological Science, Vrije Universiteit, Amsterdam,The Netherlands}
\address[address6]{Department of Forest Resources, University of Minnesota, St. Paul, USA}
\address[address7]{Dipartimento di Scienze Teoriche e Applicate (DiSTA), University of Insubria, Varese, Italy}
\address[address8]{Department of Crop Science and Plant Biology, Estonian University of Life Sciences, Kreutzwaldi, Tartu, Estonia}
\address[address9]{School of Civil and Environmental Engineering, Yonsei University, Seoul, Korea.}
\address[address10]{Centre for Research on Ecology and Forestry Applications (CREAF), Cerdanyola del Valles, Catalonia, Spain}
\address[address11]{Landscape Ecology Group, Institute of Biology and Environmental Sciences, University of Oldenburg, Oldenburg, Germany}
\address[address12]{Department of Botany, University of Wyoming, Wyoming, USA}
\address[address13]{Department of Ecology and Evolutionary Biology, University of California, California, USA}
\address[address14]{International Center for Tropical Botany (ICTB),  Department of Biological Sciences, Florida International University, Florida, USA}
\address[address15]{Conservation Biology Department, Institute of Environmental Sciences, Leiden University, Leiden, The Netherlands}
\address[address16]{Plant, Soil and Ecosystem Processes, Institute of Ecology, University of Innsbruck, Innsbruck, Austria}
\address[address17]{Jonah Ventures, Manhattan, USA}
\begin{abstract}

This paper introduces a modular processing chain to derive global high-resolution maps of leaf traits. In particular, we present global maps at 500 m resolution of specific leaf area, leaf dry matter content, leaf nitrogen and phosphorus content per dry mass, and leaf nitrogen/phosphorus ratio. The processing chain exploits machine learning techniques along with optical remote sensing data (MODIS/Landsat) and climate data for gap filling and up-scaling of in-situ measured leaf traits. The chain first uses random forests regression with surrogates to fill gaps in the database ($> 45 \% $ of missing entries)  and maximize the global representativeness of the trait dataset. Plant species are then aggregated to Plant Functional Types (PFTs). Next, the spatial abundance of PFTs at MODIS resolution (500 m) is calculated using Landsat data (30 m). Based on these PFT abundances, representative trait values are calculated for MODIS pixels with nearby trait data. Finally, different regression algorithms are applied to globally predict trait estimates from these MODIS pixels using remote sensing and climate data. The methods were compared in terms of precision, robustness and efficiency. The best model (random forests regression) shows good precision (normalized RMSE$\leq 20\%$) and goodness of fit (averaged Pearson's correlation R$=0.78$) in any considered trait. Along with the estimated global maps of leaf traits, we provide associated uncertainty estimates derived from the regression models. The process chain is modular, and can easily accommodate new traits, data streams (traits databases and remote sensing data), and methods. The machine learning techniques applied allow attribution of information gain to data input and thus provide the opportunity to understand trait-environment relationships at the plant and ecosystem scales. The new data products --the gap-filled trait matrix, a global map of PFT abundance per MODIS gridcells and the high-resolution global leaf trait maps-- are complementary to existing large-scale observations of the land surface and we therefore anticipate substantial contributions to advances in quantifying, understanding and prediction of the Earth system.
\end{abstract}

\begin{keyword}
Plant traits \sep Machine Learning \sep Random forests \sep Remote Sensing \sep Plant ecology \sep Climate \sep MODIS \sep Landsat
\end{keyword}

\end{frontmatter}

\modulolinenumbers[1]
%\linenumbers

% THE TOC AND HIGHLIGHTS
%\newpage
%\scriptsize
%\tableofcontents
%\newpage
%\begin{framed}
%{\bf Highlights}
%\begin{itemize}
%\item Presented a generalized, modular process chain for community aggregated plant trait mapping including local effects
%\item First high-resolution global maps of community aggregated plant traits with uncertainties by fusing trait data base TRY, LANDSAT and MODIS
%\item Scope for testing and parameterizing trait-enables Earth System models
%\item Implications for land management and Earth system science applications
%\end{itemize}
%\end{framed}
%\normalsize

%% THE PAPER

\newpage
\input{introduction2}
\input{methods3}
\input{results}

\input{conclusions3}

%\section*{Conflicts of Interest}
%
%The authors declare no conflict of interest.

\section*{Acknowledgements}
This research was financially supported by the NASA Earth Observing System MODIS project (grant NNX08AG87A). This work was also supported by the European Research Council (ERC) funding under the ERC Consolidator Grant 2014 SEDAL (Statistical Learning for Earth Observation Data Analysis) project under Grant Agreement 647423. We also want to gratefully acknowledge the efforts of all researchers involved at the TRY initiative on plant traits (http://www.try-db.org), hosted at the Max Planck Institute for Biogeochemistry, Jena, Germany. The authors would also like to thank the anonymous reviewers for their constructive comments on an earlier version of this manuscript.

\small
%Here the References (.bib files):
%\section*{References}
\bibliography{traits}

\clearpage
\appendix
\input{appendices2}

\end{document}

%% file: notation.tex
\def\x{{\mathbf x}}

\newcommand{\cut}[1]{} % cut out a part of the text

%% file: introduction2.tex
\section{Introduction}

In terrestrial ecosystems, environmental conditions and biogeochemical processes both influence and are influenced by plant communities. Historical processes such as evolution, migration and disturbance shape plants from the organismal to community level \citep{musavi2015imprint}. At the organismal level, plant traits, which are measurable morphological, anatomical, physiological and phenological characteristics, can influence the establishment, fitness, and survival of individuals \citep{westoby1998leaf,reich2007predicting,violle2007let,homolova2013review}. This definition has been recently updated to encompass also responses and effects attributes at broader scales such as population, community, and ecosystem \citep{reich2014world}. These traits vary widely across the $\sim$400,000 vascular plant species (\url{http://www.theplantlist.org//}), and due to acclimation and adaptation processes vary within individual species \citep{turner2006evaluation,reich2007predicting}.
Standard modelling and remote sensing approaches to estimate photosynthesis, evapotranspiration and biophysical parameters such as the fraction of absorbed photosynthetically active radiation (fAPAR) and leaf area index (LAI) use plant functional types (PFTs) to include plant traits within the model \citep{chen1999daily,myneni2002global,zhao2005improvements,krinner2005dynamic,mu2011improvements,jiang2016multi}. In so doing however, the diversity of plant communities is simplified into a relatively few categories and key variability within individual PFTs is lost \citep{running1994vegetation,wullschleger2014plant}. Subsequently, model parameters based on plant trait properties are limited by the PFT groupings, resulting in an important source of uncertainty in many biosphere models \citep{van2014fully,reich2014world,reichstein2014linking}.

In Earth system modelling, methods are being developed to improve PFT approaches, such as refining PFT categories and/or making the PFTs more spatiotemporally dynamic \citep{poulter2011plant}. An alternative approach is to model the continuous spatial variability of plant traits themselves \citep{yang2015plant,musavi2016potential,van2014fully,diaz2016global,madani2014improving}. This can be done with the use of plant trait databases through establishing empirical trait-environment relationships and trait covariation \citep{wullschleger2014plant,verheijen2015inclusion}. There are a number of global traits databases containing in-situ trait observations of a comprehensive suite of plant traits for numerous species around the globe \citep{kattge2011try,reichstein2014linking,diaz2016global}. These extensive databases are continually evolving and growing and provide the foundation for making broader and spatially explicit inferences of plant traits. Spatializing plant traits however, is not without substantial challenges. First, despite the large number of species included in trait databases, they are sparse compared to the overall richness and diversity of species globally \citep{jetz2016monitoring}. Second, the large trait databases are amalgamations of many individual datasets, and contain numerous gaps. Third, the in-situ trait observations are temporally disjointed, meaning they come from a wide range of years depending on when measurements were made. Finally, these observations are made at the individual plant scale, and not necessarily representative of the variability at coarser scales.

Attempts to spatialize plant traits fall into two general categories: biogeographical and remote sensing based approaches. Biogeographical approaches attempt to extrapolate local trait measurements across different spatial scales by relating traits to abiotic factors, assuming that these factors (i.e., climate and soils) constrain the structure and function of natural ecosystems
\citep{niinemets2001global,kattge2011try,reichstein2014linking,diaz2016global,madani2018future}. For example, \cite{van2014fully} generated global trait maps by relating traits to gridded soil and climate data. Using only these environmental drivers, they were able to explain up to 50\% of the global variation of plaint traits. These approaches, however, do not take into account actual measured vegetation dynamics and are limited by the coarser resolution of the input data. Remote sensing approaches, on the other hand, can capitalize on higher resolution observations of actual vegetation dynamics. The estimation of plant traits from optical remote sensing is often done through physical radiative transfer models (RTMs) or empirical approaches \citep{haboudane2004hyperspectral,mulla2013twenty}. RTMs attempt to explicitly define the complex interactions between the radiation and the vegetation canopy properties, these models could be inverted to retrieve biophysical variables from leaf/canopy reflectances \citep{jacquemoud1990prospect,dawson1998liberty,jacquemoud2000comparison,houborg2007combining,stuckens2009dorsiventral}. The combined use of RTMs with satellite data from airborne and satellite-based platforms \citep{liang2005quantitative,baret2008estimating,berger2018evaluation} has allowed the successful retrieval of vegetation traits at different spatial and temporal scales (e.g. chlorophyll content, \cite{zhang2005estimating,houborg2007combining}, water content, \cite{zarco2003water,houborg2007combining}, and others like leaf dry matter content and specific leaf area, \cite{ali2016estimating,feret2008prospect}). However, applying RTMs across broad spatiotemporal extents is challenging as parameterizing RTMs across a wide range of growth forms, biomes and ecosystems is challenging \citep{berger2018evaluation}. Furthermore, RTMs are generally based of single scene reflectance values and do not consider climatic variables that are valuable proxies for various plant traits. Alternatively, empirical approaches relating in-situ observations of plant traits to remote sensing data have been successful at mapping localized gradients of plant traits. These approaches have limited broader applications as in-situ data are often scarce or incomplete. Recent studies have combined remote sensing and biogeographical approaches \citep{butler2017mapping} to obtain global maps of leaf traits at a very low spatial resolution ($0.5^{\circ} \times  0.5^{\circ}$ grid). The main limitation of these approaches is that, until now, they have utilized static remote sensing PFT maps for the spatialization of traits, being restricted to the simplicity of the PFTs, and not fully exploiting the full potential of optical remote sensing data (spatial and temporal variability of spectral responses), responses that can be invaluable in the estimation of key plant traits.

In this manuscript, we present and validate a combined remote sensing and biogeographic approach to spatializing estimates of key leaf traits. We integrate plant traits databases, remotely sensed data, and climatological data resulting in spatialized global maps of leaf traits at an unprecedented spatial resolution (500 m), that can be incorporated into other Earth system's models. We capitalize on the extensiveness of traits databases, the growing archive of satellite remote sensing data at multiple resolutions through time, global climatological data, and the advent of high-performance cloud computing technologies specifically designed for remote sensing applications (e.g. Google Earth Engine), combined with machine learning models for gap filling, classification and spatializing. We develop these methods for a selected set of 5 key leaf traits: Specific Leaf  Area (SLA; ratio of leaf area per unit dry mass), Leaf Dry Matter Content (LDMC), Leaf Nitrogen Content per leaf dry mass (Leaf Nitrogen Concentration, LNC), Leaf Phosphorus Content per leaf dry mass (Leaf Phosphorus Concentration, LPC), and Leaf Nitrogen to Phosphorus ratio (LNPR). SLA is a key trait of the leaf economics spectrum reflecting the trade-off between leaf longevity and carbon gain \citep{wright2004worldwide,diaz2016global}. SLA is thus indicative for different plant life strategies with respect to fast versus slow return of carbon investments \citep{reich2014world}. Some authors have indicated that LNC and LPC correlate with SLA and could be therefore part of the leaf economics spectrum \citep{wright2004worldwide}, but these relationships are still not fully understood and are not exempt of controversy \citep{osnas2013global,osnas2018divergent}. In addition, leaves with high LDMC tend to be relatively tough, and are thus assumed to be more resistant to physical hazards (e.g. herbivory, wind, hail) than are leaves with low LDMC \citep{peirez2013new}. LNC relates to the amount of the proteins involved in the photosynthetic machinery and phosphorus is a key constituent of nucleic acids, lipid membranes and some bioenergetic molecules (e.g. ATP) directly related in cell metabolism. In contrast, LNPR, is commonly used to indicate whether a vegetation is limited by either the availability of N or P \citep{bedford1999patterns,penuelas2013human}. All together these traits influence photosynthetic capacity and plant growth rate \citep{anten2004optimal,oliva2015key}.  Because of their importance in these processes, these leaf traits are key components of many biophysical models and have been collected globally at an unprecedented coverage filling both, the geographic and climate space well \citep{reichstein2014linking}.

Our methods first involve using a random forest with surrogates technique to gap fill the largest global plant trait database available (TRY, (\url{https://www.try-db.org//})), producing a more complete trait database. The TRY initiative, started in 2007, is a growing database, currently integrating over 300 datasets. Second, we scale-up plant trait observations, by weighting species observations from the TRY database, to the community/pixel level (community weighted mean, CWM), matching the spatial scales of the local trait observations and remote sensing data. To do so, we classify plant species into functional types and estimate the abundance of functional types within a 500 m MODIS pixel using 30 m Landsat data. Finally, we generate and validate regression models, based on the weighted abundance leaf trait estimates, to calculate global trait maps using remote sensing and ancillary climate data at a 500 m spatial resolution. 

%% file: methods3.tex
\section{Materials and methods}

\begin{figure}[h!]
  \centering

  \includegraphics[width=12.5 cm]{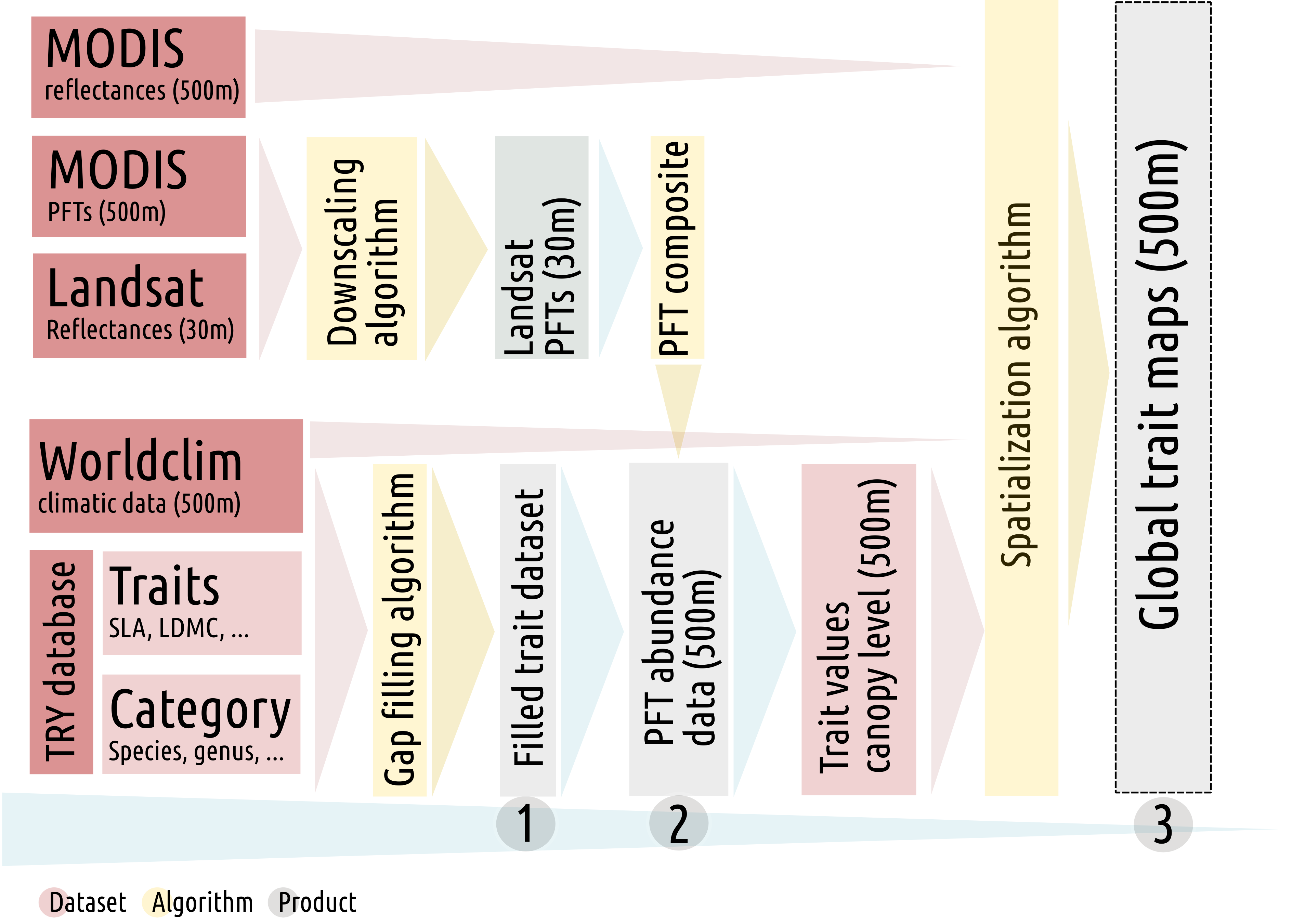}
  \caption{Flowchart of the proposed methodology for upscaling traits. The numbered boxes indicate the three main components of the methods: (1) gap filling the traits database; (2) calculating the community weighted mean (CWM) trait values at the canopy level for MODIS pixels with nearby trait observations; and (3) spatialization of CWMs to global trait maps at 500 m resolution.} %All three data products include uncertainty estimates.}
  \label{fig:TRYflowchart}
\end{figure}

We develop methods to calculate global maps of leaf traits derived from in-situ trait measurements, remote sensing data, and climatic data (see Figure \ref{fig:TRYflowchart}). The trait data was obtained from the TRY database (\url{www.try-db.org}, accession date: 2016-08-09). We integrated an assortment of remote sensing and climatic data at multiple resolutions (Table \ref{tab:RSdatadescription}), to spatialize the in-situ trait estimates. These included the full archive of the Landsat 5 top of atmosphere product  \citep{woodcock2008free}, the MODIS land cover product, MOD12Q1 \citep{friedl2010modis}, the MODIS BRDF-adjusted reflectance product, MCD43A4, the MODIS BRDF-adjusted albedo quality product, MCD43A2 \citep{strahler1999modis}, the Shuttle Radar Topography Mission (SRTM) DEM \citep{farr2007shuttle}, and the WorldClim climatic data \citep{hijmans2005very}.  Due to the global nature of this work along with the computational and storage needs associated with these datasets, we relied on the Google Earth Engine (GEE) platform for much of the remote sensing processing and analysis \citep{gorelick2017google}.

\begin{table}[!t]
\centering
\caption{General description of the data products used in this work.}
\label{tab:RSdatadescription}
\begin{tabular}{ | l | m{3.5cm}| m{1.9cm} |  m{2.6 cm} |}
\hline
\hline
Product & Description & Period & Temporal/Spatial resolutions \\
\hline
\hline
Landsat 5 TOA & Landsat 5 calibrated and orthorectified top-of-atmosphere reflectance with a quality band.  & 2000-2010  & 16 days/ 30 m \\
\hline
MOD12Q1 & MODIS Land Cover Type product and land cover type assessment. & 2001-2010  & Yearly / 500 m \\
\hline
MCD43A2 & BRDF-Albedo Quality & 2012-2015 & 16 days / 500 m \\
\hline
MCD43A4 & BRDF-Adjusted Reflectance  & 2012-2015 & 16 days / 500 m \\
\hline
STRM DEM & Elevation from The Shuttle Radar Topography Mission & - & - / 30 m \\
\hline
WorldClim & WorldClim climatic data & 1960-1990 & Monthly climatology / 1000 m \\
\hline
\hline
\end{tabular}
\end{table}

\subsection{Gap filling of the TRY database}
\label{se:gapfill}

The TRY database is a compilation of global plant trait data, sourced from thousands of disparate research activities \citep{kattge2011try}. Despite the unprecedented coverage of the TRY database, there is a number of key limitations due to the variety of underlying sources. First, only 40\% of the trait entries are georeferenced with unknown precision.  Second, the database includes measurements for a limited set (5-10\%) of known species, although the ones included are likely the most abundant species. Third, data were not necessarily collected with standardized protocols, introducing additional noise due to different measurement techniques. Finally, traits can be highly variable within canopies and across sites.  For example, plant trait values can vary at a single location, even for leaves sampled from one part of a canopy, introducing additional uncertainty in trait estimations. To overcome some of these limitations, we first removed outliers, defined by observations exceeding 1.5 standard deviations of the mean trait value for a given species \citep{kattge2011try}. Second, we discarded fern and crop species as they were scarce (less than 1\% of the total data). This also served to minimize the impact of highly managed vegetation (e.g. agriculture) in our training dataset.

Depending on the trait considered, the number of available samples and species are highly variable. For example, for SLA there are around 90,000 measurements from more than 7,000 species, while for LNPR there are 21,000 measurements from approximately 2,000 species, resulting in abundant gaps throughout the database because not all traits are measured simultaneously at each location. Gap filling methods, based on similar phylogenetic traits within taxonomic hierarchies, structural trade-offs between traits, or the relationships between traits and their environment are well established \citep{schrodt2015bhpmf,shan2012gap,wright2004worldwide,taugourdeau2014filling}. To gap fill the trait database we capitalize on these approaches, and also include trait-trait correlations. We use an ensemble of boosted random forest (RF) models with surrogates \citep{breiman2001random}. A random forest method is chosen for the gap filling (see \ref{ap:MLmethods}, for more detailed descriptions of the methods) due to three main characteristics. First, the ability of RF models to generalize is often superior to that by other machine learning algorithms due to the effect of bagging and feature selection \citep{breiman2001random}. Second, RF models are perfectly suited to dealing with mixed (categorical and continuous) data. Third, they are known to perform well with challenging data structures like high dimensionality, complex interactions and nonlinearities \citep{stekhoven2012missforest}. Finally, surrogates allow for the use of alternative decision splits at given nodes where there is missing input data through exploiting correlations between predictors \citep{feelders1999handling,friedman2009elements}. Gap filling allows us to increase the overall representativeness of the TRY database.

\subsection{Community weighted mean trait values for MODIS pixels with nearby trait observations}\label{ap:CWMs calc}

Estimates of global species abundances are necessary in order to spatialize in-situ leaf measured traits to community (canopy) level and account for the species level sampling bias in the TRY \citep{lavorel2002predicting,lavorel2008assessing,homolova2013review,van2014fully,musavi2015imprint,butler2017mapping},  but those abundances are not available globally and at required spatial resolution. We spatialize in-situ leaf level measurements to the community (canopy) level estimates by calculating a weighted mean using the relative abundances of the most dominant plants. This corresponds to the second key task indicated in the flow chart scheme (Figure \ref{fig:TRYflowchart}). To accomplish this, we utilize a lookup table of categorical traits provided through the TRY initiative that relates species name with conventional plant functional type definitions (\url{https://www.try-db.org/TryWeb/Data.php#3}). These include plant growth form (tree, shrub, grass, etc.), leaf type (needleleaf or broadleaf), and leaf phenology (evergreen or deciduous). These definitions correspond to established PFT classifications schemes such as the MODIS Land Cover Type product (MOD12Q1). This is an annually produced, 500 m land cover product that divides the terrestrial vegetated surface into seven categories: evergreen needleleaf forest (ENF), evergreen broadleaf forest (EBF), deciduous needleleaf forest (DNF), deciduous broadleaf forest (DBF), shrub lands (SHL), grasslands (GRL), and barren or sparsely vegetated \citep{friedl2010modis}. Thus, we associate each species in the TRY also with the corresponding PFT from the MOD12Q1 scheme. As an example, the categorical trait information that the TRY database provides for a plant species named 'Acer negundo' is: tree (plant growth form), broadleaved (leaf type), and deciduous (leaf phenology). Using this information, we were able to associate the PFT deciduous broadleaf type forest (DBF) to this species' name.

As we wanted to find community composition at a MODIS pixel resolution (500 m) to spatialize our leaf trait measurements, we developed a high resolution land cover map (30 m) based on the operational MODIS land cover. The high resolution land cover map allowed us to estimate the abundances of PFTs within the MODIS pixel. To calculate our land cover map, we followed a similar approach to the one proposed by \cite{friedl2010modis} for the operational MODIS land cover product and we considered similar input variables but for a higher spatial resolution satellite. The chosen input features include spectral and temporal information from Landsat 5 bands 1--7, the enhanced vegetation index (EVI) \citep{huete2002overview}, the normalized difference vegetation index (NDVI) \citep{tucker1979red}, the normalized difference water index (NDWI) \citep{gao1996ndwi}, Land Surface Temperature (LST) \citep{wan1996generalized}, and digital elevation from the Shuttle Radar Topography Mission (SRTM).

Landsat satellites have a revisit time of 16 days and on average 35\% of the images are plagued by cloud cover \citep{roy2008multi}. In order to obtain gap-free time series of Landsat 5, we computed a typical year combining 10 years of data (2000-2010). Landsat 5 was the preferred choice among other Landsat satellites due to its longer time series (1984-2012) and the absence of striped data gaps present in Landsat 7 (caused by the scan line corrector malfunctioning). Median monthly spectra were calculated using only cloud-free high quality data according to the quality assessment (QA) information available for each remote sensing scene. By taking the median value for each month in the considered period we were able to produce a more globally consistent input \citep{potapov2012quantifying,hansen2013high}. Utilizing the monthly composited spectral data, we computed temporal profiles of the vegetation indices aforementioned as well as land surface temperatures. Following the approach proposed by \cite{friedl2010modis}, we used summary variables derived from these annual time series (maximum and minimum values, annually integrated values, etc.), which allowed us to work with a reduced number of input variables (see Table \ref{tab:inputslandsat}).

The calculated input variables were used to train and test an RF classifier using the MODIS land cover product as reference data. The validation of the developed high resolution land cover map was carried out by comparing a spatially degraded version (500 m) with the original MODIS Land Cover Type product (MCD12Q1) over an independent test data set. Although PFTs remain fairly steady in natural areas, we reduced land cover related errors by estimating the mode land cover class (the class that appears most often) in the same time period as was considered for the Landsat data (2000-2010). All artificial or crop areas have been masked since most of the TRY database content is focused on natural vegetation (less than 1\% of the data are crop measurements).

\begin{table}[!h]
\small
\begin{center}
\caption{Input variables considered for the downscaling of the MODIS Land Cover Type product (MCD12Q1).*VI makes reference to a generic vegetation index. The spectral indices NDVI, EVI and NDWI have been considered in this work.}
\label{tab:inputslandsat}
\begin{tabular}{|l|l|}
\hline
\hline
Input variables & Description \\
\hline
\hline
B1med                                          & Median value of the reflectance band 1 (0.45-0.52 $\mu m$).           \\
B2med                                          & Median value of the reflectance band 2 (0.52-0.60 $\mu m$).              \\
B3med                                          & Median value of the reflectance band 3 (0.63-0.69 $\mu m$).                \\
B4med                                          & Median value of the reflectance band 4 (0.76-0.90 $\mu m$).             \\
B5med                                          & Median value of the reflectance band 5 (1.55-1.75 $\mu m$).           \\
B7med                                          & Median value of the reflectance band 7 (2.08-2.35 $\mu m$).         \\
LSTmed                                         & Median value of the land surface temperature (10.40-12.50 $\mu m$).  \\
VI*max                                         & Maximum value of the vegetation index during the year.  \\
VI*min                                         & Minimum value of the vegetation index during the year.     \\
VI*std                                         & Standard deviation of the vegetation index during the year.\\
VI*sum                                         & Accumulation of the vegetation index during the year.    \\
LSTmax                                         & Maximum value of the LST during the year.           \\
LSTmin                                         & Minimum value of the LST during the year.                 \\
LSTstd                                         & Standard deviation of the LST during the year.      \\
LSTsum                                         & Accumulation of the LST during the year.                 \\
Elevation                                      & Elevation from The Shuttle Radar Topography Mission. \\
\hline
\hline
\end{tabular}
\end{center}

\end{table}

To calculate community weighted mean (CWM) trait estimates, we used the developed high resolution land cover map by extracting the PFT abundances around a 500 m area for each geographical coordinate of the respective in-situ trait measurement. As a preliminary step, for each provided location, trait data which do not correspond to any PFT composing the considered pixel were automatically discarded. This step assumes that these PFT do not largely contribute to the CWM as their abundance is too small.  With the rest of data, the procedure to assign the value for each pixel was carried out employing the following three steps: First, all TRY data were sorted by distance to the selected pixel for each PFT. Secondly, the mean value for each pixel-specific PFT trait estimate was calculated from the selected closest nearby leaf trait observations corresponding with that PFT. This approach maximizes local species representativeness for each pixel while capturing spatial heterogeneities by weighting in-situ measurements according to their relative abundances. Thirdly, the assigned trait value for the considered MODIS pixel was calculated as the weighted mean (according to the community abundance) of the pixel-specific PFT trait estimate. 

The calculation of CWM trait estimates using only in-situ measurements within a MODIS pixel (500m) is often difficult because the representation of trait observations in spatial context is limited, but this can be addressed to attain a minimum representation. The number of available measurements is heterogeneous at a global scale and could potentially not be representative of the dominant PFTs `compositing'  the pixel or correspond with very different environmental conditions. So, in order to deal with that problem, for each location,  a maximum distance threshold (100 km) and a limited number of neighbors  (ten closest in spatial distance) were used to compute the community weighted means. These values are the result of an heuristic approach, we tried out different values and selected a sensible value that provided the most stable and reasonable results for all considered leaf traits. Of course, one could further optimize, but results were not much affected when increasing the threshold beyond a maximum distance of 100 km, so we fixed it to derive all our results. Recent studies have demonstrated that these thresholds are adequate for the spatialization global plant trait databases \citep{datta2016hierarchical,butler2017mapping}. In addition, we computed the percentage of trait measurements which match the PFTs present within a MODIS pixel. Pixels with trait measurements which were not representative of more than 50\% of the estimated PFTs composition were discarded.

\subsection{Spatializing community weighted mean trait values to global trait maps}\label{ap:MODISspatial}

Spatializing the CWM trait estimates was based on remote sensing and climatological data. Remote sensing provides great sensitivity to canopy phenology, structure, and chemical components while climatological data allow the models to capture how climatic constraints shape the structure and function of natural ecosystems. More precisely, we used the MODIS Reflectance product MCD43A4 \citep{strahler1999modis} and the WorldClim climatic data (described in more detail in \ref{ap:climatedata}). The MODIS reflectance product has global coverage, a temporal resolution of 16 days, and a spatial resolution of 500 m. It combines the MODIS sensor in TERRA and AQUA platforms, providing the highest probability for quality input data and designating it as an MCD (meaning combined) product. Nadir equivalent reflectance is derived by using a bidirectional reflectance function and multiple observations. Reflectance data acquired between 2012-2015 was used to obtain monthly gap free reflectance estimates by means of the QA information stored as a separate product (MCD43A2). The higher revisit times of the MODIS satellites, in addition to the use of a combined reflectance product (MCD), allowed us to obtain gap free monthly reflectance measurements with a shorter compositing period than Landsat.

Leaf trait values vary with many factors but they are clearly stratified by plant functional types \citep{abelleira2016scaling} which can be discriminated using multitemporal remote sensing. To address this, we considered input variables potentially sensitive to PFT stratification and chemical components of vegetation and therefore we repeated the Landsat processing scheme with MODIS observations. As above mentioned, the approach consisted on extracting a reduced set of input variables (summary variables) derived from annual time series of vegetation indices and the median spectrum (Table \ref{tab:inputslandsat}).

In order to extrapolate the trait measurements, we compared different machine learning algorithms: neural networks, kernel methods and decision trees. We observed a high consistency of RF predictions across all leaf traits and precision measures. For this reason, we selected random forests as the preferred default option for approximating the five different leaf traits considered, while also capitalizing on the speed and parallelization at the test phase. The detailed description and precision comparison of regression methods tested is provided in \ref{ap:MLmethods}.

%% file: results.tex
\section{Results}

This section presents the three main results obtained in the different steps of the processing chain (see Figure \ref{fig:TRYflowchart}). Results are evaluated both quantitatively and qualitatively; studying bias and accuracy of performance, and comparing the maps to expected values according to recent studies. More experimental results and comparisons are given in the corresponding appendices for the interested reader.

\subsection{Gap filling of the TRY database} \label{gap filling}

For each trait, the parameters of the RF (number of trees, learning rate, and maximum number of splits) were optimized following a cross-validation methodology. The optimization process involved a $10$-fold cross-validation for the estimation of the errors. In Table \ref{tab:gapfillingresults}, we show the error estimates in the gap filling process for each trait considered.

\begin{table}[!h]
\centering
\caption{Statistics of the gap filling approach per trait. The mean error (ME) accounts for the bias of the estimates, the root mean square error (RMSE) for the accuracy, and the Pearson's correlation coefficient R for the goodness-of-fit.}
\label{tab:gapfillingresults}
\begin{tabular}{l|c|c|c|c|c|}
\cline{2-6}
                                                & \multicolumn{1}{l|}{\textbf{SLA}} & {\textbf{LDMC}} & {\textbf{LNC}} & {\textbf{LPC}} & \textbf{LNPR} \\ \hline
\multicolumn{1}{|l|}{\textbf{ME}}               & 0.01                              & 0.00                               & 0.07                              & 0.00                              & 0.12          \\ \cline{1-1}
%\multicolumn{1}{|l|}{\textbf{MAE}}              & 2.22                              & 0.02                               & 2.34                              & 0.18                              & 1.10          \\ \cline{1-1}
\multicolumn{1}{|l|}{\textbf{RMSE}}             & 3.13                              & 0.02                               & 3.28                              & 0.27                              & 1.97          \\ \cline{1-1}
\multicolumn{1}{|l|}{\textbf{R}}                & 0.96                              & 0.96                               & 0.90                              & 0.86                              & 0.95          \\ \hline
\multicolumn{1}{|l|}{\textbf{Samples}}          & 89,355                             & 73,958                              & 54,036                             & 32,290                             & 21,407          \\ \cline{1-1}
\multicolumn{1}{|l|}{\textbf{Missing data [\%]}} & 47                                & 45                                 & 68                                & 74                                & 84            \\ \cline{1-1}
\multicolumn{1}{|l|}{\textbf{Mean value}}       & 16.87                             & 0.27                               & 20.38                             & 1.16                              & 13.55         \\ \hline
\end{tabular}
\end{table}

The accuracy of the gap-filling algorithm was very high for all five traits considered in this work. These results match those recently published in the literature \citep{schrodt2015bhpmf,shan2012gap,taugourdeau2014filling}, and in some cases ours are superior. In contrast to the method developed by \cite{shan2012gap}, who facilitate close to normal trait distribution of transformed traits, trait-trait covariances, and taxonomic hierarchy, our method is able to easily make use of additional explanatory variables. We additionally introduced climate and plant growth form. These two types of explanatory variables were amongst the five most relevant variables for gap-filling, as can be seen in the results from the sensitivity analysis shown in \ref{ap:senGF}.

\subsection{Abundance of PFTs at a MODIS pixel level}

We calculated the abundance of PFTs at a MODIS pixel level (500 m) using Landsat spectral information (30 m) as input and the operational MODIS PFTs map as a reference. The MODIS land cover product provides a land-cover type assessment that was used for an optimal selection of training samples (more than 85\% reliable according to the MODIS quality band). Taking into account this quality band, we sampled the globe randomly to build up the training and validation datasets (around 2000 pixels for each considered PFT). The considered inputs include Landsat median monthly TOA reflectances and LST values which were resampled to match MODIS (500 m) pixel size. The classification algorithm was carried out by a calibrated RF model on a global dataset in the GEE platform. The RF was trained and optimized with 50\% of the dataset and its accuracy was tested with the remaining data.

The achieved overall accuracy and Cohen's Kappa coefficient of agreement were 96\% and 0.85, respectively. Those results indicate the suitability of the model and its capability to generalize on unseen test data. Table \ref{tbl:confussionmatrix} shows the confusion matrix for each PFT over the test data set. The confusion matrix indicates that the most confounded PFTs are shrubs and grasses but still the corresponding accuracy is high.

\begin{table}[!h]
\small
\centering
\caption{Confusion matrix obtained when upscaling the global PFT map to a 30 m Landsat spatial resolution. The considered land cover types are: evergreen needle leaf forests (ENF), evergreen broad leaf forests (EBF), deciduous needle leaf forests (DNF), deciduous broad leaf forests (DBF), shrub lands (SHL), grass lands (GRL), and barren or sparsely vegetated (BARREN). Results correspond to an out-of-sample test set.}
\label{tbl:confussionmatrix}
\begin{tabular}{c|c|c|c|c|c|c|c|}
\cline{2-8}
       & {\textbf{ENF}} & {\textbf{EBF}} & {\textbf{DNF}} & {\textbf{DBF}} & \multicolumn{1}{l|}{\textbf{SHL}} & {\textbf{GRL}} & \textbf{BARREN} \\
\hline
\multicolumn{1}{|c|}{\textbf{ENF}}    & 870                               & 5                                 & 7                                 & 4                                 & 3                                 & 2                                 & 0               \\ \cline{1-1}
\multicolumn{1}{|c|}{\textbf{EBF}}    & 3                                 & 971                               & 0                                 & 2                                 & 0                                 & 0                                 & 0               \\ \cline{1-1}
\multicolumn{1}{|c|}{\textbf{DNF}}    & 15                                & 1                                 & 406                               & 0                                 & 0                                 & 2                                 & 0               \\ \cline{1-1}
\multicolumn{1}{|c|}{\textbf{DBF}}    & 3                                 & 2                                 & 1                                 & 992                               & 0                                 & 2                                 & 0               \\ \cline{1-1}
\multicolumn{1}{|c|}{\textbf{SHL}}    & 4                                 & 4                                 & 10                                & 4                                 & 737                               & 78                                & 15              \\ \cline{1-1}
\multicolumn{1}{|c|}{\textbf{GRL}}    & 1                                 & 1                                 & 4                                 & 4                                 & 30                                & 934                               & 20              \\ \cline{1-1}
\multicolumn{1}{|c|}{\textbf{BARREN}} & 0                                 & 0                                 & 0                                 & 0                                 & 4                                 & 17                                & 965             \\ \hline
\end{tabular}
\end{table}

\begin{figure}
  \centering
   \includegraphics[width=12cm,trim={0 1cm 0 0}]{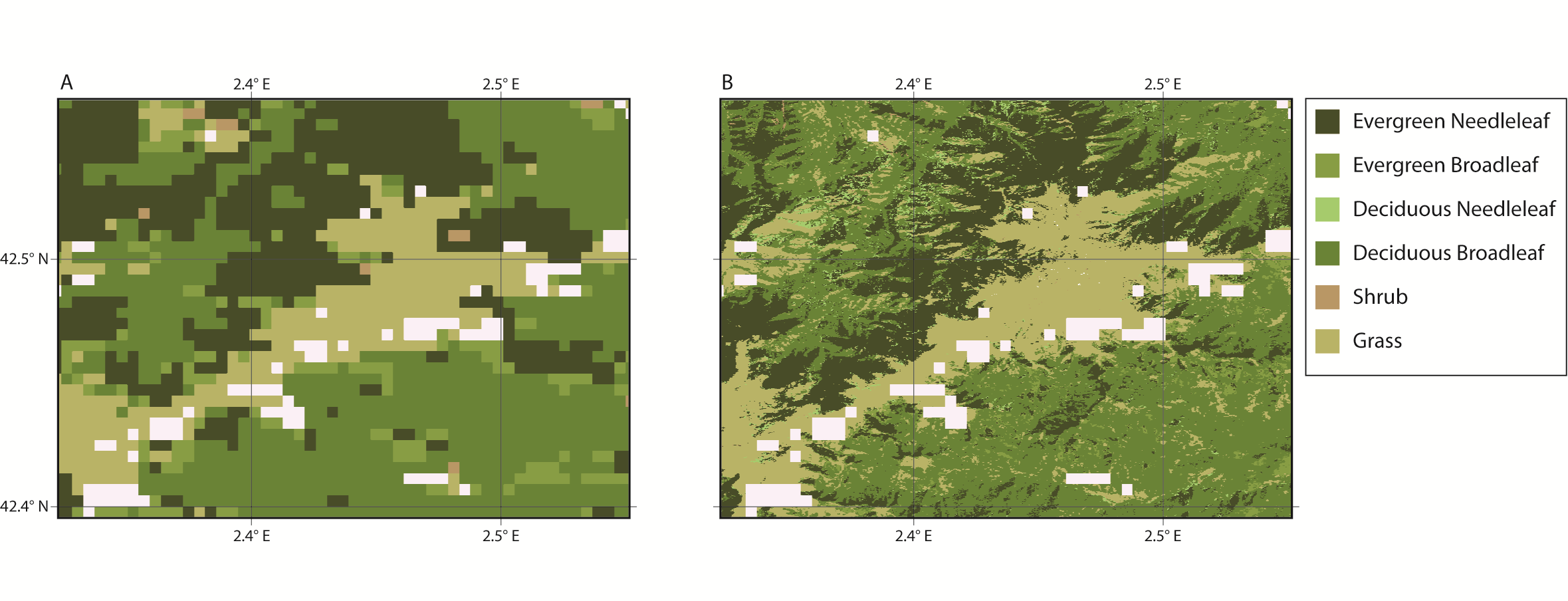}
  \caption{Comparison of the original MODIS PFT Type 5 classification scheme (a) and the RF-generated map using high-resolution landsat spectral information (b).}\label{fig:HRlandcover}
\end{figure}

As an example, Figure \ref{fig:HRlandcover} shows the original MODIS land cover and our estimate from RF in a selected heterogeneous mountain region in the northern part of the Iberian peninsula. The visual comparison of both images confirms the ability of the developed classifier to reproduce the original MODIS data. More importantly, the developed classification map allows us to extract the proportions of the different PFTs composing the 500 m MODIS pixel by exploiting the finer spatial resolution of Landsat imagery.

\subsection{Precision of the trait prediction models}

All the input features (remote sensing, elevation and climatic predictors) detailed previously were standardized \citep{friedman2009elements} before model training. RF model parameters were optimized to minimize cross-validation error following the above mentioned approach (section \ref{gap filling}). We split the data into a cross-validation set containing 80\% of the data to select model parameters, and the remaining 20\% acted as an independent, out-of-sample test set where we evaluate model's performance. \ref{ap:MLmethods} includes a comparison and analysis of robustness among different regression models also considered in the present paper. 

RF results for all plant traits are given in Table~\ref{tab:accuracyRF}. Results show an overall low biased for the RF method.
We also tested the model's performance in \ref{ap:MLmethods} with more difficult scenarios in which a reduced number of training samples were used. RF performed similarly and reported higher accuracies across all the reduced-sized training data rates. These results are in agreement with recent literature which reveals random forests to be highly efficient in other remote sensing and geosciences problems~\citep{Tramontana2015360,Tramontana16bg}. RF also have additional advantages over competing methods: they are fast to train and test, they can be easily implemented in parallel, and can work with missing data and features naturally.

\begin{table}[t!]
\small
\caption{Results in the cross-validation set, scores, and plant traits. 80\% of the data were used to select model parameters, and the remaining 20\% acted as an independent, out-of-sample test set to evaluate model's performance.}
\label{tab:accuracyRF}
\begin{center}
\renewcommand{\arraystretch}{0.8}
\begin{tabular}{|p{2cm}|p{2cm}|p{2cm}|}
\hline
\hline
 {\bf ME}	 & {\bf RMSE} & {\bf R} \\
\hline
\hline
\multicolumn{3}{|l|}{Specific Leaf Area (SLA), $n=4407$} \\
\hline
 -0.031 & { 3.185}	 & { 0.763} \\
\hline
\multicolumn{3}{|l|}{Leaf Nitrogen Concentration (LNC), $n=4422$} \\
\hline
 { -0.029} & 2.298	 & 0.734 \\
\hline
\multicolumn{3}{|l|}{Leaf Phosphorus Concentration (LPC),  $n=3851$} \\
\hline
 {0.001}	 & { 0.132}	 & { 0.778} \\
\hline
\multicolumn{3}{|l|}{Leaf Nitrogen-Phosphorus ratio (LNPR), $n=2074$} \\
\hline
 0.016	 & { 1.806}	 & { 0.781} \\
\hline
\multicolumn{3}{|l|}{Leaf Dry Matter Content (LDMC), $n=1842$}\\
\hline
 { 0.000}	 & { 0.038}	 & 0.718 \\
\hline
%Fresh mass (FM),  $n=1086$  &  & & \\
%\hline
%RLR	 & {\bf 0.006}	 & 0.408	 & 0.528 \\
%RF	 	 & -0.008 & 0.408	 & 0.586 \\
%ELM	 & 0.009	 & 0.424	 & 0.480 \\
%KRR	 & 0.009	 & {\bf 0.398}	 & {\bf 0.598} \\
%GPR	 & 0.025	 & 0.435	 & 0.405 \\
%\hline
\hline
\end{tabular}
\end{center}
\end{table}

\begin{figure}[h!]
\small
\begin{center}
\setlength{\tabcolsep}{5pt}
\begin{tabular}{ccc}
SLA, R=0.76$\pm$0.01   & LNC, R=0.75$\pm$0.01   &  LPC, R=0.76$\pm$0.02 \\
\includegraphics[width=4cm,trim={3cm 0cm 1.5cm 0}]{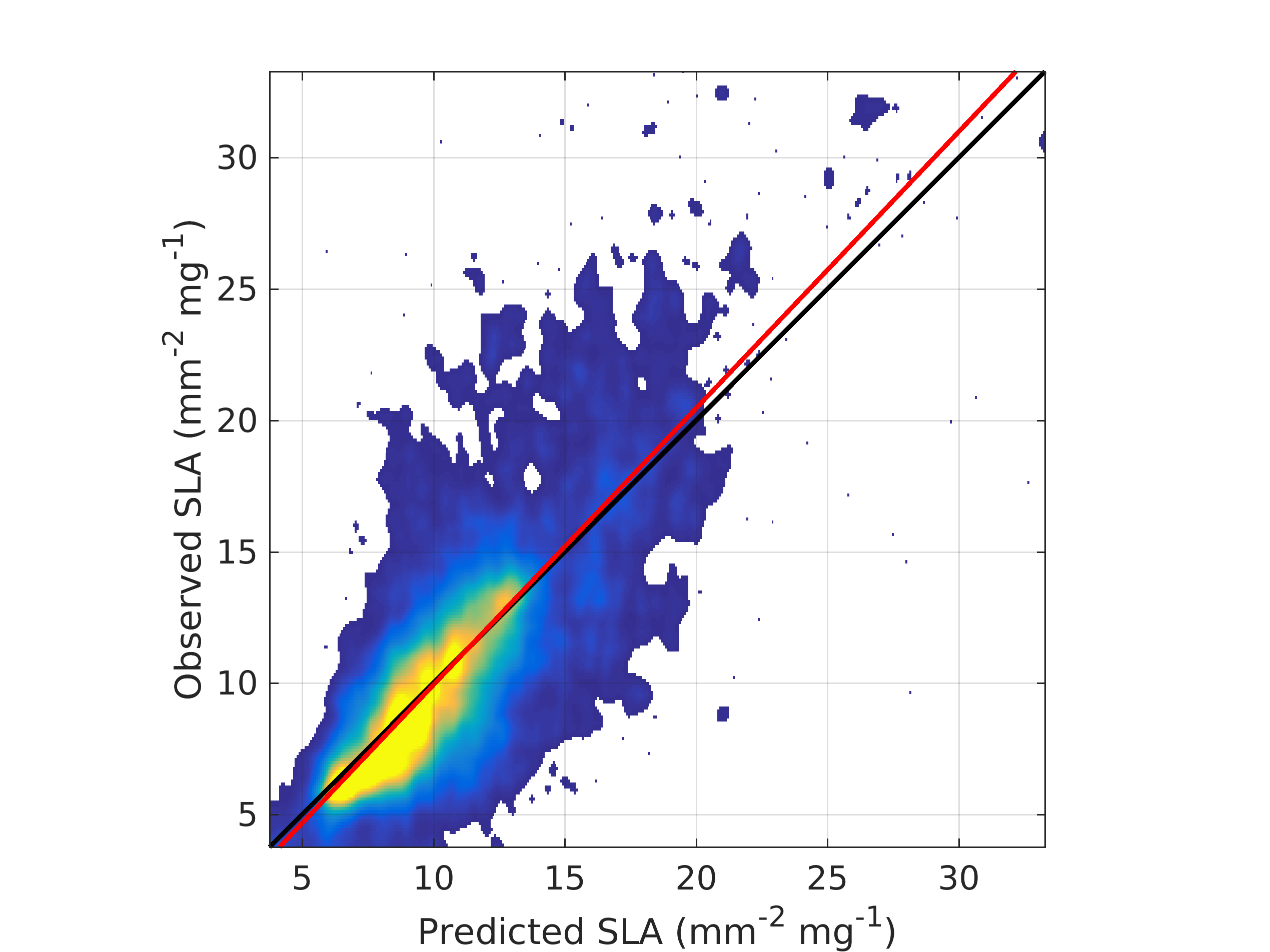} &
\includegraphics[width=4cm,trim={3cm 0cm 1.5cm 0}]{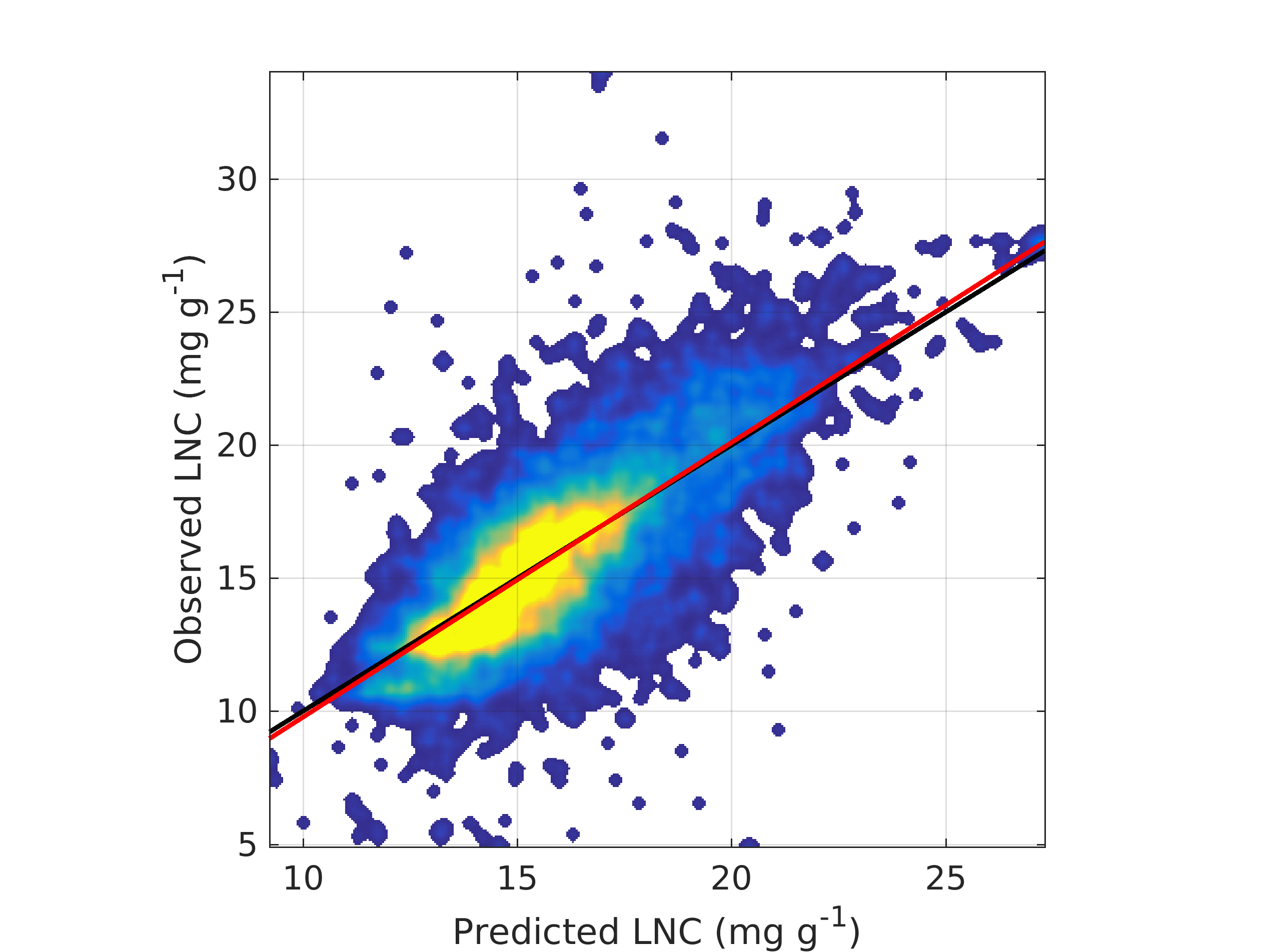} &
\includegraphics[width=4cm,trim={3cm 0cm 1.5cm 0}]{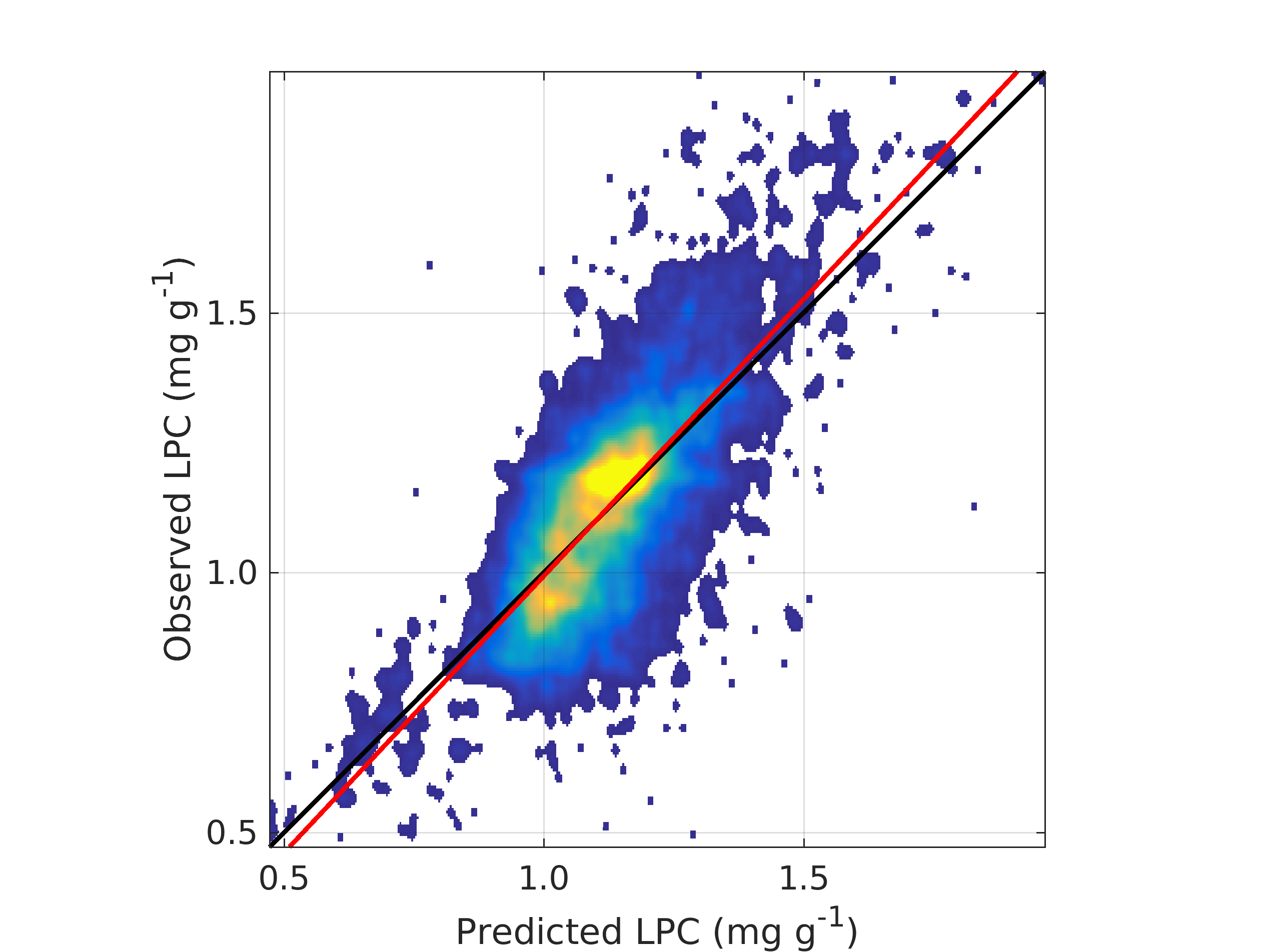} \\
LNPR, R=0.77$\pm$0.02  & LDMC, R=0.76$\pm$0.02  & \\%{\bf FMNC, R=0.76$\pm$0.02} \\
\includegraphics[width=4cm,trim={3cm 0cm 1.5cm 0}]{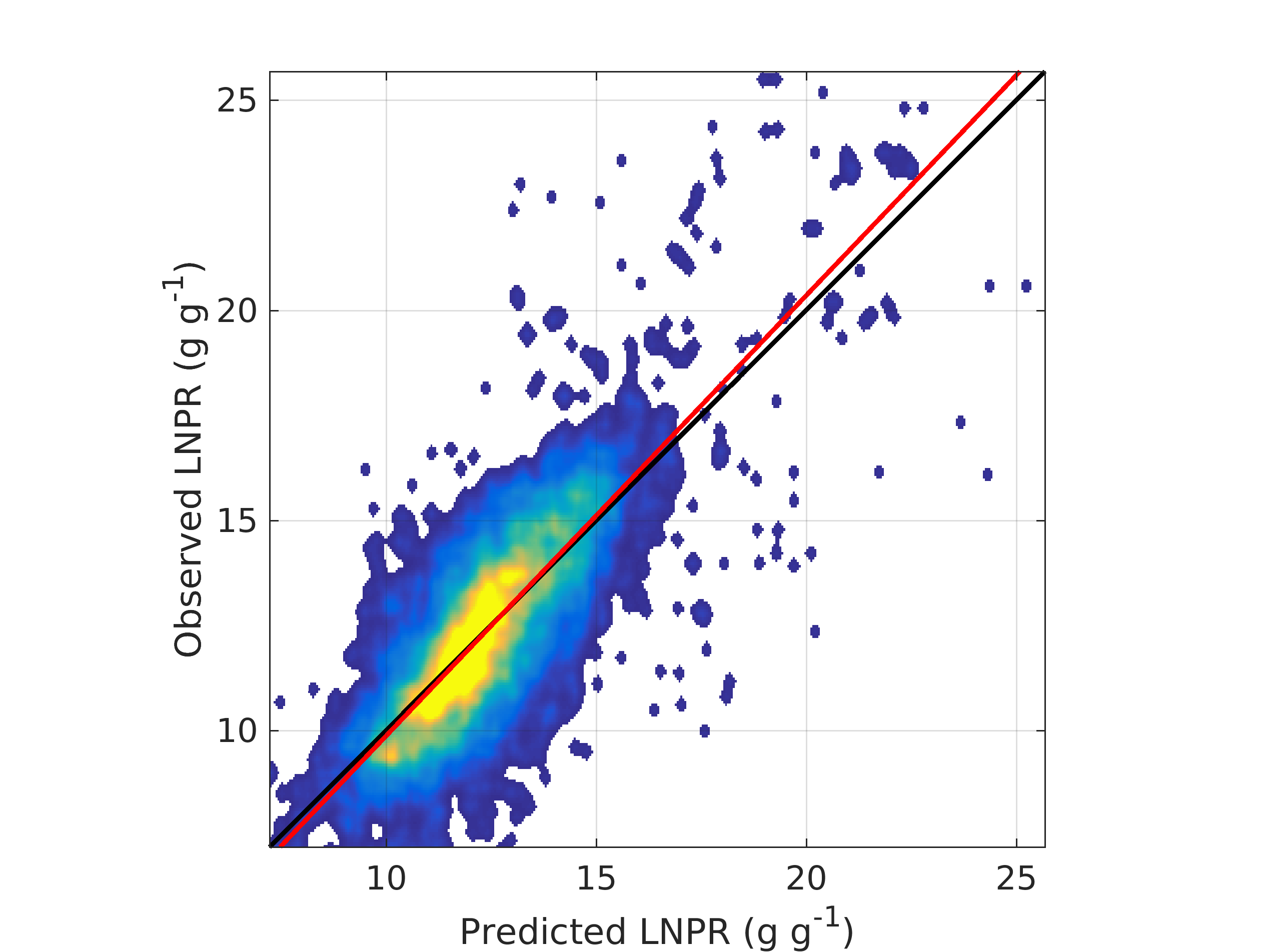}  &
\includegraphics[width=4cm,trim={3cm 0cm 1.5cm 0}]{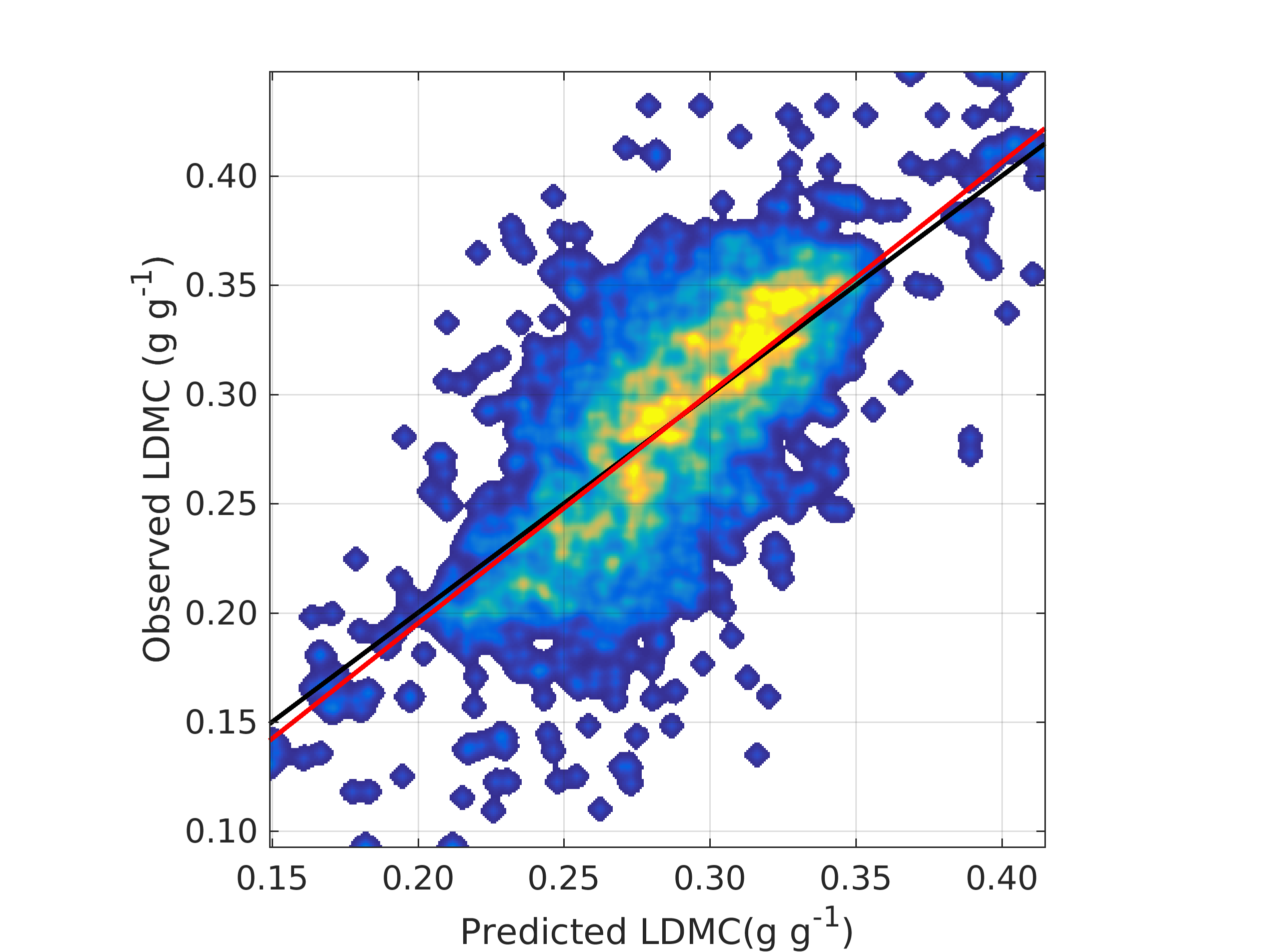} &
\end{tabular}
\end{center}
\vspace{-0.25cm}
\caption{Scatter plots of the predicted versus observed pixel trait values obtained by the RF regression models. The best linear fit (red) and the one-to-one line (black) are shown for reference. We also give the average and standard deviation Pearson's correlation coefficient R over $20$ realizations in the test set.}\label{fig:scatter}
\end{figure}

Figure~\ref{fig:scatter} shows the predicted-versus-observed scatter plots obtained by the RF model in the cross-validation sets. Good correlations and virtually no bias is observed for all trait models. It should be noted that the best linear fit (red) and the one-to-one line (black) are almost coincident for all traits. Our tests also revealed little variation per realization and a low standard deviation of the correlation coefficient over the different realizations, corroborating a great consistency in the chosen modelling approach.

\begin{figure}[t!]
\small
\begin{center}
\setlength{\tabcolsep}{10pt}
\begin{tabular}{ccc}
\rotatebox{90}{SLA} & \includegraphics[width=3.7cm]{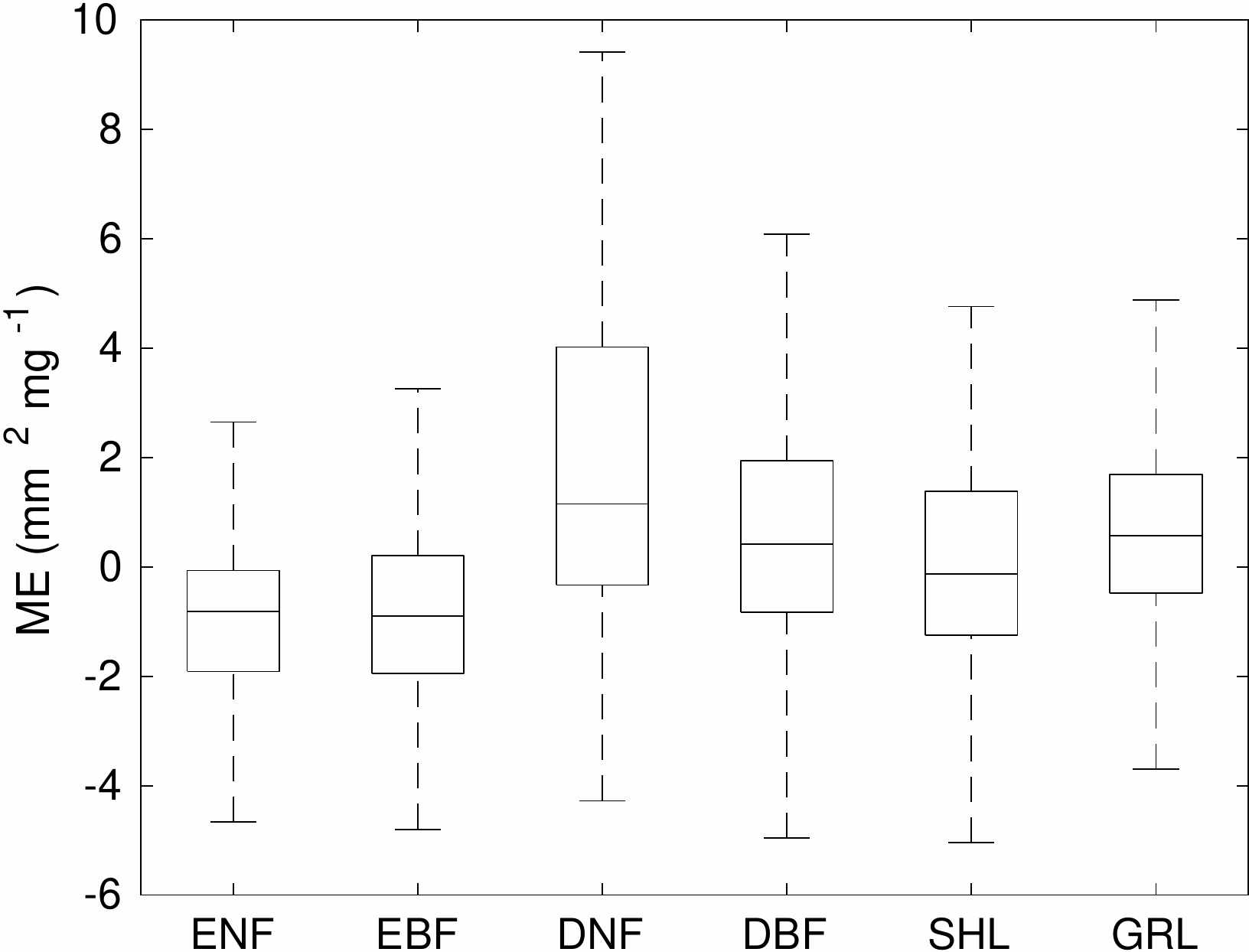}  &
\includegraphics[width=3.7cm]{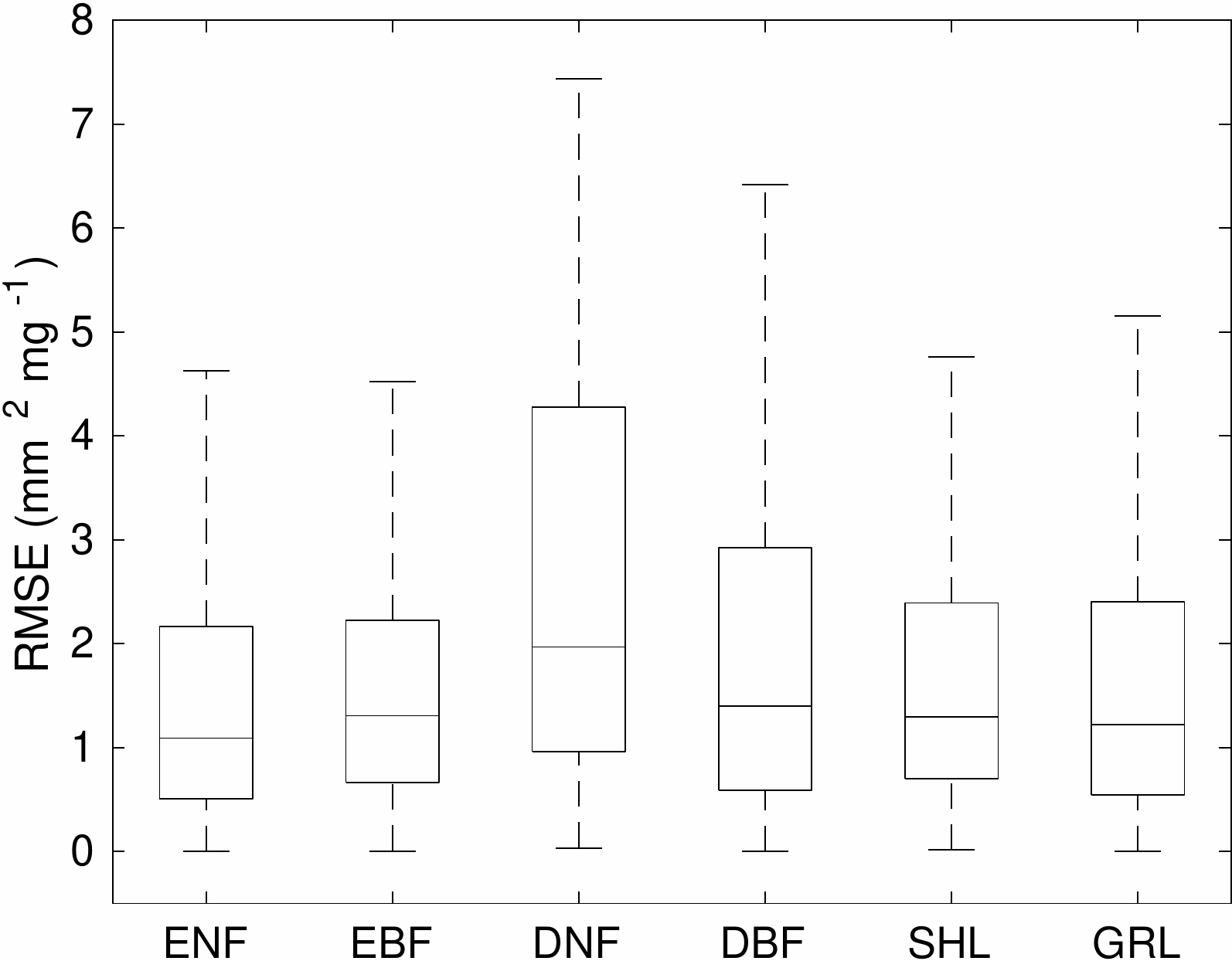} \\
\rotatebox{90}{LNC} & \includegraphics[width=3.7cm]{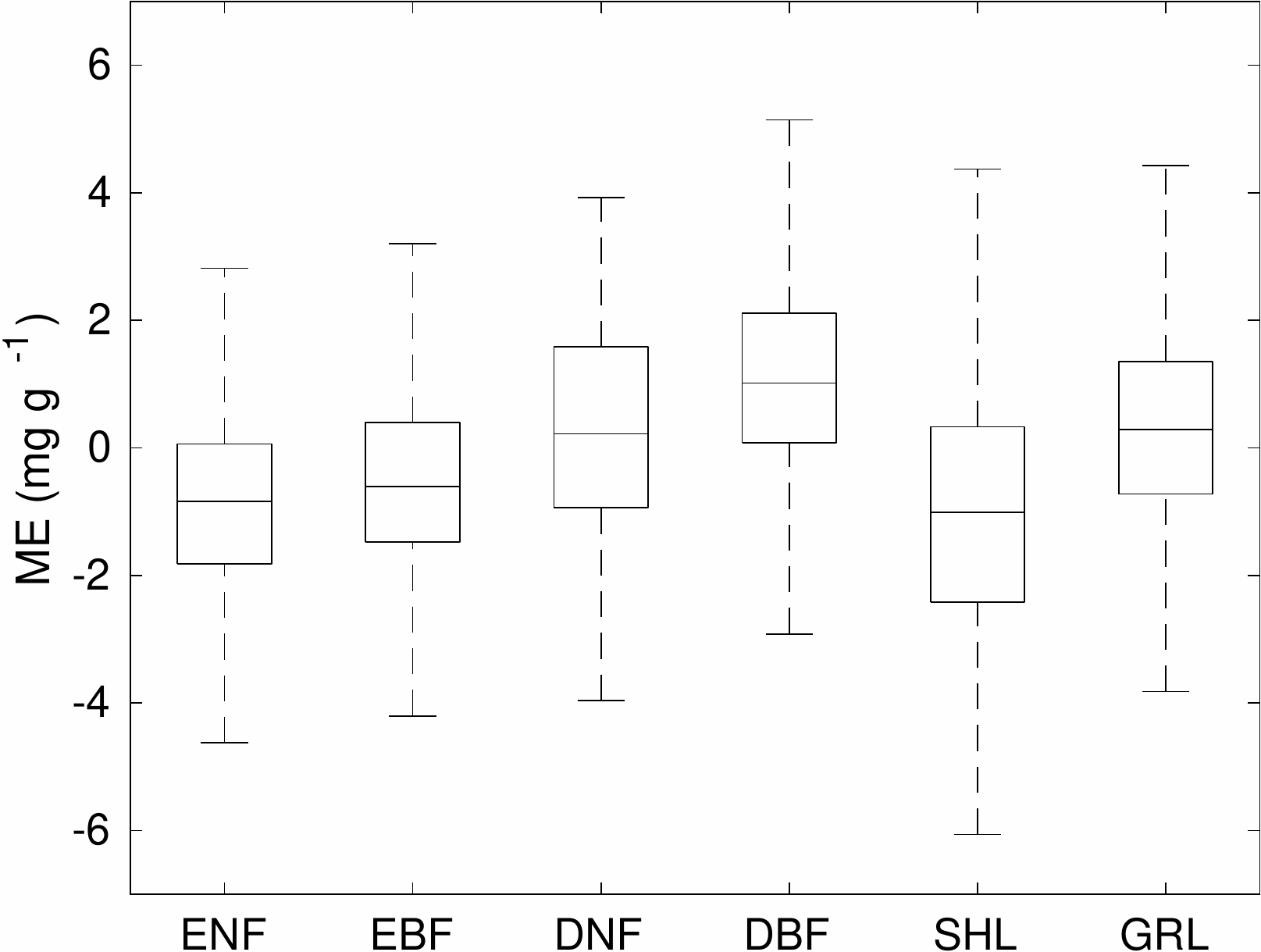}  &
\includegraphics[width=3.7cm]{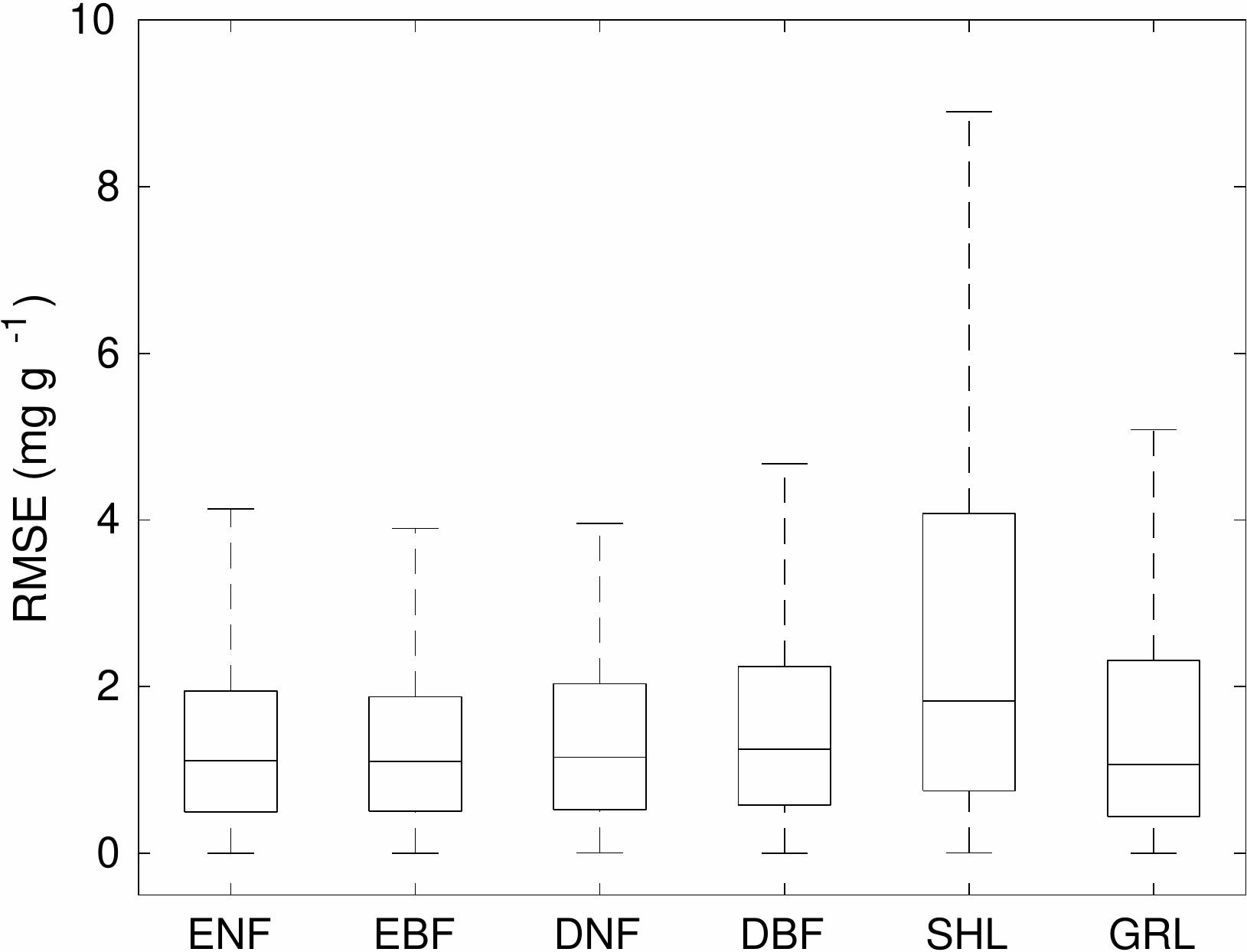} \\
\rotatebox{90}{LPC} & \includegraphics[width=3.7cm]{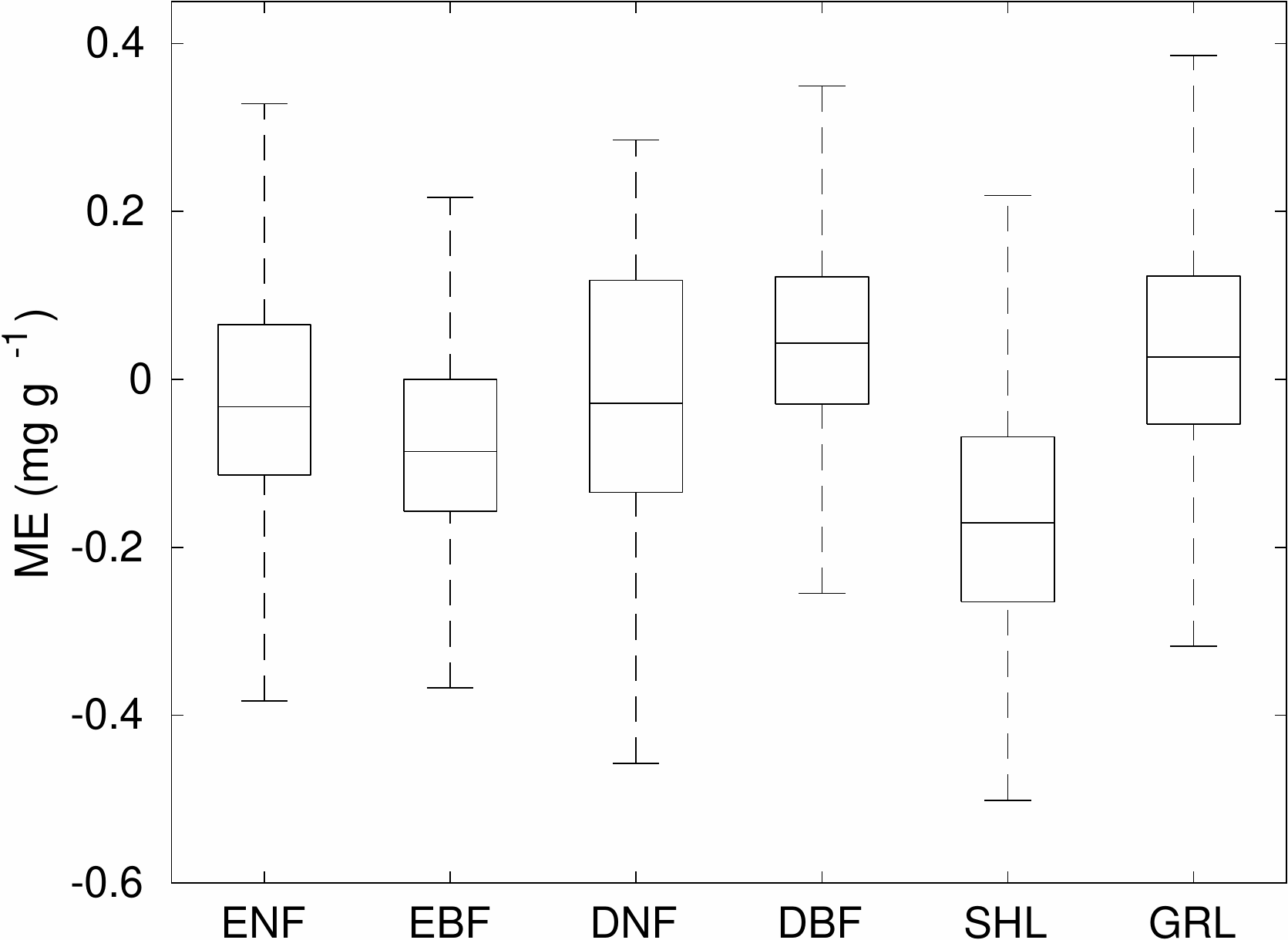}  &
\includegraphics[width=3.7cm]{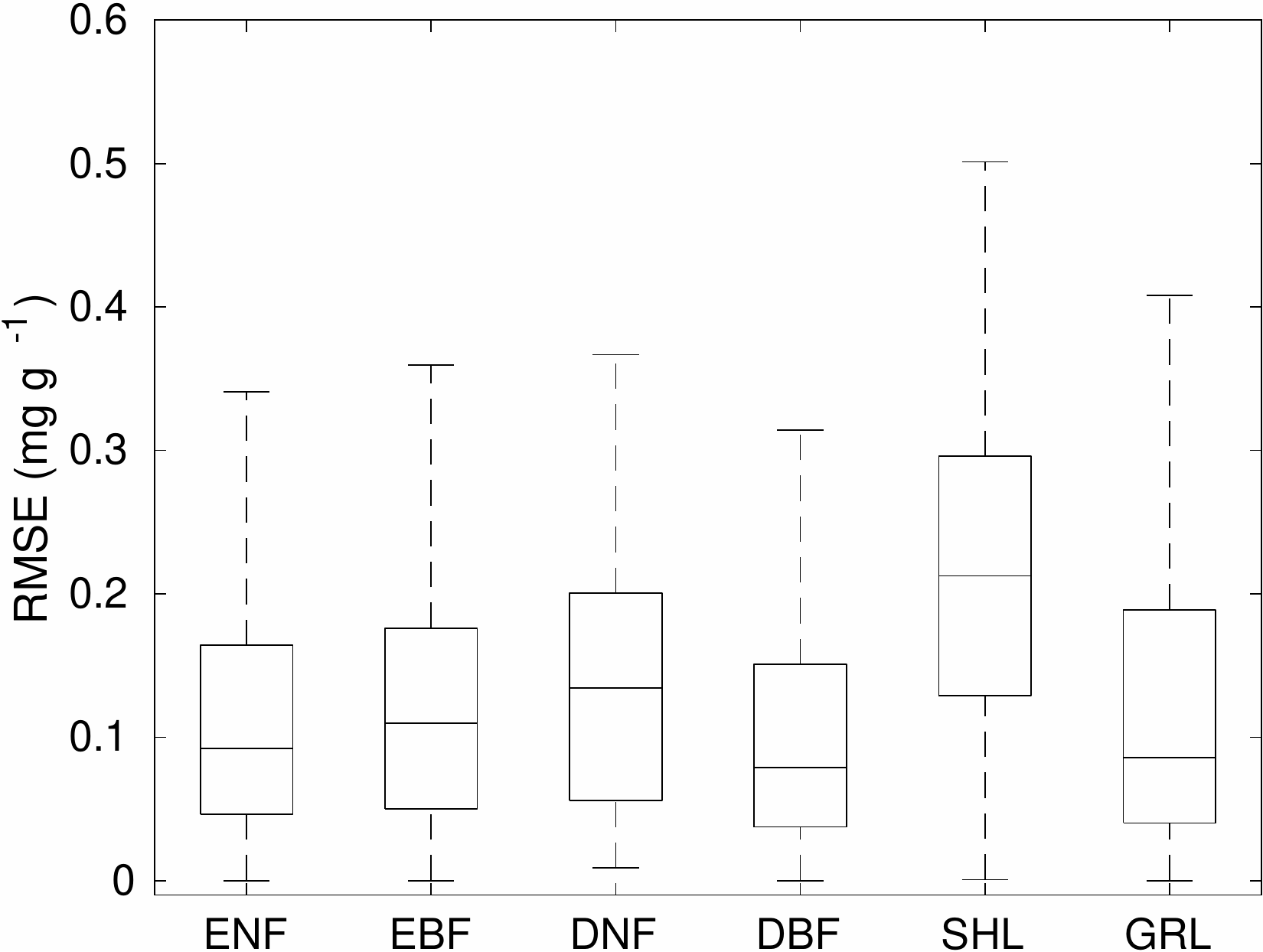} \\
\rotatebox{90}{LDMC} & \includegraphics[width=3.7cm]{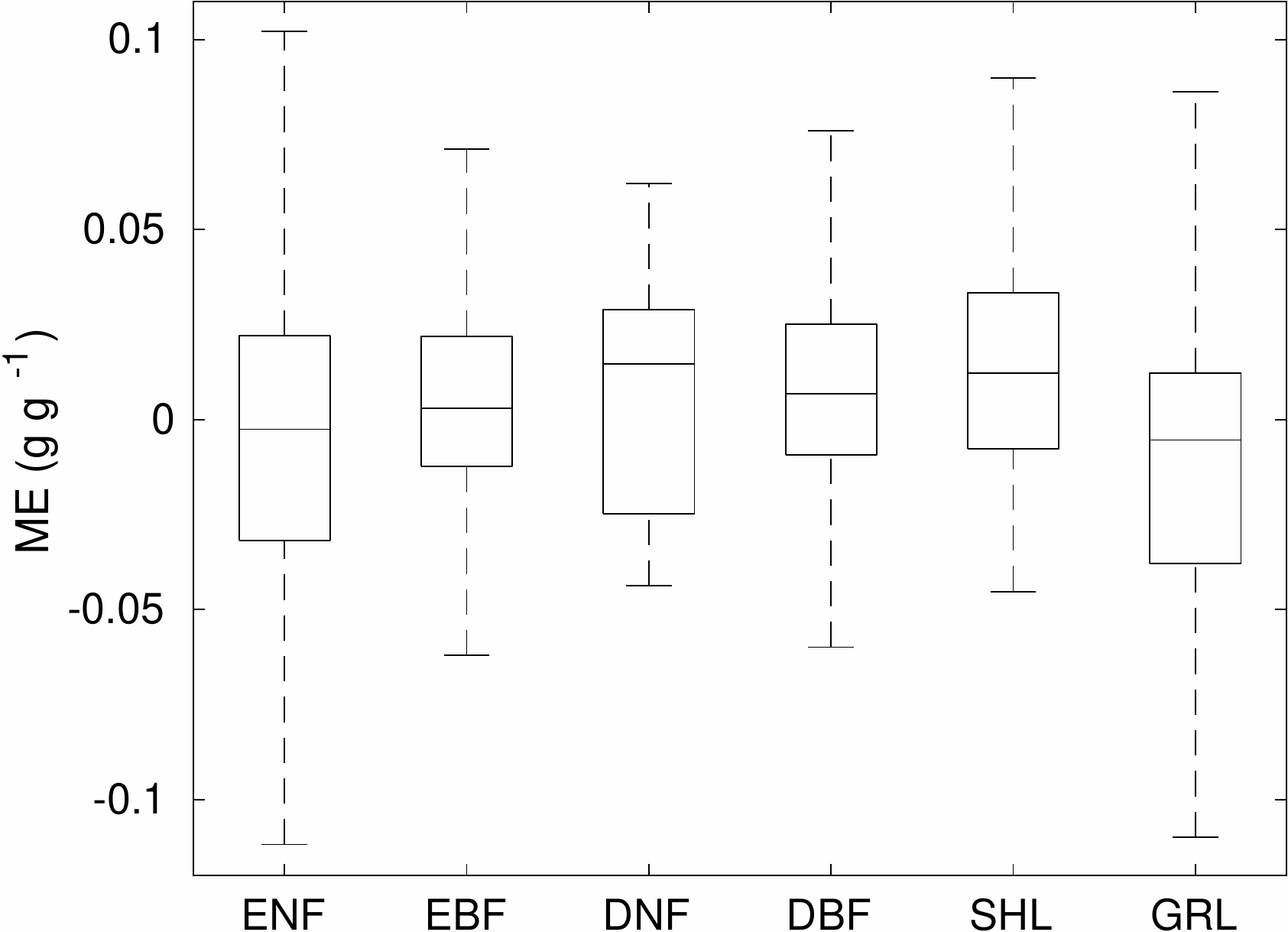}  &
\includegraphics[width=3.7cm]{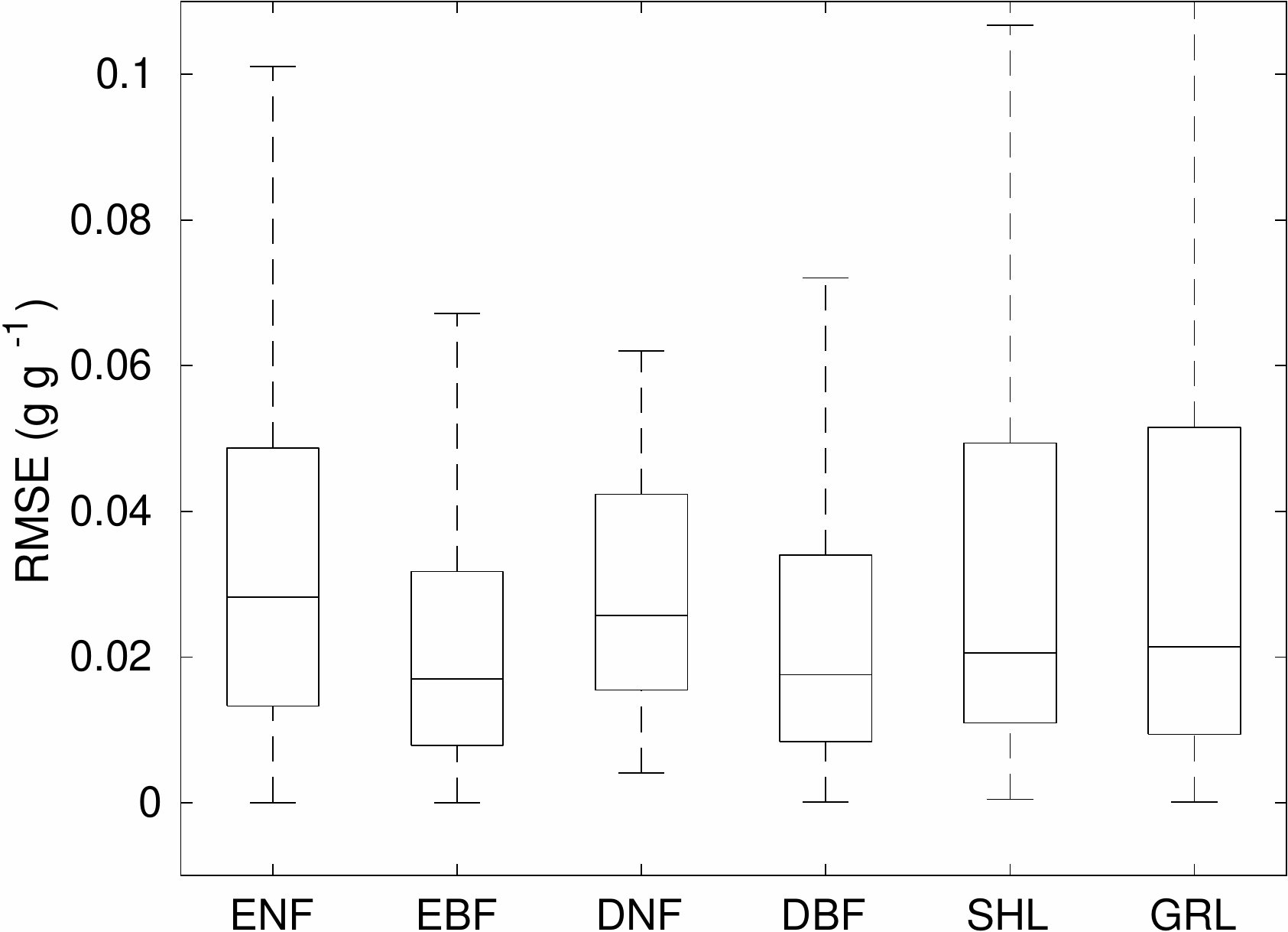} \\
\rotatebox{90}{LNPR} & \includegraphics[width=3.7cm]{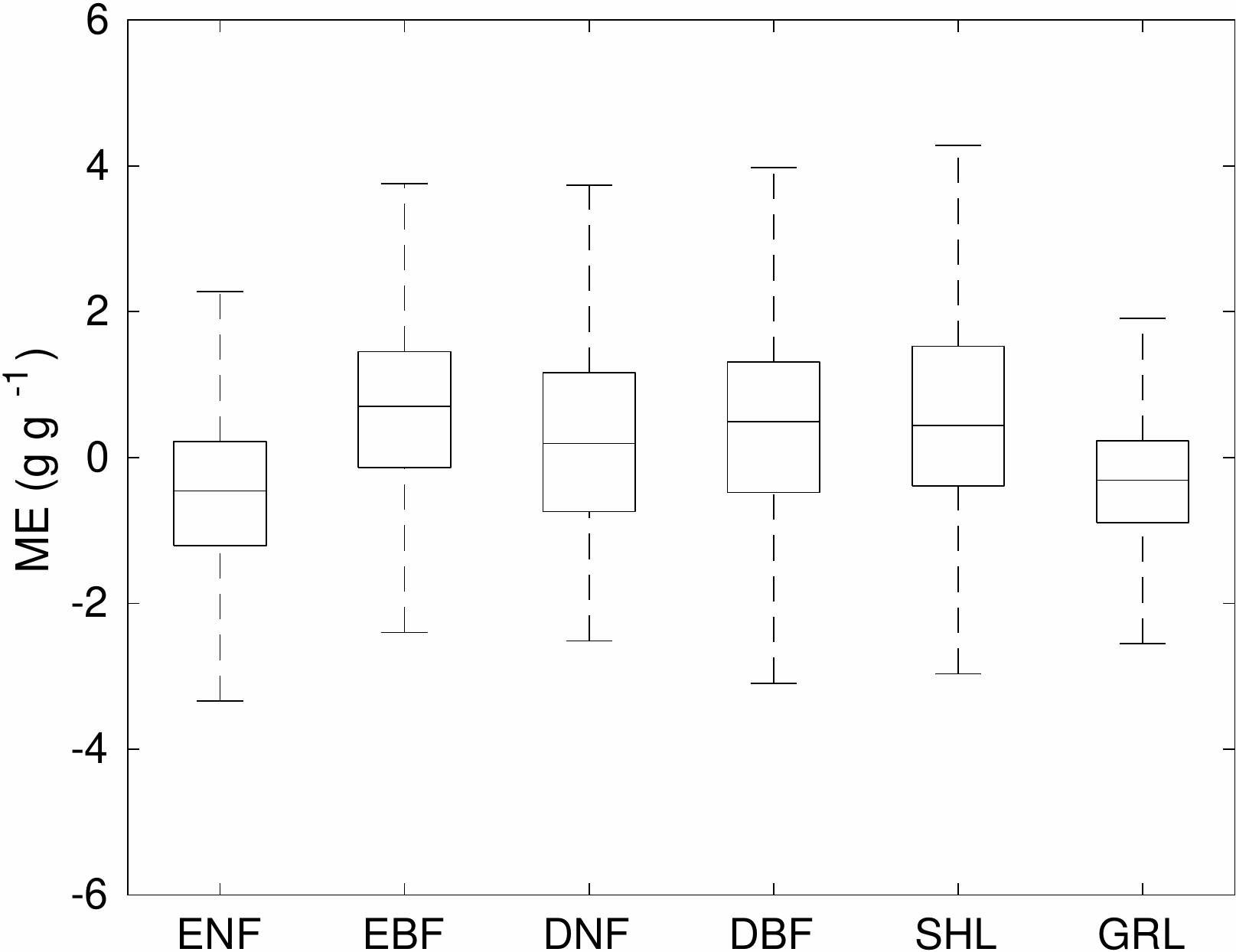}  &
\includegraphics[width=3.7cm]{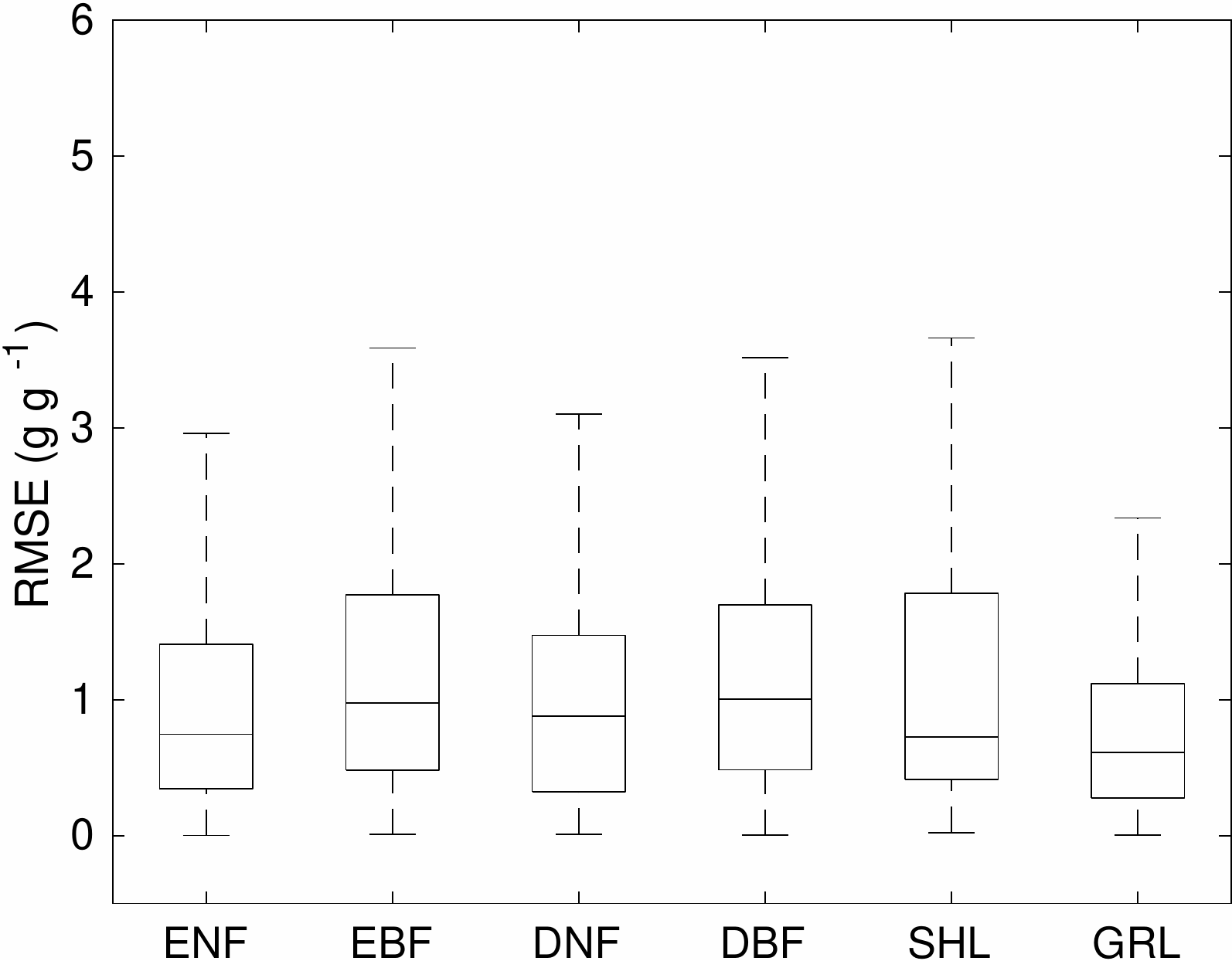} \\
\end{tabular}
\end{center}
\caption{Mean error (left) and root mean squared error (right) for each PFT in the training dataset and plant trait.}\label{fig:boxplots}
\end{figure}

In Figure \ref{fig:boxplots} we present the error estimates for the dominant PFT over the training dataset computed by means of the above mentioned cross-validation process. The median values of the boxplots of the residuals are distributed around zero indicating that the regression algorithm is performing adequately. Evergreen PFTs present the lowest errors in SLA estimates while deciduous needle-leaf forest has the highest bias and RMSE for that trait. LNC and LPC errors tend to be low for forest PFTs.  In contrast, the corresponding shrub-land error estimates are higher according to the statistics shown. LDMC residuals have the largest uncertainties in evergreen needle-leaf forests and grasslands while LNPR errors show the opposite behavior over the same PFTs. The small range of residuals, about 10\% on average (Figure \ref{fig:boxplots} right side), and the lack of a consistent pattern of the residuals among the different PFTs for the considered traits (Figure \ref{fig:boxplots} left side) suggests that the proposed methodology overall produces good estimates independently of the vegetation types present. However, we note that for some PFTs the non-symmetric distribution of residuals around zero indicates that the link between PFT, spectra and climatic variables may introduce potentially significant biases. This is an expected limitation of the proposed modelling approach, which was designed to be global and no PFT-specific.  Apart from uncertainties in the data, the PFT attribution may also contribute to this error.

\subsection{Global trait maps visualization}

\begin{figure}[p!]
\small
\begin{center}
\setlength{\tabcolsep}{10pt}
\begin{tabular}{ccc}
\rotatebox[origin=c]{90}{SLA} & \includegraphics[width=5.5cm,trim={3cm 2cm 0cm 0}]{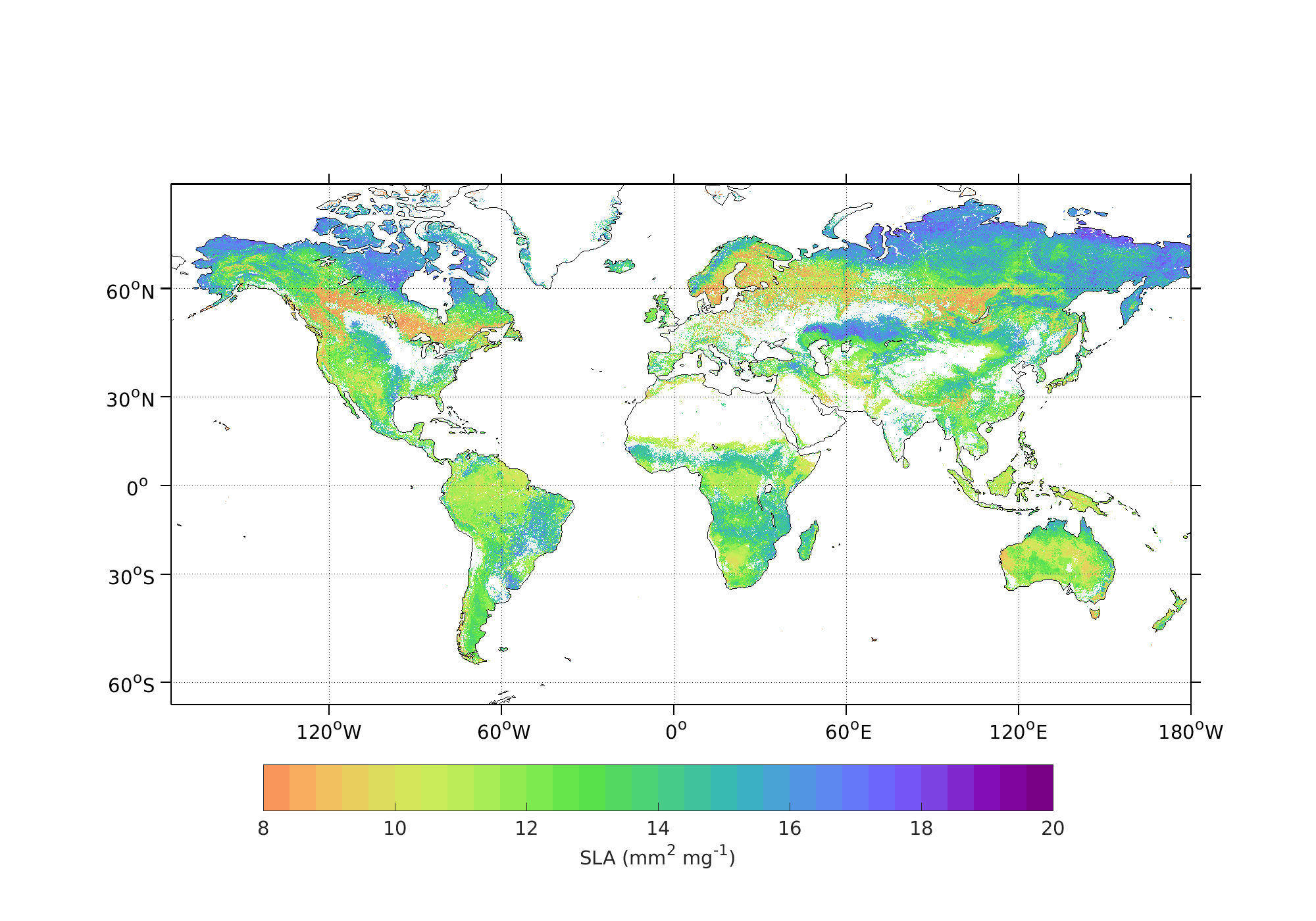} &
\includegraphics[width=5.5cm,trim={3cm 2cm 0cm 0}]{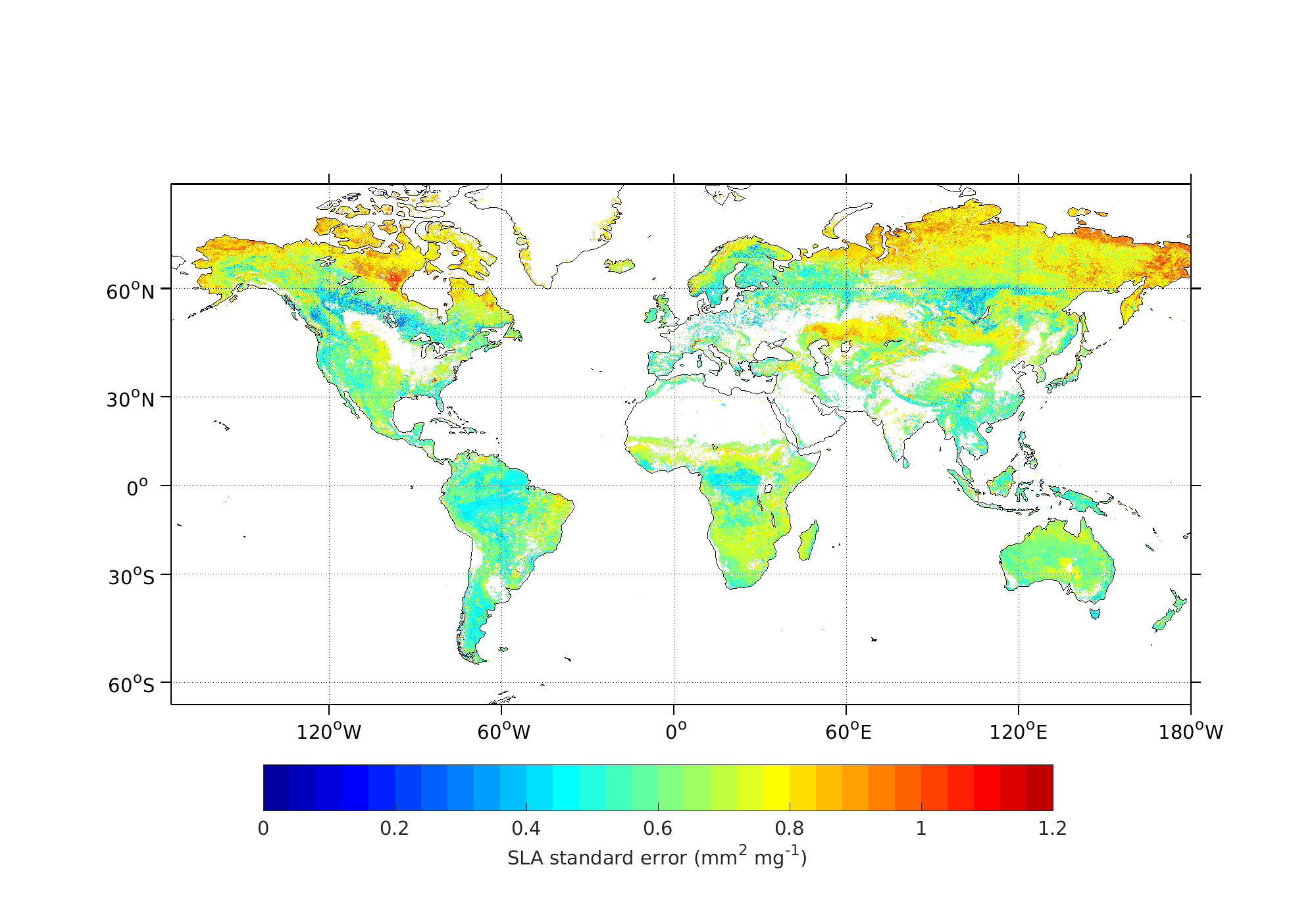} \\
\rotatebox{90}{LNC} & \includegraphics[width=5.5cm,trim={3cm 2cm 0cm 0}]{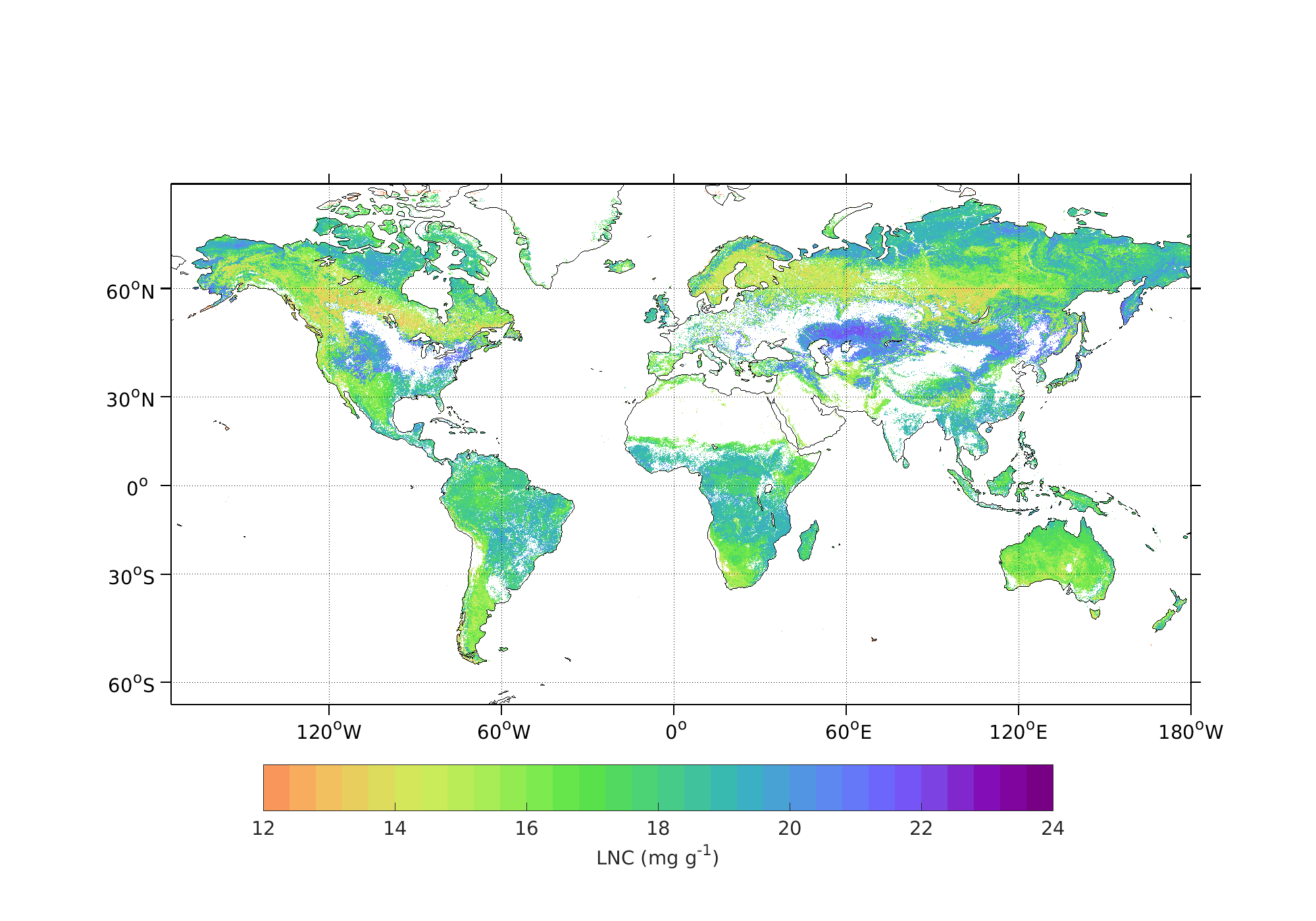} &
\includegraphics[width=5.5cm,trim={3cm 2cm 0cm 0}]{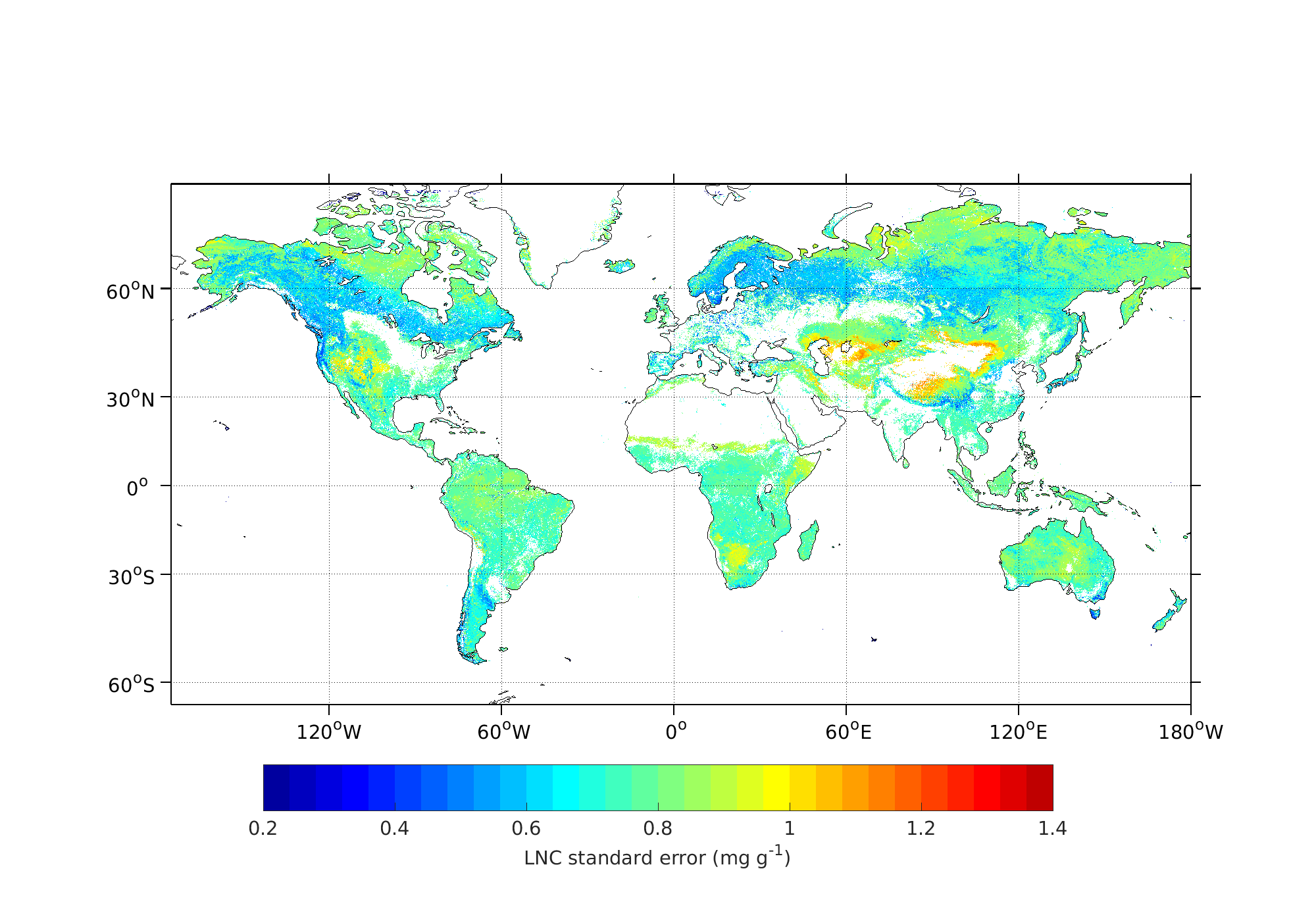} \\
\rotatebox{90}{LPC} & \includegraphics[width=5.5cm,trim={3cm 2cm 0cm 0}]{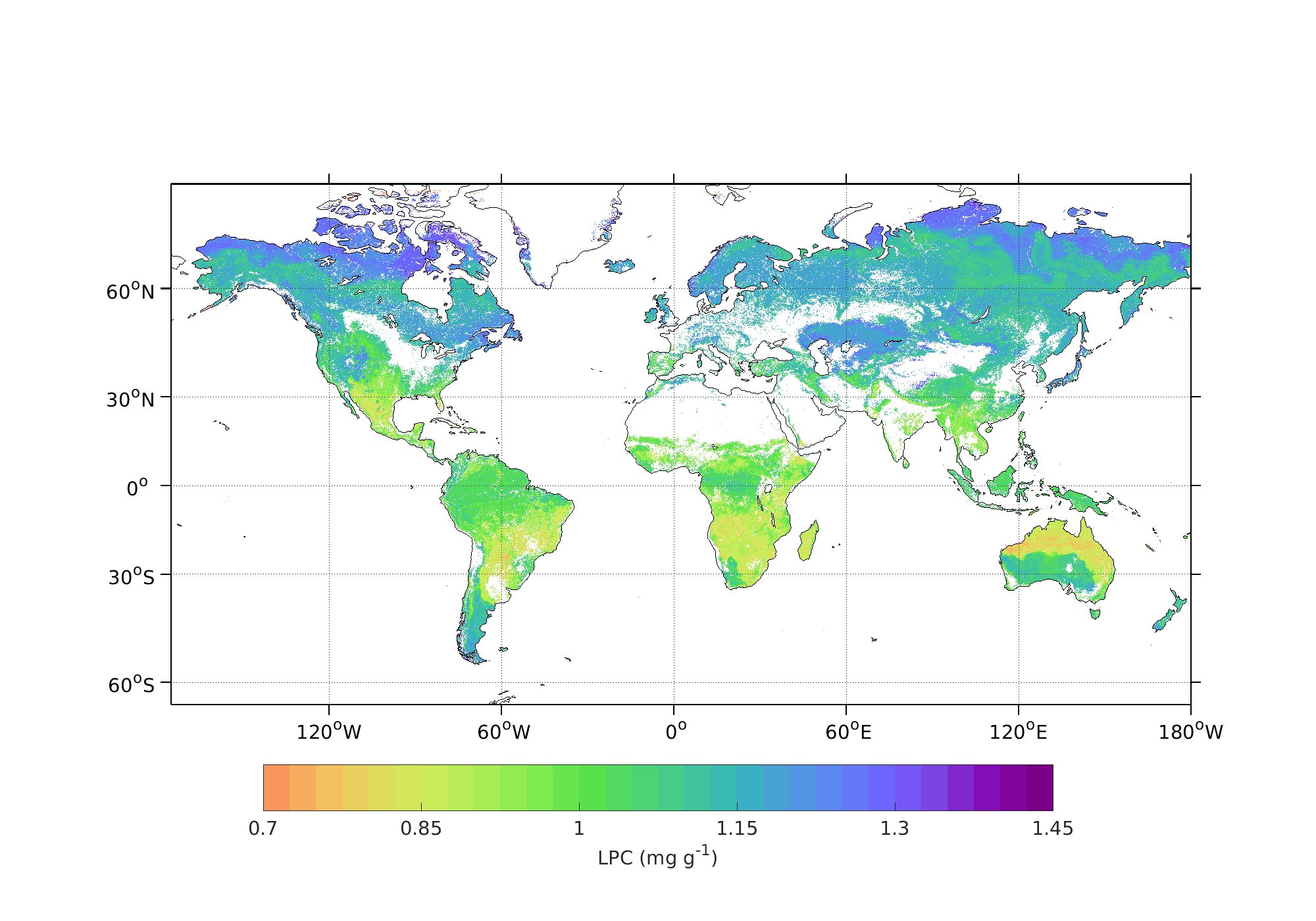} &
\includegraphics[width=5.5cm,trim={3cm 2cm 0cm 0}]{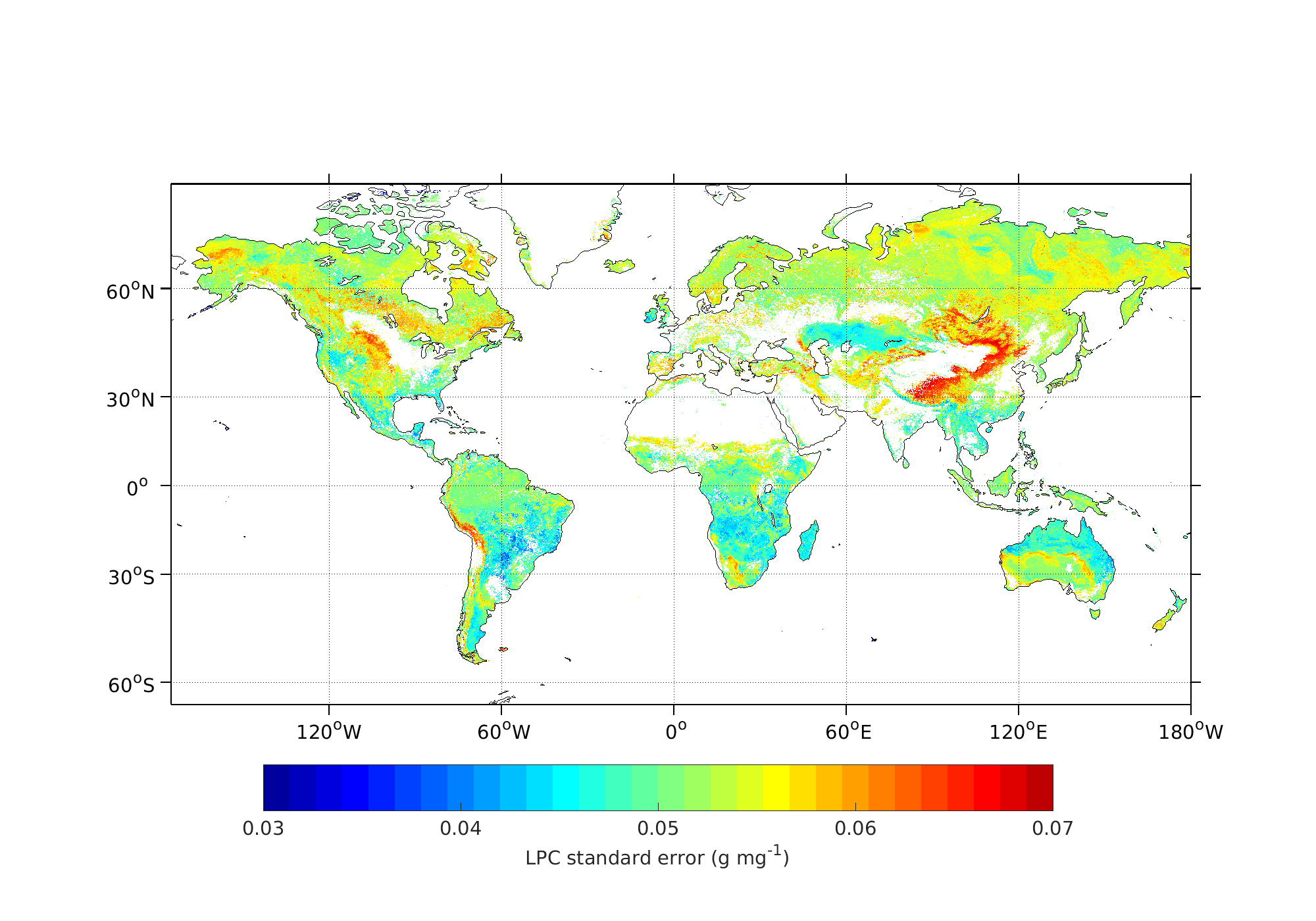} \\
\rotatebox{90}{LDMC} &\includegraphics[width=5.5cm,trim={3cm 2cm 0cm 0}]{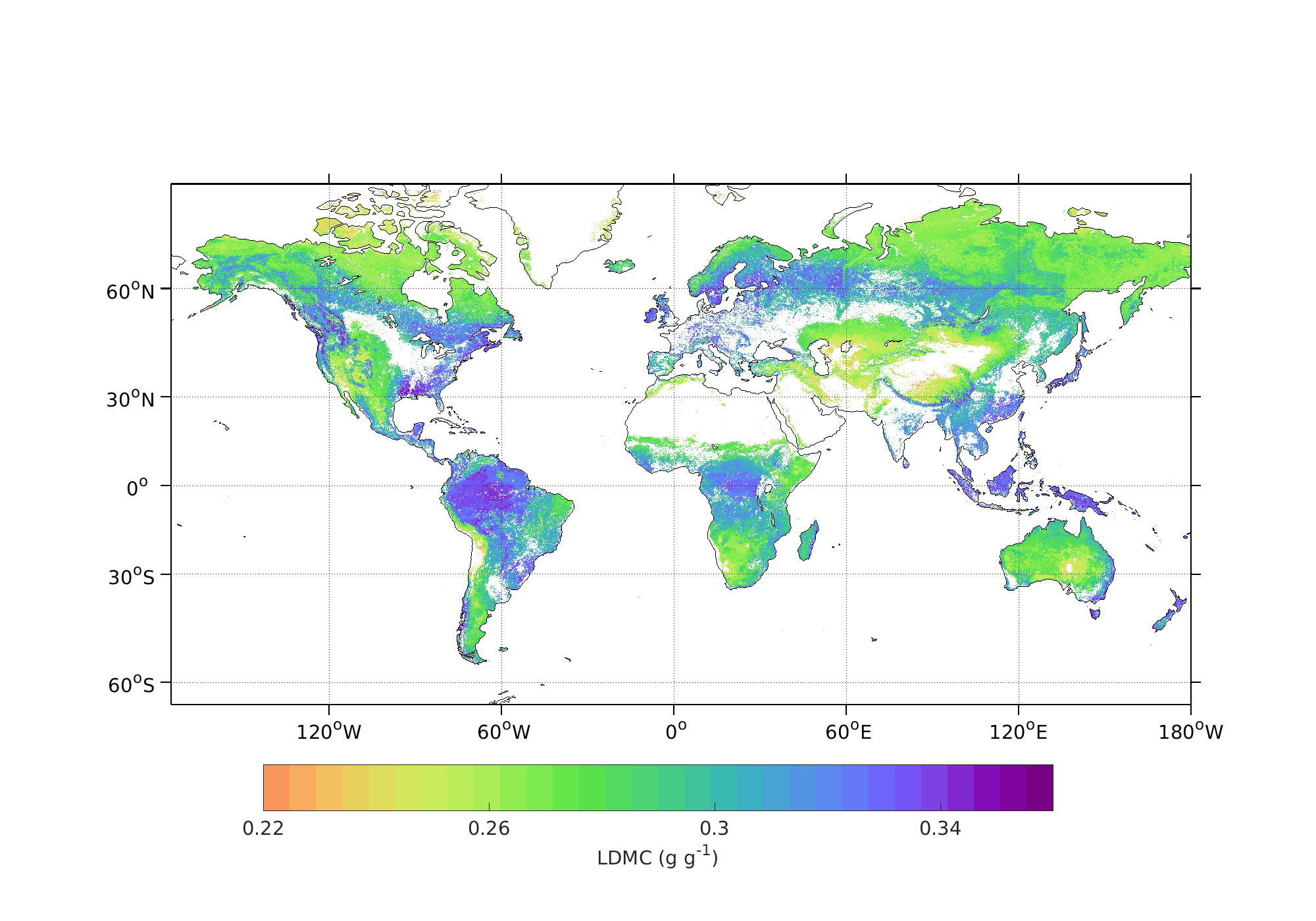} &
\includegraphics[width=5.5cm,trim={3cm 2cm 0cm 0}]{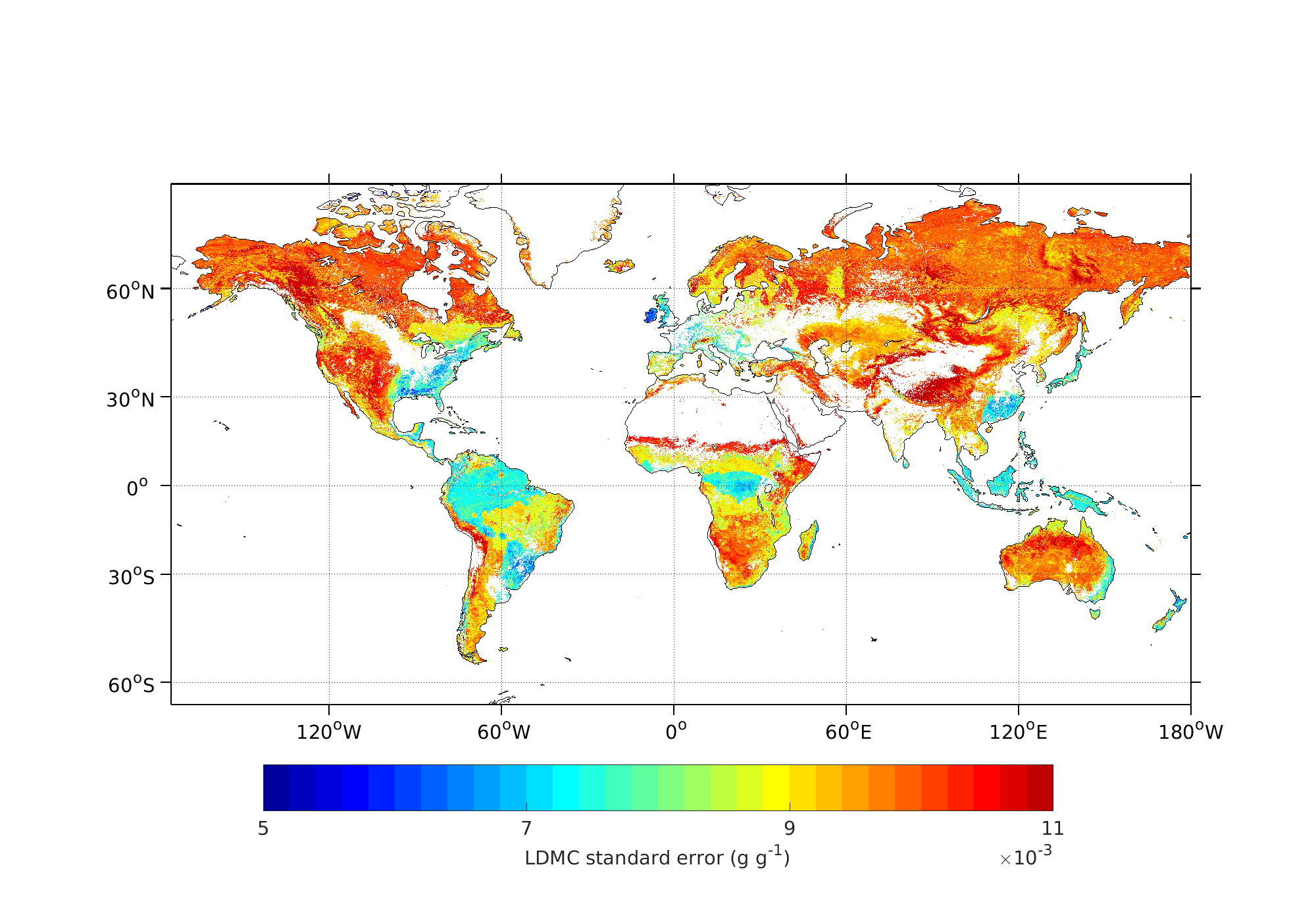} \\
\rotatebox{90}{LNPR} &\includegraphics[width=5.5cm,trim={3cm 2cm 0cm 0}]{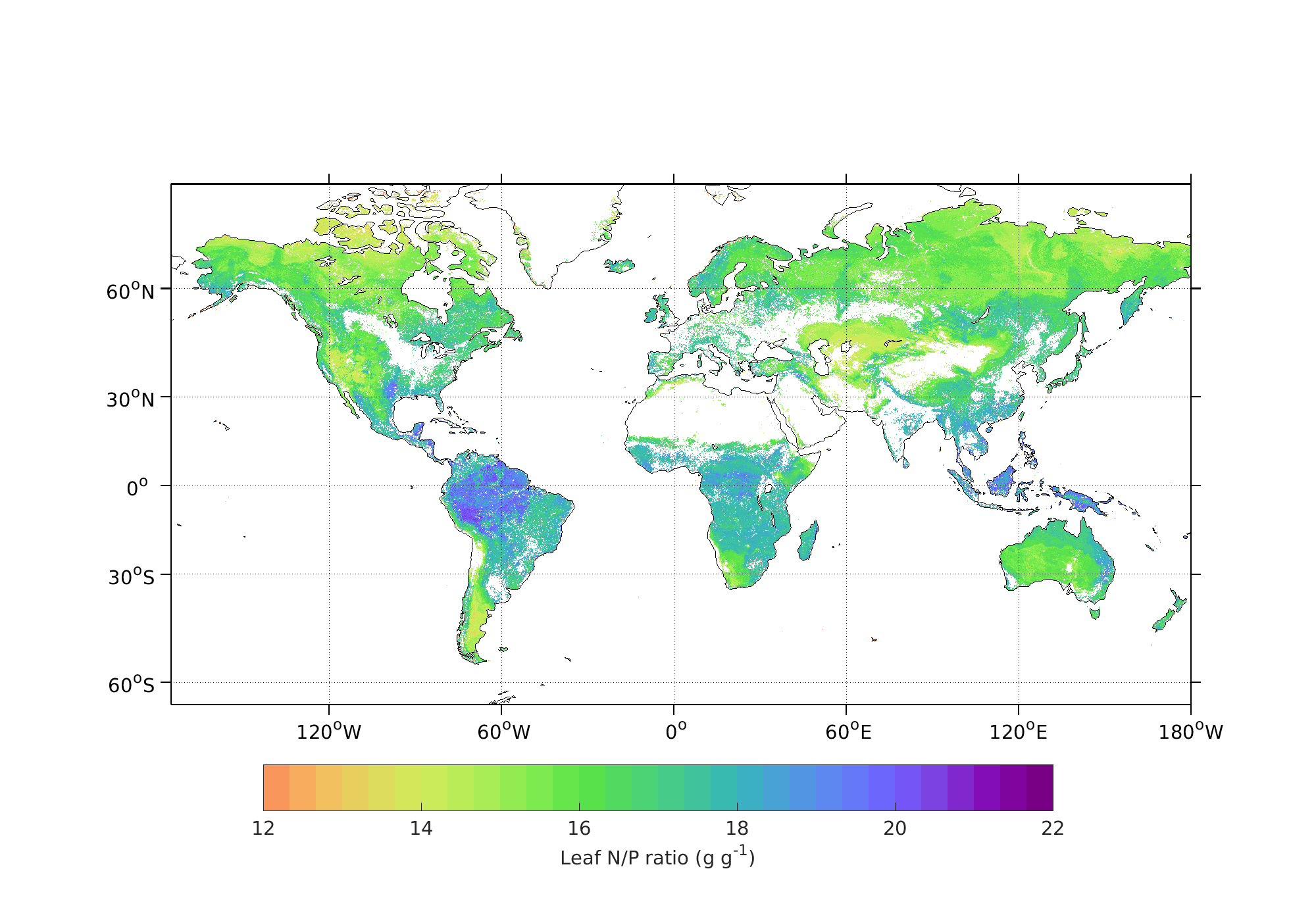} &
\includegraphics[width=5.5cm,trim={3cm 2cm 0cm 0}]{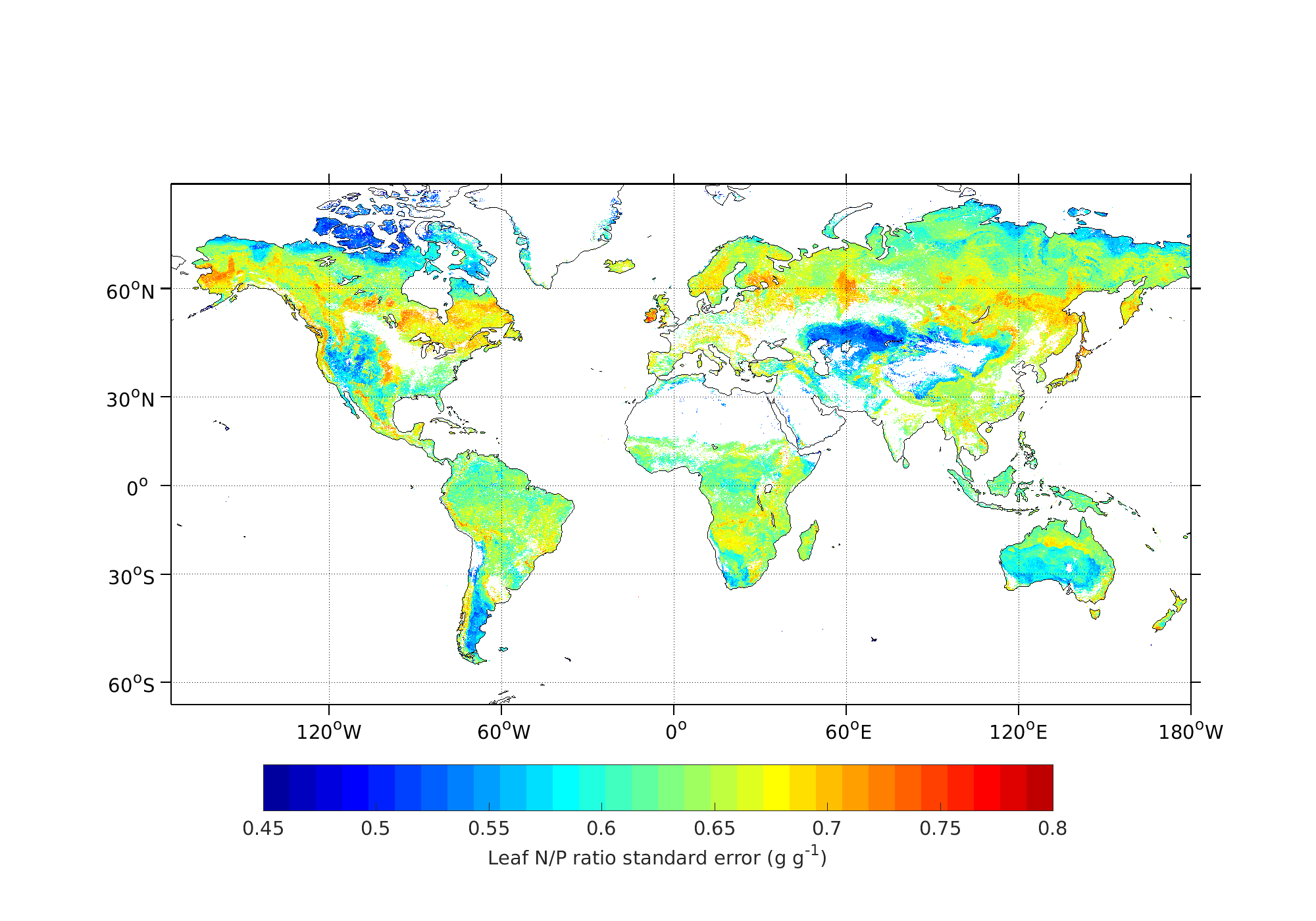} \\
\end{tabular}
\end{center}
\caption{Global estimates of plant traits (left) and predicted standard error (right) for different leaf traits.}\label{fig:globalmaps}
\end{figure}

In Figure~\ref{fig:globalmaps} we show the global trait maps resulting from applying the trained models. In addition to the trait estimates, we also include in the same figure the estimated predictive standard error maps obtained from the RF models. We computed this ancillary information layer for model evaluation.

According to our maps, tree species cover a range of SLA values between 7-16 $mm^{2} \, mg^{-1}$. Grasslands and savannas present the highest values (16-20). Evergreen broadleaved forests in the tropical rainforests present intermediate to low values (9-13) in our images. These forests occur in a belt around the Equator, with the largest areas in the Amazon basin of South America, the Congo basin of central Africa, and parts of the Malay Archipelago. On the other hand, the lowest values were found in the needle leaf evergreen trees (7-9) located in boreal forests. They occur in the more southern parts of the Taiga ecoregion that spreads across the northern parts of the world. Finally, note that shrub-lands present the largest variability and extreme values (9-20). The lowest values occur in Australia, South Africa and Mexico (9-13) and the highest are in the tundra region (16-20). Tundra vegetation is mainly composed of herbaceous plants, mosses, lichens, and broadleaved deciduous shrubs. We hypothesize that values for shrublands in tundra are higher because shrubs are mostly deciduous and because of the pervasive presence of grasses. SLA is inversely related with the dry mass cost to deploy new leaf area which intercepts light. Leaf longevity could be understood as the duration over which photosynthetic investment is returned. From a leaf economics perspective, our results corroborate the fact that SLA and leaf longevity are inversely related \citep{reich2004global}. Thus, in the calculated maps, evergreen vegetation (long leaf longevity) shows the lowest SLA  values (highest investment) while areas with annual grasses and deciduous vegetation (low leaf longevity) present the highest SLA (lowest investment). The obtained value ranges for the different PFTs are in good agreement with published results of other authors at leaf level \citep{poorter2009causes,wright2004worldwide,kattge2011try,wright2005assessing}.

LDMC maps show a high correlation with tree cover maps as expected. Note that leaves with high LDMC tend to be relatively tough, and are thus assumed to be more resistant to mechanical impacts (e.g. herbivores and wind). The estimated LPC map shows lower phosphorus concentration in vegetation closer to the Equator. In fact, broadleaved evergreen forests and non-deciduous shrub-lands are mainly located in those areas. In contrast, boreal and tundra areas present the highest values, which are mostly populated with high LPC concentration PFTs. Leaf P is lowest at warmer temperatures ($>15 ºC$ mean annual temperature) and increases with increasing absolute latitudes \citep{reich2004global}. The LPC increase poleward is possibly related with the effect of glaciations which deliver rocks rich in P and other mineral nutrients to the soil profile \citep{walker1976fate}.
The LNC map is more homogeneous, but the coniferous forests clearly stand out (boreal area). Our map also shows the highest values for grasslands (24 $mg \, g^{-1}$) in the northern hemisphere(central America and Kazkhstan more precisely) and the lowest (around 16 $mg \, g^{-1}$) in Africa close to the Sahara dessert. The significant variability is in agreement with recent studies and has been reported in global studies with field data \citep{kattge2011try,wright2005assessing}.
Leaf Nitrogen to Phosphorus ratio (LNPR) has been used widely in the ecological stoichiometry literature to understand nutrient limitation in plants. Thus, for example, warm (i.e., tropical) habitats are more P- than N-limited \citep{reich2004global}. This is likely because of substrate age and higher rates of leaching associated with higher rainfall; whereas plants in temperate soils, which are typically younger and less leached, are mostly N-limited. \cite{walker1976fate} predicted also that P limitation should be stronger than N limitation in equatorial regions, due to effect of soil age and climate. Our LNPR map is capturing adequately this documented result with the highest values of LNPR (P limited) occurring in the Equator area and low to medium LNPR (N limited) values found in temperate and dry areas.

Random forests are also a popular and straightforward method for ranking the importance of a set of predictors. Using this feature, we ranked the importance of the input variables of the RF model for the calculation of the mapped canopy traits (\ref{ap:sentraitmaps}). The results show that remote sensing data play a crucial role explaining the spatial variability of all traits. Among them, EVIstd and EVImax are the most influential ones (see Table \ref{tab:inputslandsat} for the definition of the variables). EVIstd reflects leaf longevity while EVImax is sensitive to maximum present vegetation during the year and green biomass. The median albedo for MODIS bands 2(0.84-0.88 $\mu m$) , 5 (1.23-1.25 $\mu m$)  and 6 (1.63-1.65 $\mu m$) have also a very significant explanatory power in SLA, LDMC and LNC. \cite{ollinger2008canopy} used remote sensing data to spatialize LNC in temperate and boreal forests in north America and they also found a strong and consistent response of increasing reflectance with increasing LNC in similar spectral bands. In addition, as expected, climatological data also exhibited a great importance in the specialization of all traits. Among them, bioclimatic variables related with water availability are the most influential ones (BIO12-17).  Some temperature related bioclimatic variables are also included in the ranking and those are mostly related with the maximum annual temperatures and isothermality (BIO5 and BIO 3). Although these variables are generally in the lower part of the ranking of all traits, they are still significantly influential and explicative of all considered traits' spatial variability. Recent papers have also highlighted the importance of environmental factors like elevation in the total variance explained for SLA and LNC with airborne imaging spectroscopy at the Peru scale \citep{asner2017airborne}. Our findings are in concordance with that, and show that elevation (and thus temperature) plays a key role in the specialization of certain traits like SLA and LDMC (both traits closely related to each other).

\section{Discussion}\label{ap:discussion}

Due to a lack of availability of global maps of this kind, there is little opportunity to compare to independent maps. For this reason, in this section we make a qualitative assessment of plausibility against other available data such as typical values from look up tables and other approaches for some traits when available.

\subsection{Comparison based on in-situ leaf level measurements}

\begin{figure}[t!]
\small
\begin{center}
\setlength{\tabcolsep}{10pt}
\begin{tabular}{cc}
\includegraphics[width=4.7cm]{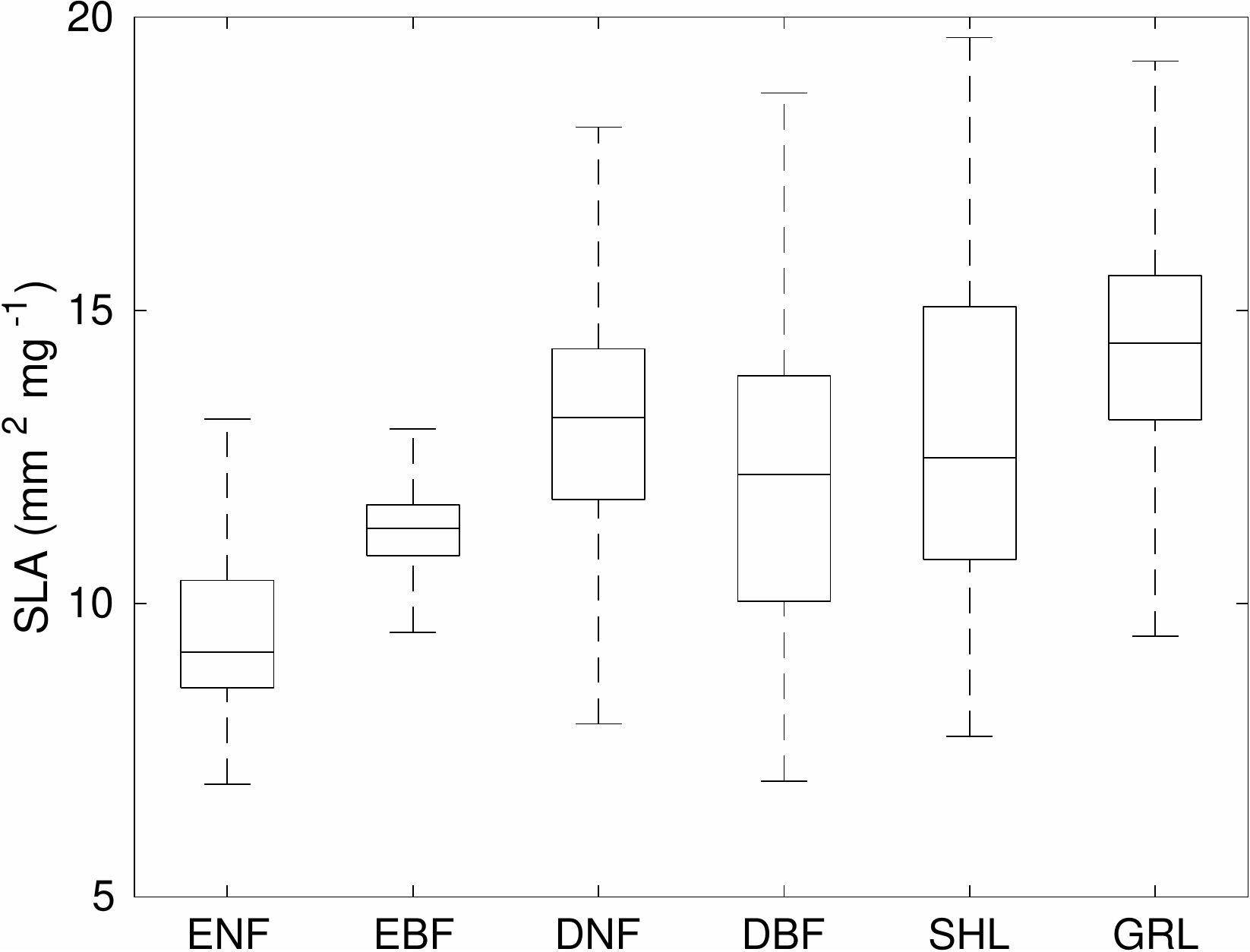}  &  \includegraphics[width=4.7cm]{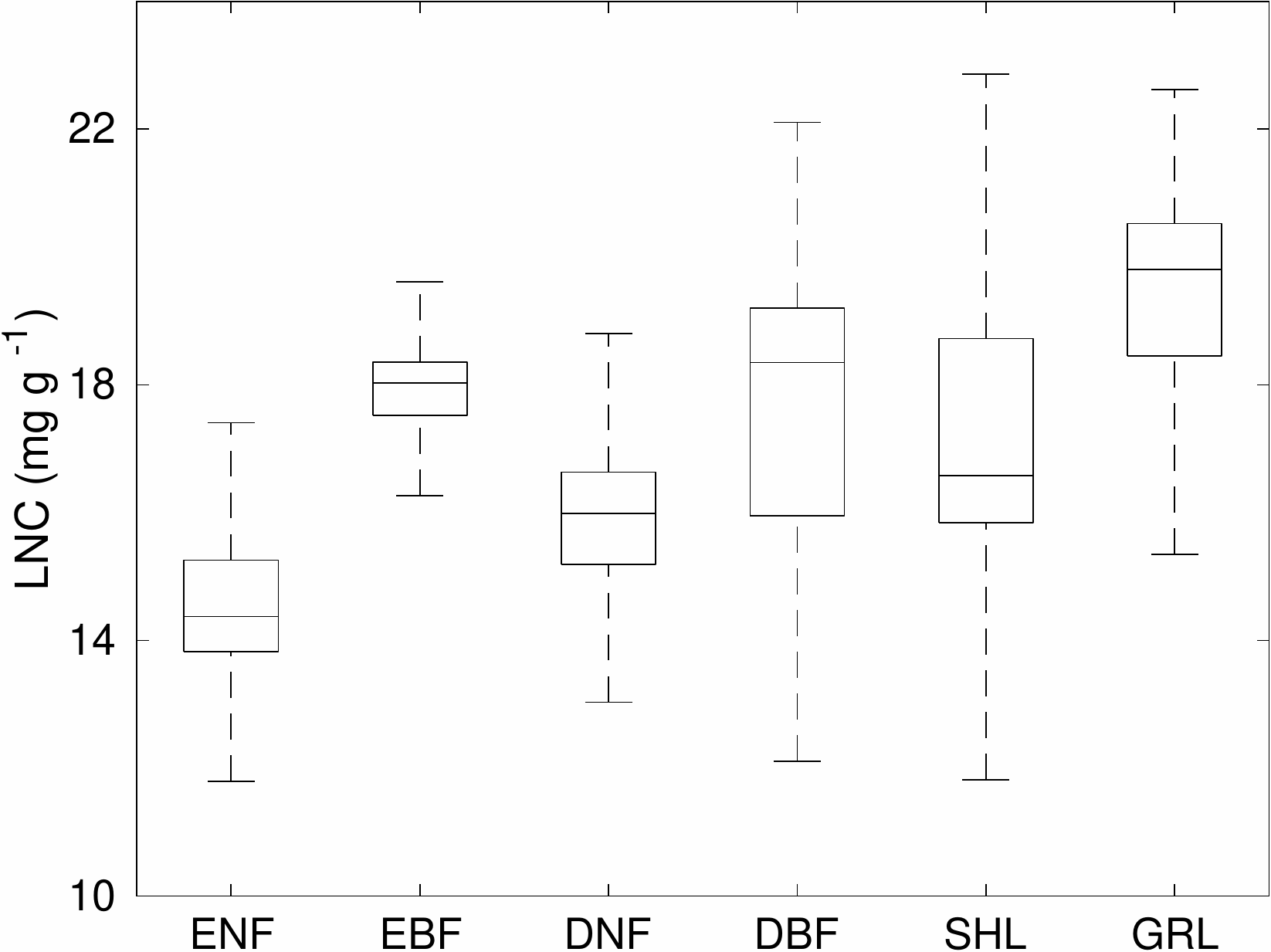} \\
\includegraphics[width=4.7cm]{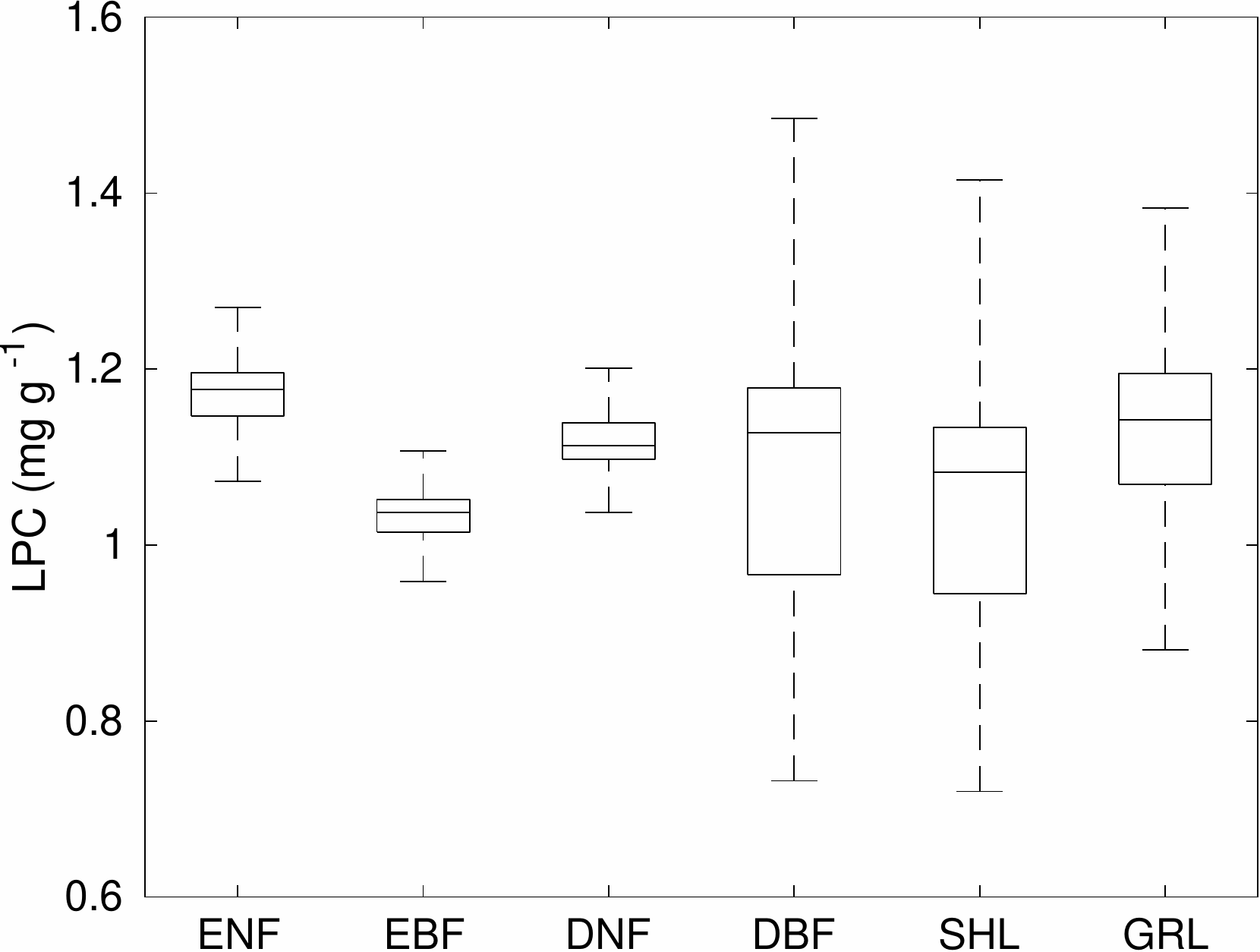}  &  \includegraphics[width=4.7cm]{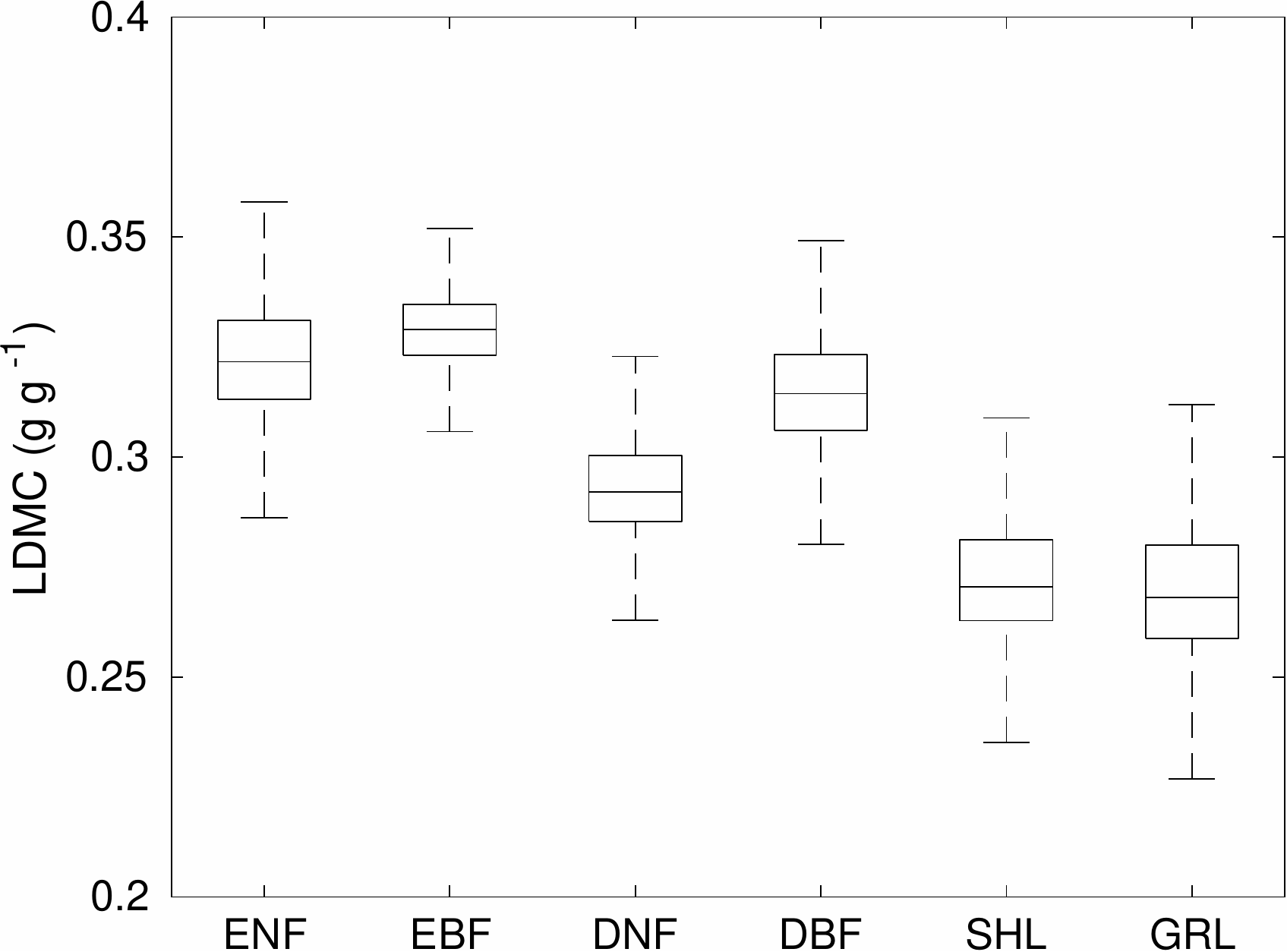} \\
\includegraphics[width=4.7cm]{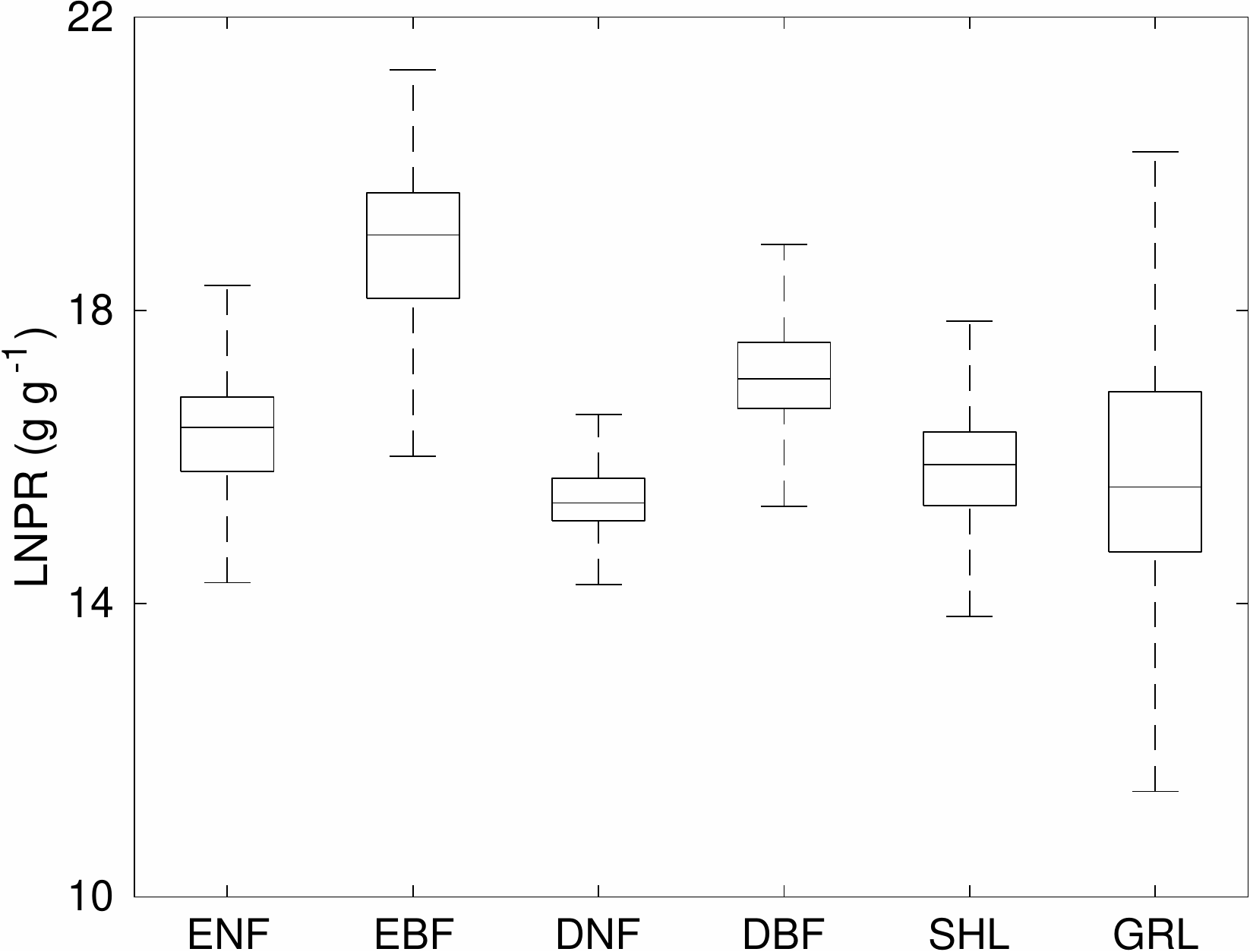}  &   \\
\end{tabular}
\end{center}
\caption{Box plots of trait values per PFT in the training dataset for the different traits considered in this work.}\label{fig:boxplotsMeanvaluesperPFT}
\end{figure}

In order to provide an effective way to display if our global trait estimates capture trait variability among the different PFTs, we analyzed our maps by means of box plots. To calculate the boxplots, we have stratified our global trait maps for each PFT using the operational MODIS land cover (MOD12). Figure~\ref{fig:boxplotsMeanvaluesperPFT} shows box plots diagrams for all considered traits and the different PFTs. The observed variability within PFTs is not surprising and can be attributed to three main factors: First, the above mentioned intrinsic variability of traits within individual PFT. Secondly, discrepancies between the discrete categorical MODIS land cover classes and our mixture of PFTs approach which provides the abundance weighted mean trait value of vegetation types per pixel. Thirdly, uncertainties in the MODIS land cover map and our trait estimates.

We compared between and within PFT trait variances in our estimates (Figure~\ref{fig:boxplotsMeanvaluesperPFT}) with leaf level measurements shown in \ref{ap:TRYtypicalvalues}. In the majority of cases, the calculated trait maps respect reasonably well the expected means and ranges when they are compared with look up table values found in the literature and with the mean values per PFT measured at leaf level. Obviously, as we have used the TRY database to compute our maps, we expected to obtain similar variability between and within PFTs trait estimates when we compare with in-situ leaf measurements. Nevertheless, this comparison is helpful to check if our processing chain is capable to spatialize efficiently leaf level trait measurements to a pixel level and capturing at a global scale PFTs trait typical values and variances. However, it should be noted that some discrepancies are noticeable and, as in the previous comparison, they can be attributed to the different scales between in-situ leaf level trait measurements and our CWM trait estimates at a pixel level. 

\begin{figure}[h!]
\small
\begin{center}
\setlength{\tabcolsep}{5pt}
\begin{tabular}{ccc}
SLA   & LNC   &  LPC \\
\includegraphics[width=3.7cm,trim={1cm 0cm 0cm 0}]{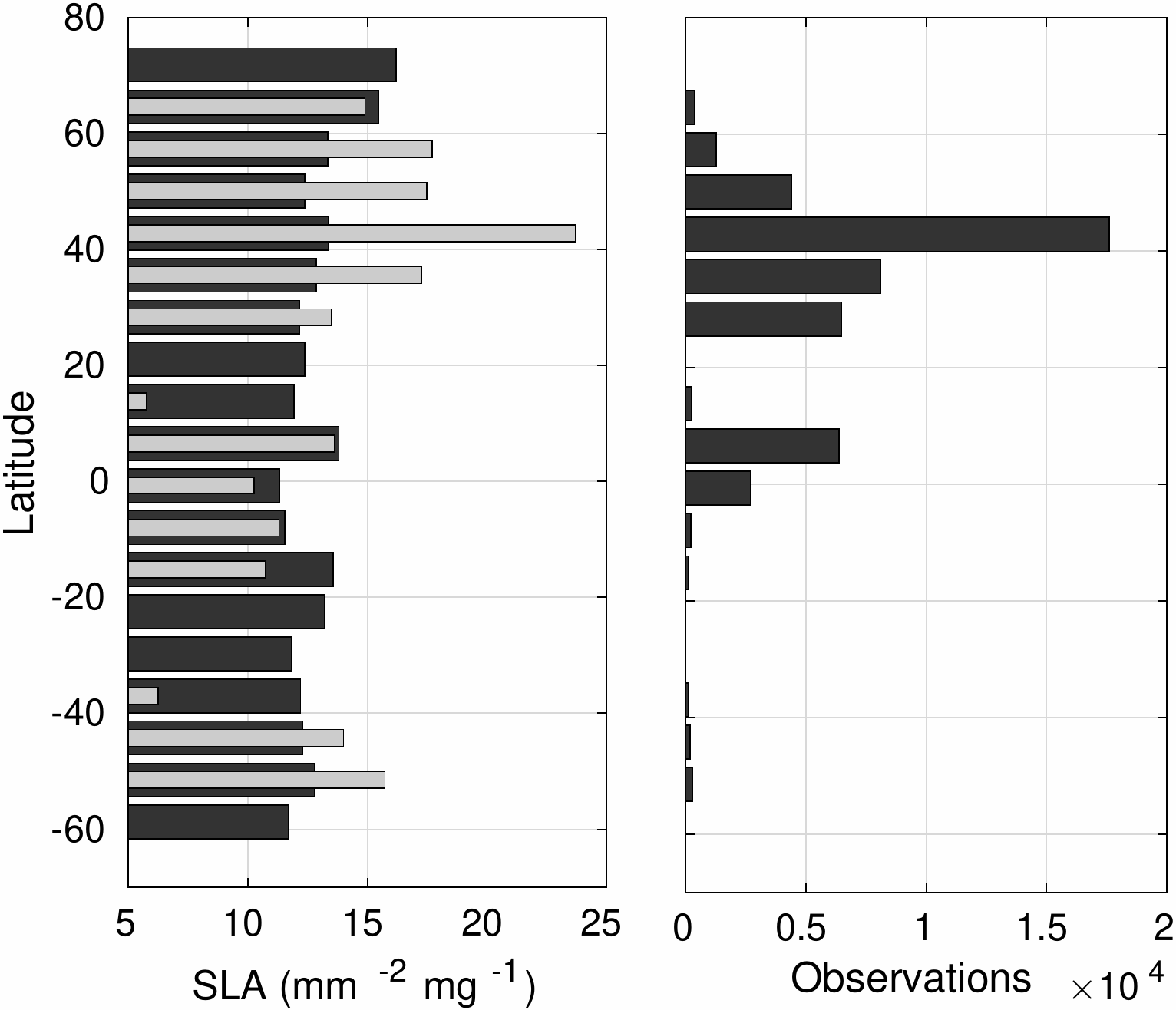} &
\includegraphics[width=3.7cm,trim={1cm 0cm 0cm 0}]{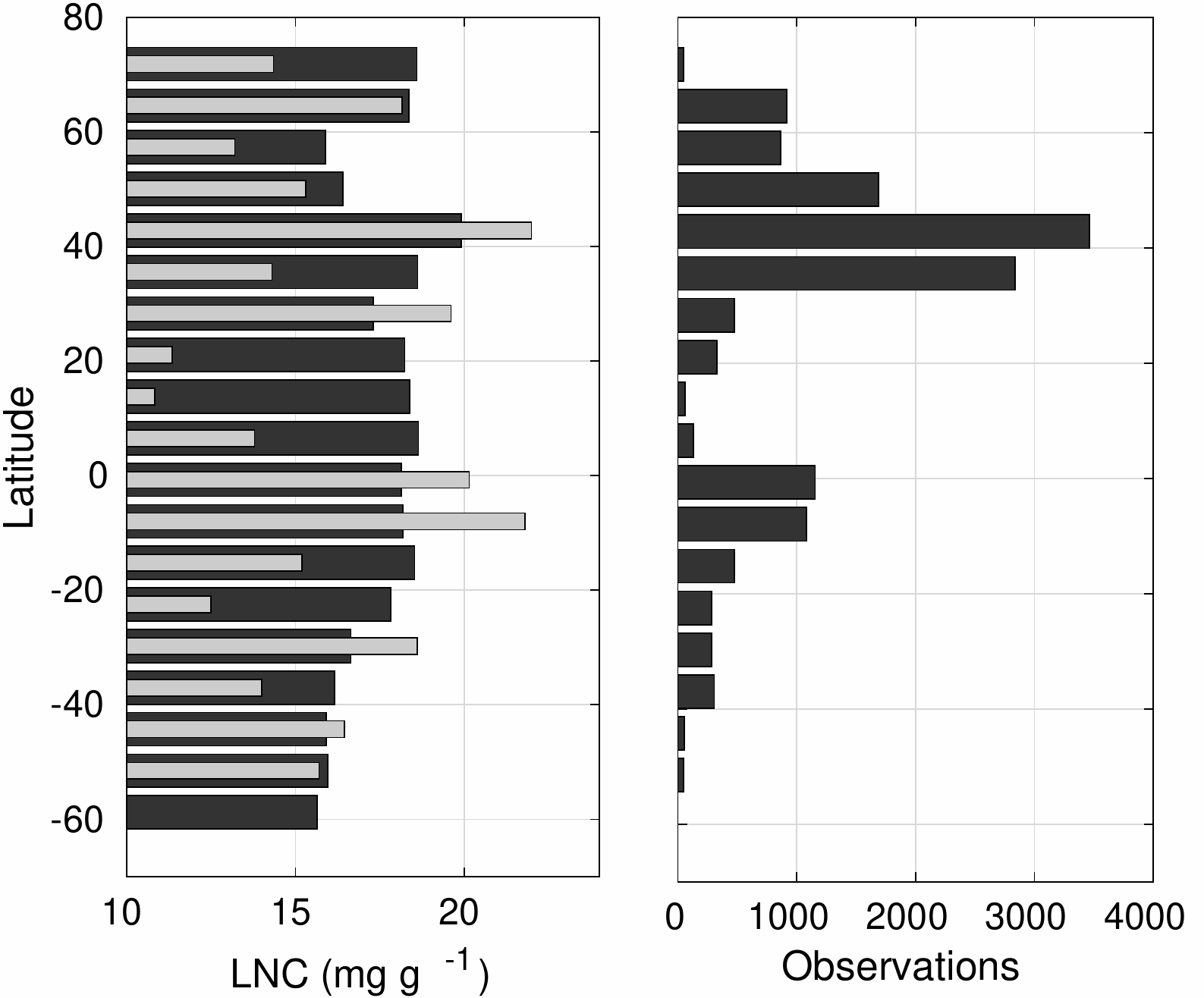} &
\includegraphics[width=3.7cm,trim={1cm 0cm 0cm 0}]{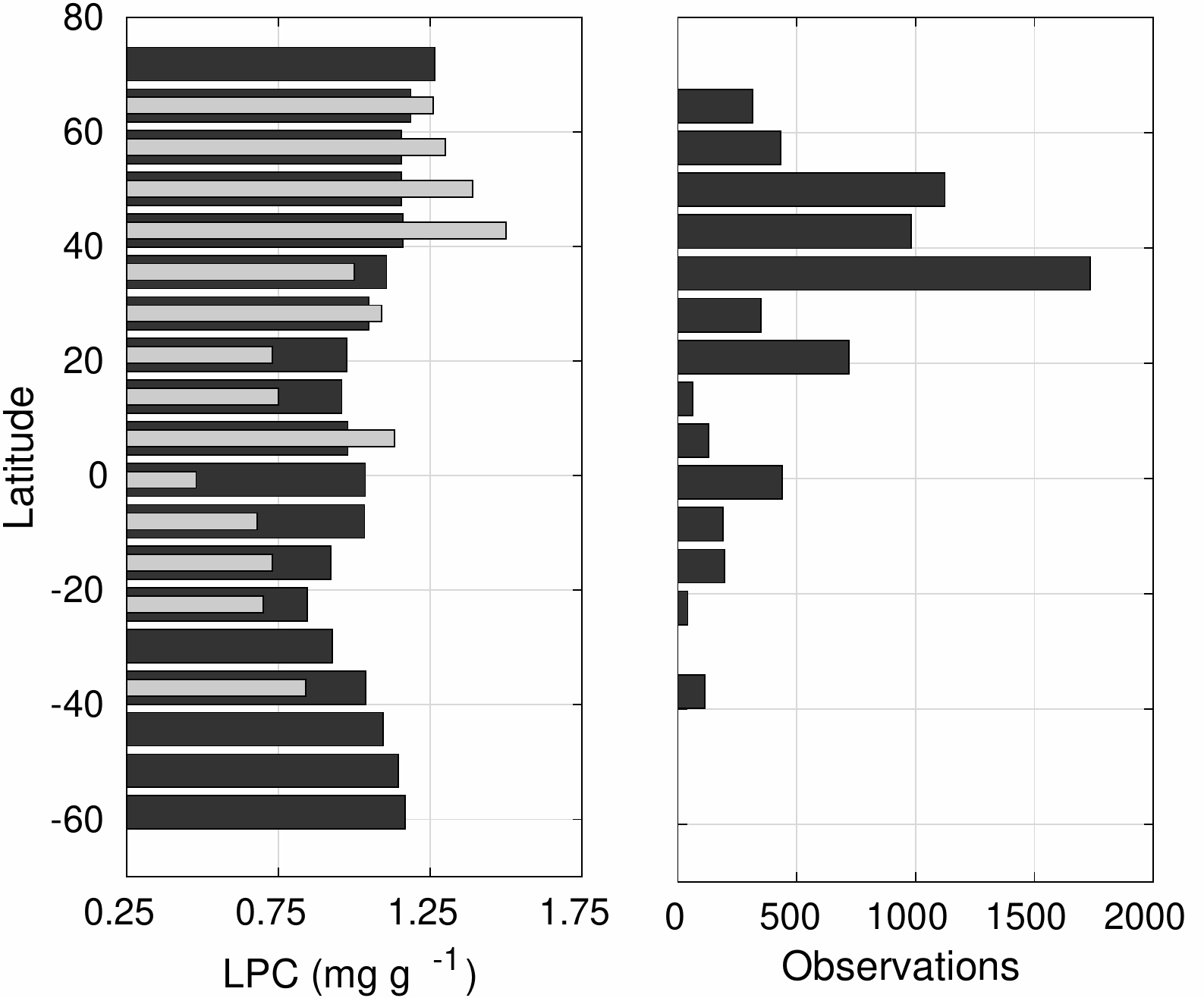} \\
LNPR  & LDMC  & \\%{\bf FMNC, R=0.76$\pm$0.02} \\
\includegraphics[width=3.7cm,trim={1cm 0cm 0cm 0}]{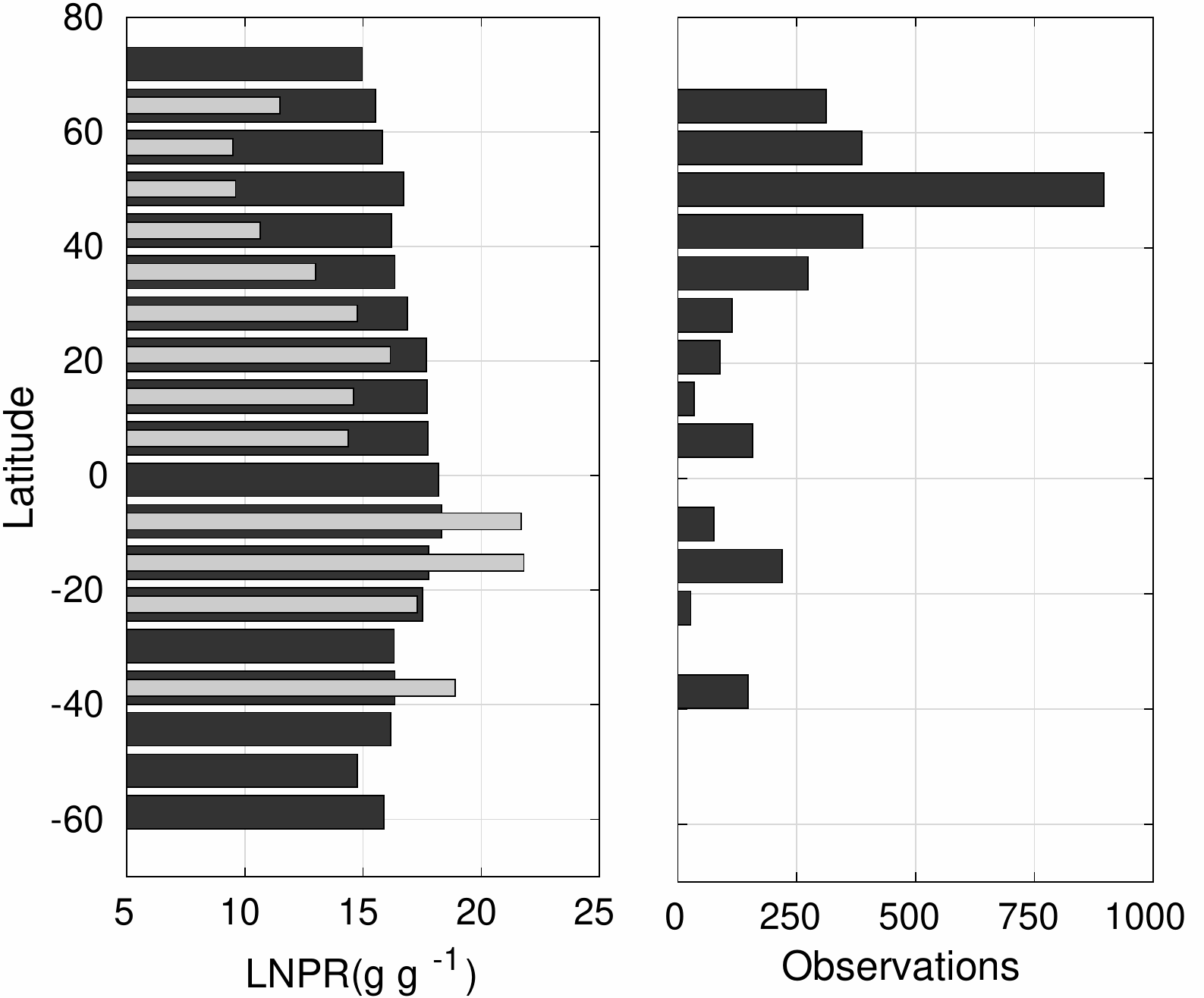}  &
\includegraphics[width=3.7cm,trim={1cm 0cm 0cm 0}]{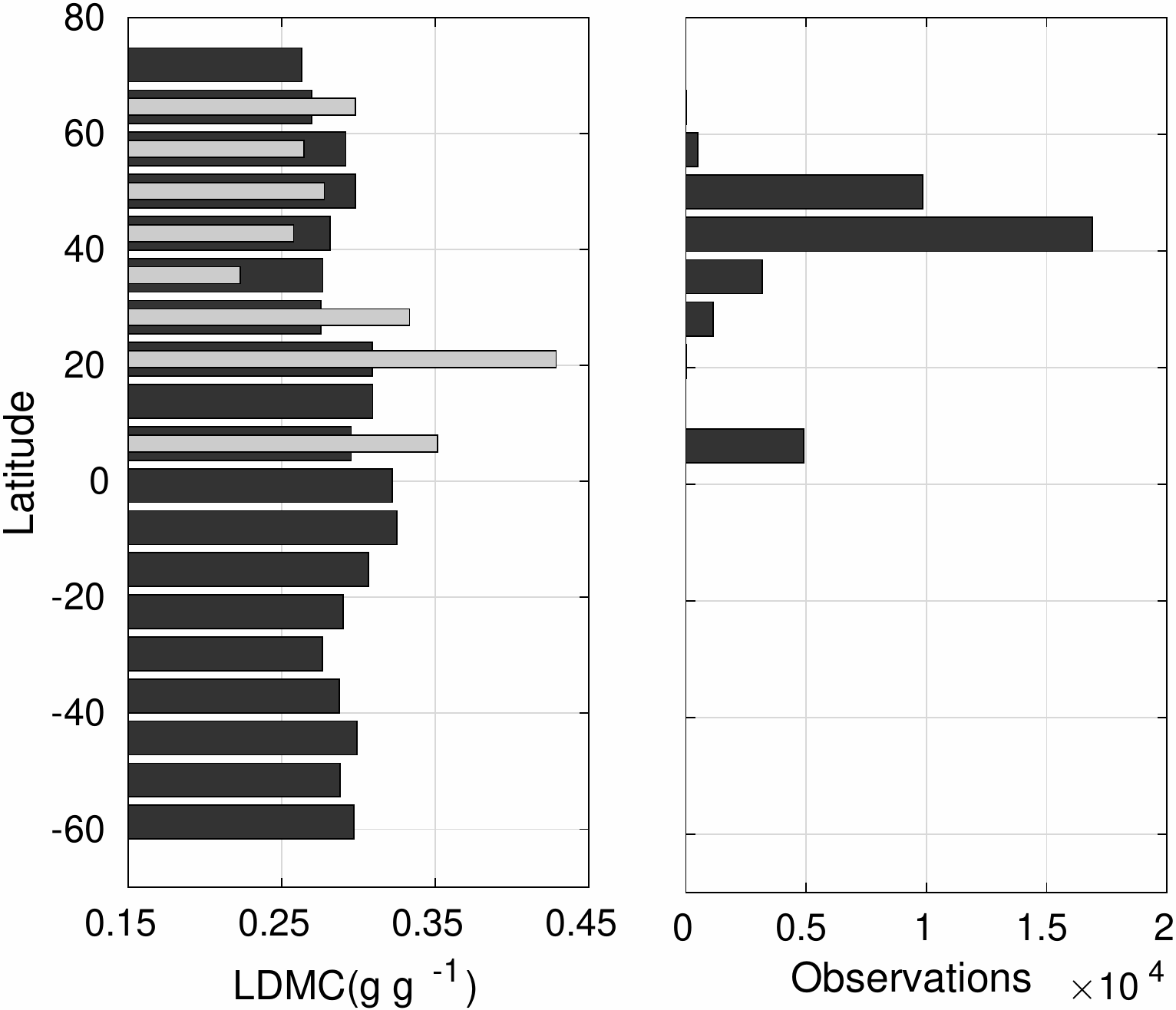} &
\end{tabular}
\end{center}
\vspace{-0.25cm}
\caption{Comparison of latitudinal mean values between the TRY leaf measurements (light gray) and the calculated trait maps (dark grey).}\label{fig:latitudehistograms}
\end{figure}

In Figure \ref{fig:latitudehistograms} mean latitudinal trait values are shown and compared with the leaf measurements used. The number of available observations is also shown and clearly indicates a very significant bias towards the Northern Hemisphere,  despite the much larger land area in the Northern Hemisphere. Europe has the highest density of measurements. However, there are obvious gaps in boreal regions, the tropics, northern and central Africa, parts of South America, and southern and western Asia. In tropical South America, the sites fall in relatively few grid cells, but there are high numbers of entries per cell \citep{kattge2011try}.

Although a comparison of both distributions gives an idea about the variability of the considered traits (and similar patterns are observed between our maps and in-situ measurements), distributions need not to be coincident because trait entries in the database are not abundance-weighted with respect to natural occurrence. In-situ data represent the variation of single measurements, while our maps produce pixel representative estimates \citep{kattge2011try}. It is recognized that all major biome types occur in both hemispheres except the boreal forests and analogous coniferous forests of the northern hemisphere. The main climatic variables which correlate with, and appear to limit, the analogous biome types and vegetation regions, also appear to be similar in the two hemispheres and thus of global validity \citep{box2002vegetation}. This implies that although the TRY database is sampled spatially in an unbalanced way, our maps could potentially extrapolate effectively; given that the most important biomes are sampled and the developed models are generic. 

Besides, comparing the trait mean values in latitudinal data, along with the differences between the TRY leaf measurements and our calculated trait maps, interesting areas of discrepancy can be identified. These discrepancies could be used as an indicator to discover areas where the most abundant species are not being adequately sampled, allowing resources to be optimized by only collecting in-situ data where it is really needed.

\subsection{Qualitative comparison with previous works}

\cite{van2014fully} calculated global maps of the inverse of SLA (Leaf Mass Area, LMA) using a combination of soil and climate variables. The qualitative comparison of their maps and the followed approach in the present work indicated that there is a general agreement in the global distribution of the estimated values. However, significant differences are still noticeable, particularly in areas where \cite{van2014fully} reported that their uncertainties were larger (wet tropical and boreal forests). At the spatial resolution considered in this study (500 m), climatic drivers can only provide partial and broad information about global vegetation distribution and they are insensitive to disturbances which are an important driver of many traits \citep{van2012going}. For example, the occurrence of different PFTs could be equally probable under similar environmental conditions and although environmental gradients tend to be gradual and smooth, it is known that the spatial distribution of PFTs is usually patchy. \cite{van2014fully} pointed out that problem when they evaluated the robustness of their approach predicting vegetation distribution using trait estimates. \cite{swenson2010plant} combined a continental-scale forest inventory data set (18000 plots) with trait data to generate a community trait distributions for LNC and other traits in eastern North American tree communities.  In their study, they computed mean LNC values for each grid cell using abundance weighting estimates. The results presented show a similar spatial variability to ours.  \cite{ollinger2008canopy} estimated LNC maps using MODIS data in North American temperate and boreal forests with a reduced in-situ training dataset consisting of 181 field plot observations. The qualitative comparison of their estimates and ours for the coincident study area shows a very good agreement between both approaches. Moreover, \cite{ollinger2008canopy} also observed similar spatial patterns related to dominant PFTs (higher values in the eastern deciduous forests and lower values in northern evergreen forets).

%% file: conclusions3.tex
\section{Summary and conclusions}

This paper introduced a processing chain to derive spatially explicit global maps of leaf traits from in-situ measurements using remote sensing and climatological side information. We focused on well represented leaf traits: SLA, LDMC, leaf nitrogen and phosphorus concentrations, as well as nitrogen-phosphorus ratios. The generated global maps are at 500 m spatial resolution. 
The proposed methodology first considers filling the gaps of the TRY database. While many methods are available for matrix completion, in either statistics and machine learning communities, we proposed here a simple yet very powerful approach: we used random forests regression with surrogates for completing the missing entries via prediction. The second step of the process provides abundance estimation of PFTs at the MODIS pixel resolution. For this we developed a classifier of PFTs at the Landsat pixel resolution. This map at a finer resolution allowed us to adequately resample in-situ observations. Finally, we developed one model for each considered trait using remote sensing, canopy level trait values, and ancillary climate data. The final models were implemented using random forests and showed good precision and robustness in all traits. Models were run globally and provided not only estimated traits per pixel but also associated uncertainty maps.

The proposed chain is modular, flexible, and allows for improvements and updates in almost all steps. This is an important feature of our approach as the maps can be both improved and updated as more data are available. The remote sensing and climatic data used could be also replaced with new datasets. For example, the European Space Agency (ESA) is developing a new family of missions called Sentinels \citep{aschbacher2012european}. These missions provide new satellites and sensors with better specifications than the ones used in the present work. This new Sentinel data could replace the Landsat and MODIS data and potentially lead to significant improvements in our estimates through better temporal, spectral, and spatial resolutions.

Even though the TRY database provides an unprecedented number of in-situ measurements of plant traits, at a global scale this kind of data is inherently sparse and irregular spatiotemporally, especially when considering species richness and intra-specific variability. In this work, we estimated community composition at a MODIS pixel resolution (500 m) to spatialize local trait measurements from leaf to canopy level. % The computed high resolution land cover map allowed us to estimate the abundances of PFTs within MODIS pixels.
Although the followed approach is not the definitive solution to address species level sampling bias of global plant trait databases, it mitigates the PFT level bias at each training location due to the lack of information about local species assemblages at the required spatial resolution \citep{butler2017mapping}. Further work is needed to assess the quality of the computed community weighted mean trait values, including improvements related to the proposed heuristic approach to select nearby leaf trait observations. The Global Biodiversity Information Facility (GBIF) database \citep{telenius2011biodiversity}, which includes hundreds of millions of species occurrence records, could be a promising complementary source of information to compare and validate our estimates in the future.

The use of in-situ measurements to provide global trait maps entails transforming the information from plant organ to canopy level and regional scales. Remote sensing provides continuous global coverage, in space and time, to evaluate and assess how plants diversify and function while facilitating an enticing way to scale up from leaves to landscapes \citep{asner2016spectranomics,homolova2013review,ollinger2008canopy}. New space-based observations including hyperspectal observations could complement in-situ measurements, providing required quantitative ecosystem information to track changes in plant functional diversity around the globe \citep{jetz2016monitoring,schimel2015observing}. However, hyperspectral data for plant traits are not (yet) available globally, restricting its applicability to local scales. Future missions like the Environmental Mapping and Analysis Program (EnMAP) German imaging spectroscopy mission \citep{guanter2015enmap} or the NASA Hyperspectral InfraRed Imager (HyspIRI) mission \citep{lee2015introduction} could overcome some of the limitations of sensors with coarse spectral resolution like MODIS and Landsat used in this work \citep{jetz2016monitoring}. These missions will use high fidelity imaging spectroscopy sensors and they will provide a more direct path for the estimation of plant functional biodiversity. Yet, not all plant traits can be observed from space. For these traits, e.g. wood density, upscaling of in-situ measurements (as presented here) will probably be the method of choice for some time. It summarizes the knowledge presently available from in-situ measured plant traits and combines it with remote sensing and climate information using advanced machine learning. In addition, almost all steps of the modular workflow aim at improving the representativeness of in-situ information at plant organ level for canopy level and MODIS pixel scale. The plausibility checks against independent data presented above indicate that this approach seems promising.

Different authors have highlighted the importance, utility, and potential of using trait information to explain long-term (e.g. annual) patterns underlying carbon, water, energy fluxes, and biodiversity globally \citep{musavi2016potential,maron2015agree}.
The provided maps could replace the current static PFT maps while eventually improving models of maximum photosynthetic capacity and fluorescence. As some continuous leaf traits (LNPR) might be better predictors of vegetation responses to nutrient availability \citep{ordonez2009global}, they could also be an alternative way to infer soil fertility in natural vegetation. Initiatives such as the Group on Earth Observations Biodiversity Observation Network (GEOBON) \citep{scholes2008toward} work in the selection and implementation of essential variables required to study biodiversity worldwide. In this context, the produced trait estimates reflect functional properties of vegetation globally that could be a valuable addition for the understanding and monitoring of the biosphere.

%% file: appendices2.tex
%\section{Considerations about the TRY database}
\setcounter{figure}{0}
\setcounter{table}{0}
\setcounter{equation}{0}

\section{Climate data}\label{ap:climatedata}

The included climatological data in the TRY database have a lack of a consistent structure. In the present study,  climatological data have been used in two key steps: the gap filling of the database and the spatialization of pixel representative trait estimates. For these purposes, we have used the WorldClim database (Table \ref{tab:inputsworldclim}) which includes interpolated bioclimatic variables for current conditions \citep{hijmans2005very}. The WorldClim interpolated climate data are composed by major climate databases including the Global Historical Climatology Network (GHCN), the FAO, the WMO, the International Center for Tropical Agriculture (CIAT), RHYdronet, and other minor more local databases. WorldClim data were restricted to records comprising the 1950–2000 period to be representative of the recent climate at a 1 km spatial resolution.

In the present paper, we resampled them to a 500 m spatial resolution by means of a bilinear interpolation to solve the inconsistency in spatial resolution with the rest of products considered and to avoid steep gradients in WorldClim coarser pixels.
The WorldClim bioclimatic variables represent annual trends, seasonality, and extreme or limiting environmental factors.  Examples of each include: mean annual temperature and precipitation, annual range in temperature and precipitation, and the temperature of the coldest and warmest months as well as the precipitation of the wettest and driest quarters. \citep{hijmans2005very}.

\begin{table}[t!]
\small
\centering
\caption{Bioclimatic variables considered in this work. A quarter corresponds with a period of three months.}
\label{tab:inputsworldclim}
\begin{tabular}{|l|l|}
\hline
\hline
Variable & Description        \\
\hline
\hline
BIO1                                        & Annual mean temperature.          \\
BIO2                                          & Mean Diurnal Range.              \\
BIO3                                           & Isothermality (BIO2/BIO7) (*100).                \\
BIO4                                          & Temperature Seasonality (standard deviation *100).             \\
BIO5                                        & Max Temperature of Warmest Month.           \\
BIO6                                          & Min Temperature of Coldest Month.         \\
BIO7                                          & Temperature Annual Range (BIO5,BIO6).  \\
BIO8                                        & Mean Temperature of Wettest Quarter.  \\
BIO9                                         & Mean Temperature of Driest Quarter.     \\
BIO10                                         & Mean Temperature of Warmest Quarter.\\
BIO11                                         & Mean Temperature of Coldest Quarter.    \\
BIO12                                          & Annual Precipitation.           \\
BIO13                                         & Precipitation of Wettest Month.                 \\
BIO14                                          & Precipitation of Driest Month.      \\
BIO15                                          & Precipitation Seasonality (Coefficient of Variation).                 \\
BIO16                                          & Precipitation of Wettest Quarter. \\
BIO17                                          & Precipitation of Driest Quarter. \\
BIO18                                          & Precipitation of Warmest Quarter. \\
BIO19                                          & Precipitation of Coldest Quarter. \\
\hline
\hline
\end{tabular}
\end{table}

\section{Description and comparison of machine learning regression methods}
\label{ap:MLmethods}

This section summarizes the theory underlying the machine learning methods used, and their numerical comparison in terms of precision, fit and bias, as well as robustness to the number of training data.

\subsection{Machine learning methods}
\setcounter{figure}{0}
\setcounter{table}{0}
\setcounter{equation}{0}
\paragraph{Regularized Linear Regression}

In multivariate (or multiple) linear regression (LR) the output variable $y$ (plant trait) is assumed to be a weighted sum of $F$ input variables (or features), $\x:=[x_1,\ldots,x_F]^\top$, that is $\hat y=\x^\top{\bf w}$. Maximizing the likelihood is equivalent to minimizing the sum of squared errors, and hence one can estimate the weights ${\bf w}=[w_1,\ldots,w_F]^\top$ by least squares minimization.
Very often one imposes some smoothness constraints to the model and also minimizes the weights energy, $\|{\bf w}\|^2$, thus leading to the regularized linear regression (RLR) method that we used in this work.

\paragraph{Random Forest}

A RF model is an ensemble learning method for regression that operates by constructing a multitude of decision trees at training time and outputting the mean prediction of the individual trees~\citep{Breiman85}. They combine many decision trees working with different subsets of features. The RF strategy is very beneficial by alleviating the often reported overfitting problem of simple decision trees. In addition, RFs are quite robust; handling a large number of input variables, excelling in the presence of missing entries, dealing with heterogeneous variables, and can be easily parallelized to tackle large scale problems. RF classification and regression have been applied in different areas of concern within forest ecology:  modeling the gradient of coniferous species \citep{Evans09}, the occurrence of fire in Mediterranean  regions \citep{Oliveira12}, the classification of species or land cover type \citep{Gislason06,Cutler07}, and the analysis of the relative importance of the proposed drivers \citep{Cutler07} or the selection of drivers \citep{Genuer10,Gislason06,Jung13}.

\paragraph{Neural networks}

Neural networks are nonlinear, nonparametric regression methods. Their base operational unit is the neuron, where nonlinear regression functions are applied. The neurons are interconnected and organized in layers. The outputs of all neurons in a given layer are the inputs for neurons of the next layer. In a network, each neuron performs a linear regression followed by a nonlinear (sigmoid-like) function. Neurons of different layers are interconnected by weights that are  adjusted during the training~\citep{Haykin99}. In the standard neural network, the weights are typically adjusted using a backpropagation algorithm of the error. This requires tuning several hyperparameters, such as the learning rate, number of epochs, network structure, momentum and dropout terms, which makes training computationally demanding. As an alternative we used a fast version of neural network known as Extreme Learning Machine (ELM)~\citep{Huang06}: here the network structure is fixed, the weights connecting the input to the hidden layer are randomly selected, and one only optimizes the weights from the hidden to the output layer via least squares. The cost is then drastically reduced at the expense of loss in precision over standard neural nets.

\paragraph{Kernel methods}

Kernel methods constitute a family of successful methods for regression~\citep{CampsValls09wiley}. We aim to incorporate two instantiations: (1) the KRR (Kernel Ridge Regression) is considered as the (non-linear) version of the LR~\citep{Shawetaylor04}; and (2) GPR (Gaussian Process Regression) is a probabilistic approximation to nonparametric kernel-based regression, where both a predictive mean and predictive variance can be derived \citep{CampsValls16grsm}. Kernel methods offer the same explicit form of the predictive model, which establishes a relation between the input (i.e., computed explanatory variables) $\x\in\mathbb{R}^{B}$ and the output variable (i.e. the particular plant trait) is denoted as $y\in\mathbb{R}$. The prediction for a new radiance vector $\x_*$ can be obtained as:
\begin{equation}
\hat{y} = f(\x)= \sum\limits_{i=1}^{N} \alpha_i K_\theta (\x_i, \x_*) + \alpha_o,
\end{equation}
where $\{\x_i\}_{i=1}^{N}$ are the spectra used in the training phase, $\alpha_i$ is the weight assigned to each one of them, $\alpha_o$ is the bias in the regression function, and $K_\theta$ is a kernel or covariance function (parametrized by a set of hyperparameters $\boldsymbol{\theta}$) that evaluates the similarity between the test spectrum $\x_*$ and all $N$ training spectra. We used the automatic relevance determination (ARD) kernel function:

\begin{equation}
K(\x,\x')=\nu\exp\bigg(-\sum_{f=1}^F(x_f-x_f')^2/(2\sigma_f^2)\bigg)+\sigma_n^2\delta_{ij}
\end{equation}
and we learned the hyperparameters $\boldsymbol{\theta}=[\nu,\sigma_1,\ldots,\sigma_F,\sigma_n]$ by marginal likelihood maximization.

\subsection{Accuracy and robustness of all considered regression models}
\label{ap:RMcomparisosn}
\setcounter{figure}{0}
\setcounter{table}{0}
\setcounter{equation}{0}
All input features were standardized before model training. We followed a standard cross-validation approach for all the methods: we split the data into a cross-validation set containing 80\% of the data to select the hyperparameters, and the remaining 20\% acted as an independent, out-of-sample test set where we evaluate model's performance.  In addition, in order to assess model's robustness we trained models with a varying number of cross-validation data between 10\% and 80\%. The whole procedure was repeated for 20 realizations and the average results are reported.

\subsubsection{Model evaluation: measuring precision and bias}
Models parameters were optimized to minimize cross-validation error.  In particular several scores were used to evaluate model's performance: the mean error (ME) as a measure of bias of the estimations, the root-mean-square-error as a measure of precision, and the Pearson's correlation coefficient (R) as a measure of goodness-of-fit of the models.

\subsubsection{Numerical comparison: precision, bias and goodness-of-fit}

The best average results for all the regression method and leaf traits are given in Table~\ref{tab:accuracy}. Results reveal that both random forests and kernel methods are the most precise and less biased methods, outperforming the regularized linear regression and the extreme learning machine in all cases. RF outperforms the rest for prediction of SLA, LPC and LNPR, while kernel machines (either KRR or GPR) excel in predicting LNC, and LDMC. It is worth noting, however, that numerical differences in R and RMSE were not significant between RF and kernel machines.

\begin{table}[t!]
\small
\caption{Results in the cross-validation set for all methods, scores, and leaf traits. We highlight the best results in bold faced font. }
\label{tab:accuracy}
\begin{center}
\renewcommand{\arraystretch}{0.8}
\begin{tabular}{|l|c|c|c|}
\hline
\hline
{\bf METHOD}	 & {\bf ME}	 & {\bf RMSE} & {\bf R} \\
\hline
\hline
\multicolumn{4}{|l|}{Specific Leaf Area (SLA), $n=4407$} \\
\hline
RLR	 & 0.064	 & 3.819	 & 0.629 \\
RF	 	 & -0.031 & {\bf 3.185}	 & {\bf 0.763} \\
ELM	 & {\bf 0.022}	 & 3.583	 & 0.696 \\
KRR	 & 0.158	 & 3.365	 & 0.741 \\
GPR	 & 0.200	 & 3.215	 & 0.753 \\
\hline
\multicolumn{4}{|l|}{Leaf Nitrogen Concentration (LNC), $n=4422$} \\
\hline
RLR	 & 0.049	 & 2.642	 & 0.621 \\
RF	 	 & {\bf -0.029} & 2.298	 & 0.734 \\
ELM	 & 0.032	 & 2.537	 & 0.674 \\
KRR	 & 0.040	 & {\bf 2.249}	 & {\bf 0.742} \\
GPR	 & 0.037	 & 2.269	 & 0.734 \\
\hline
\multicolumn{4}{|l|}{Leaf Phosphorus Concentration (LPC),  $n=3851$} \\
\hline
RLR	 & 0.001	 & 0.158	 & 0.639 \\
RF	 	 & {\bf 0.001}	 & {\bf 0.132}	 & {\bf 0.778} \\
ELM	 & 0.001	 & 0.150	 & 0.694 \\
KRR	 & 0.001	 & 0.135	 & 0.742 \\
GPR	 & 0.001	 & 0.133	 & 0.763 \\
\hline
\multicolumn{4}{|l|}{Leaf Nitrogen-Phosphorus Ratio (LNPR), $n=2074$} \\
\hline
RLR	 & -0.014	 & 2.087	 & 0.706 \\
RF	 	 & 0.016	 & {\bf 1.806}	 & {\bf 0.781} \\
ELM	 & {\bf -0.012} & 1.933	 & 0.754 \\
KRR	 & -0.022	 & 1.832	 & 0.772 \\
GPR	 & -0.039	 & 1.890	 & 0.767 \\
\hline
\multicolumn{4}{|l|}{Leaf Dry Matter Content (LDMC), $n=1842$}\\
\hline
RLR	 & 0.001	 & 0.048	 & 0.528 \\
RF	 	 & {\bf 0.000}	 & {\bf 0.038}	 & 0.718 \\
ELM	 & 0.000	 & 0.043	 & 0.661 \\
KRR	 & -0.001	 & {\bf 0.038}	 & 0.728 \\
GPR	 & -0.001	 & {\bf 0.038}	 & {\bf 0.731} \\
\hline
%Fresh mass (FM),  $n=1086$  &  & & \\
%\hline
%RLR	 & {\bf 0.006}	 & 0.408	 & 0.528 \\
%RF	 	 & -0.008 & 0.408	 & 0.586 \\
%ELM	 & 0.009	 & 0.424	 & 0.480 \\
%KRR	 & 0.009	 & {\bf 0.398}	 & {\bf 0.598} \\
%GPR	 & 0.025	 & 0.435	 & 0.405 \\
%\hline
\hline
\end{tabular}
\end{center}
\end{table}

\subsubsection{Models' robustness to number of training samples}
\setcounter{figure}{0}
\setcounter{table}{0}
\setcounter{equation}{0}
We also tested model performance in more difficult scenarios in which a reduced number of training samples was used. Results in both R and RMSE are given in Figure ~\ref{fig:rates} for all methods and leaf traits. It is worth noting that two groups of curves (methods) can be easily identified for all traits and scores: on the one hand, RLR and ELM perform poorly, and on the other hand RF and kernel machines perform similarly and report higher precisions across all the reduced-sized training data rates. We observed a high consistency of RF predictions across all plant traits and precision measures, which makes random forests the preferred default option.

\begin{figure}[t!]
\begin{center}
\setlength{\tabcolsep}{5pt}
\begin{tabular}{ccc}
{\bf } & {\bf R}  & {\bf RMSE}\\
\rotatebox{90}{\bf SLA}   & \includegraphics[width=3.7cm]{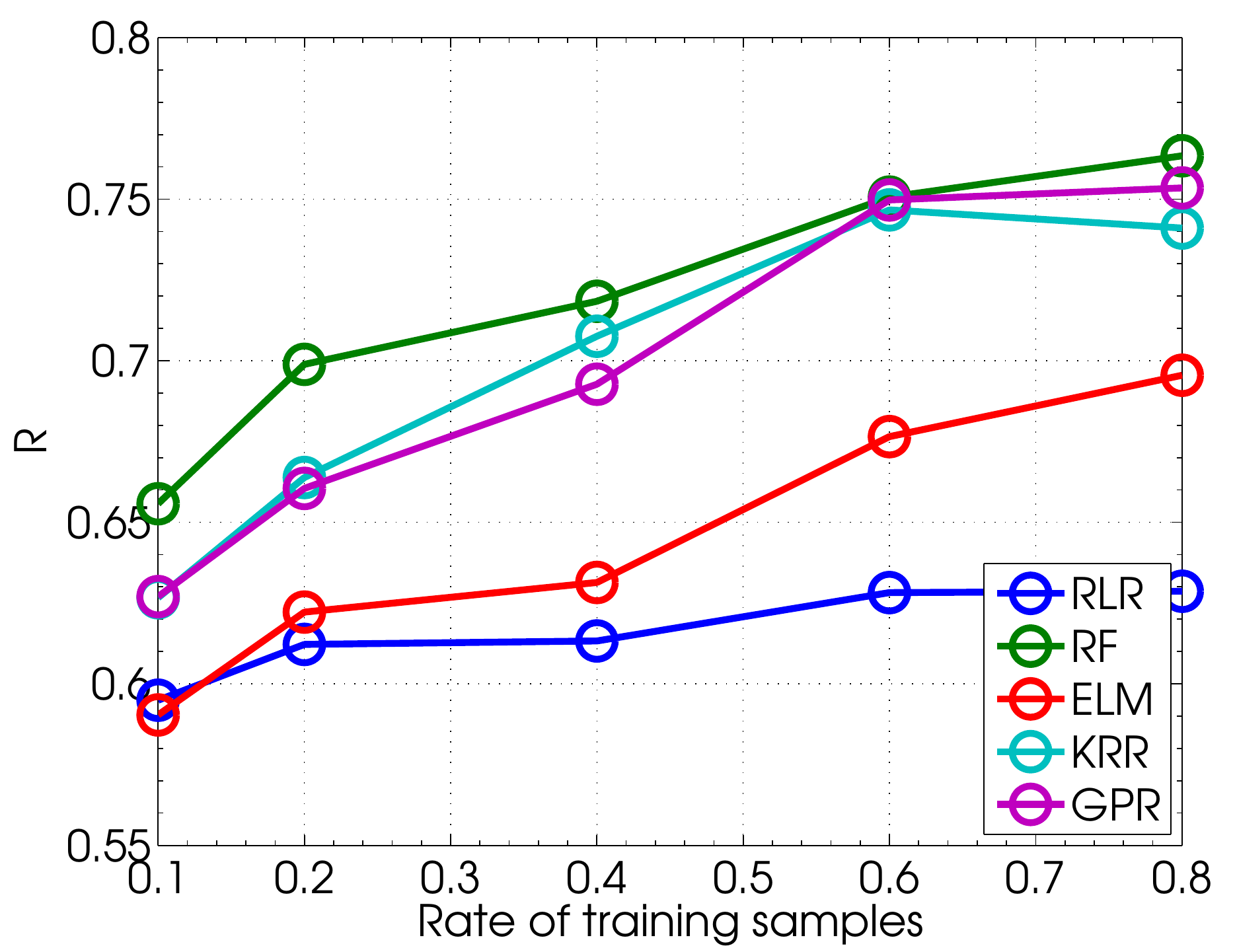} & \includegraphics[width=3.7cm]{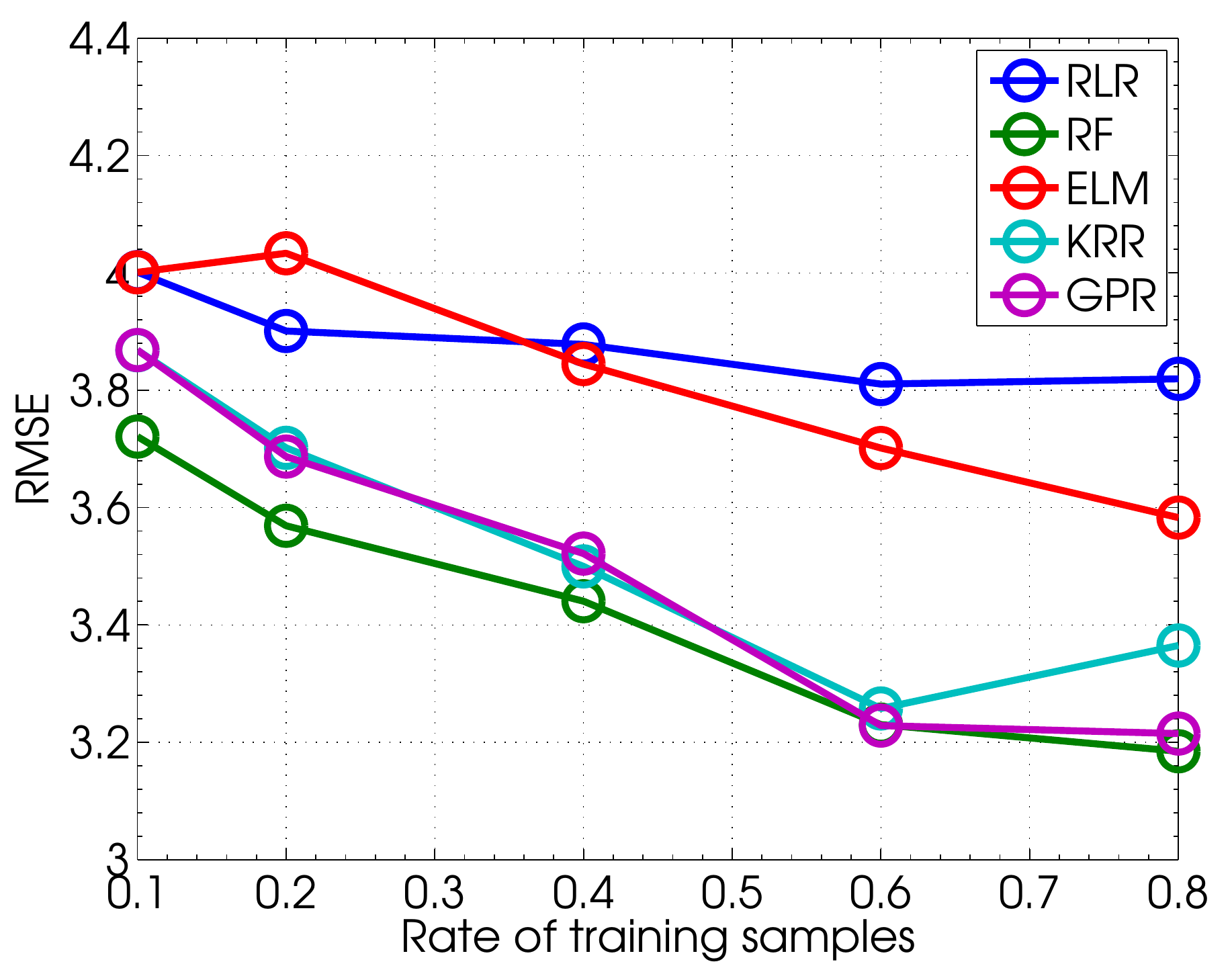} \\
\rotatebox{90}{\bf LNC}    & \includegraphics[width=3.7cm]{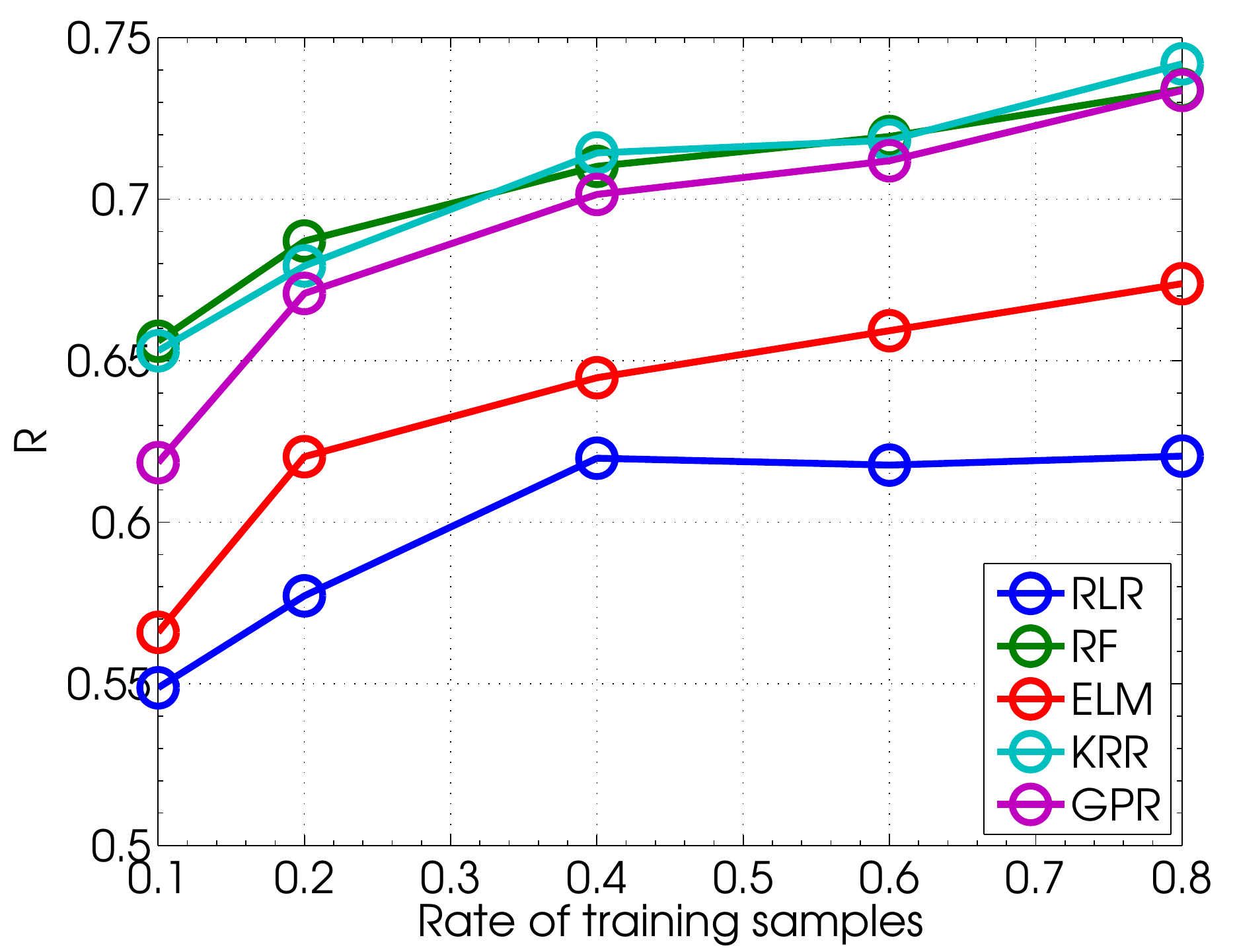} & \includegraphics[width=3.7cm]{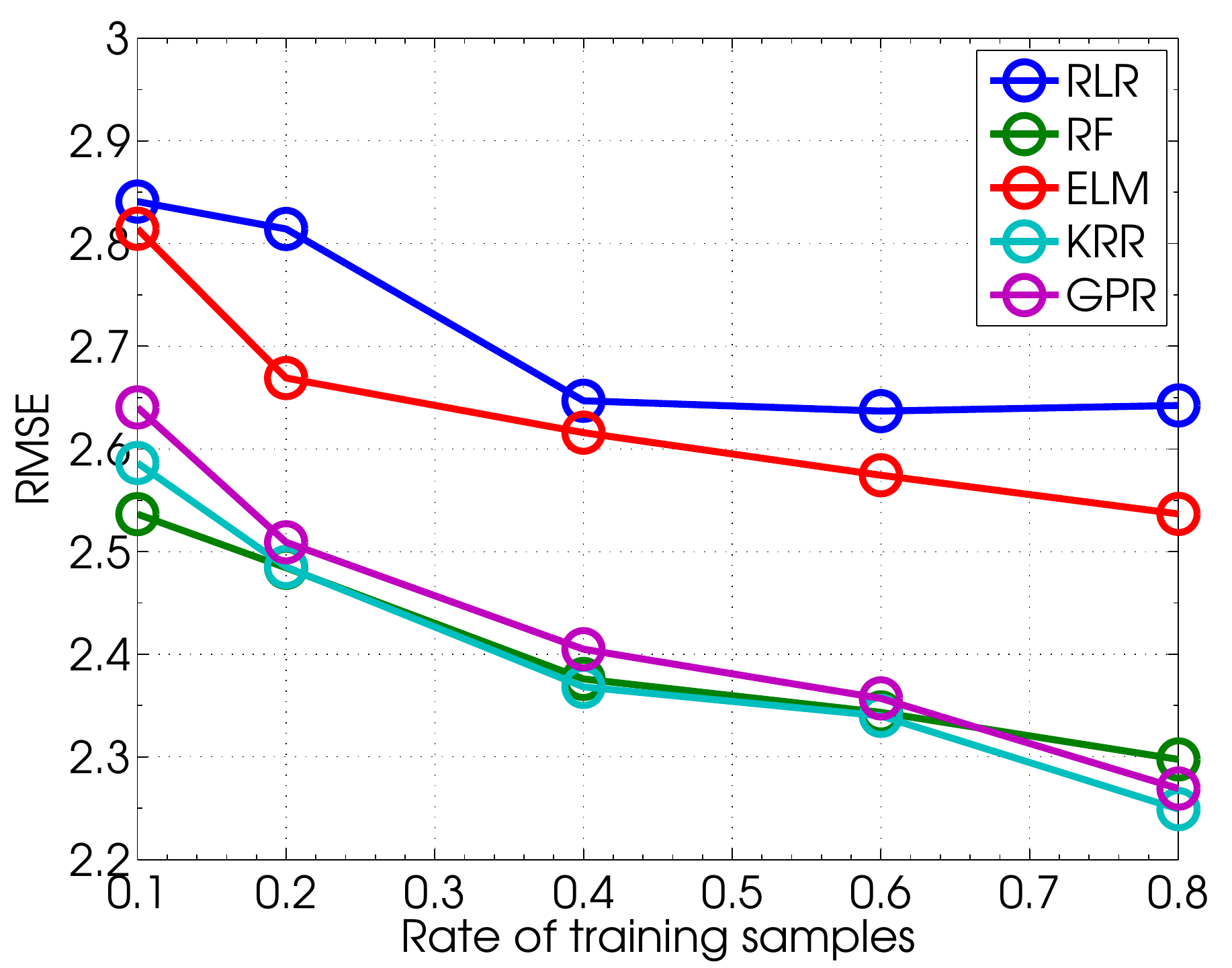} \\
\rotatebox{90}{\bf LPC}    & \includegraphics[width=3.7cm]{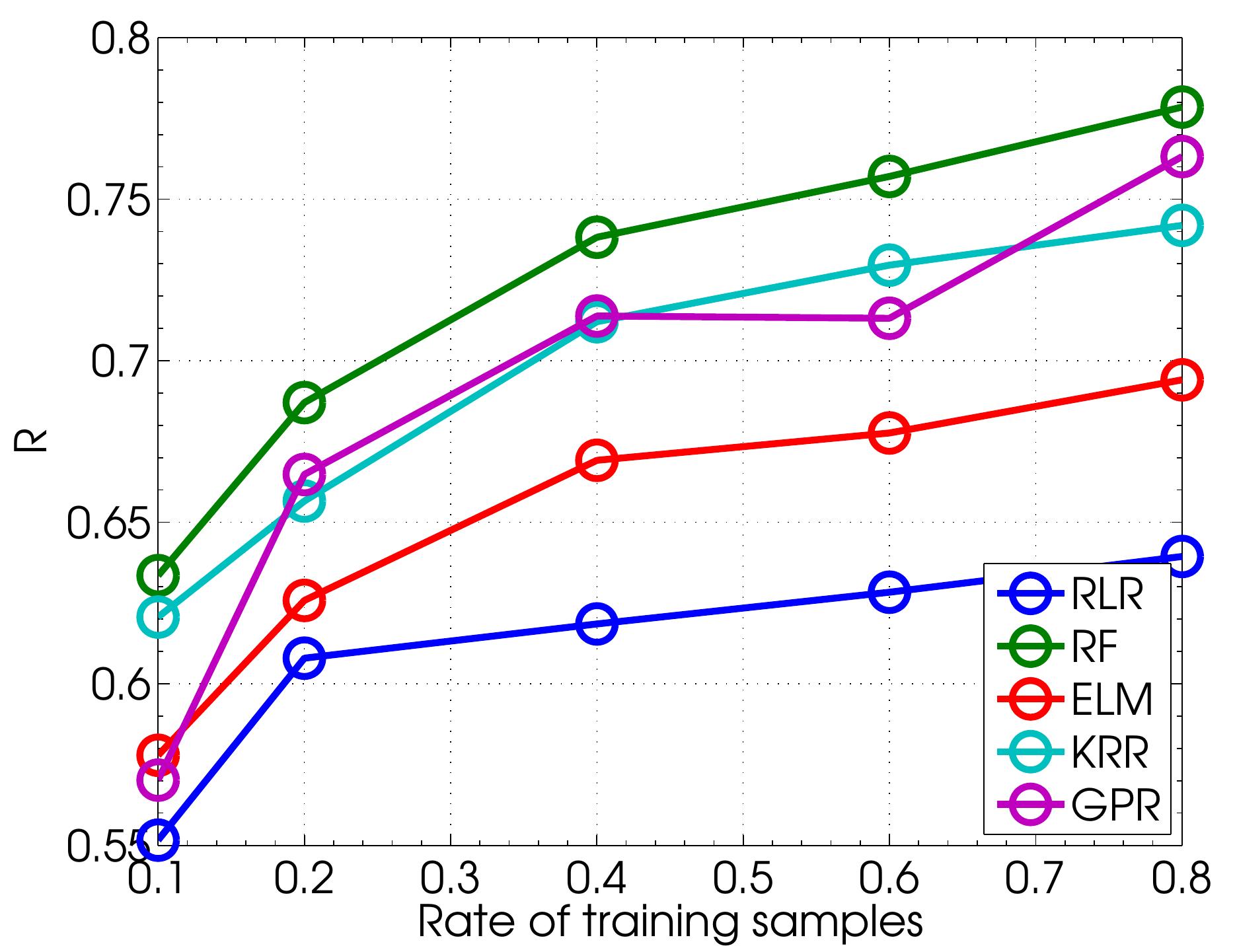} & \includegraphics[width=3.7cm]{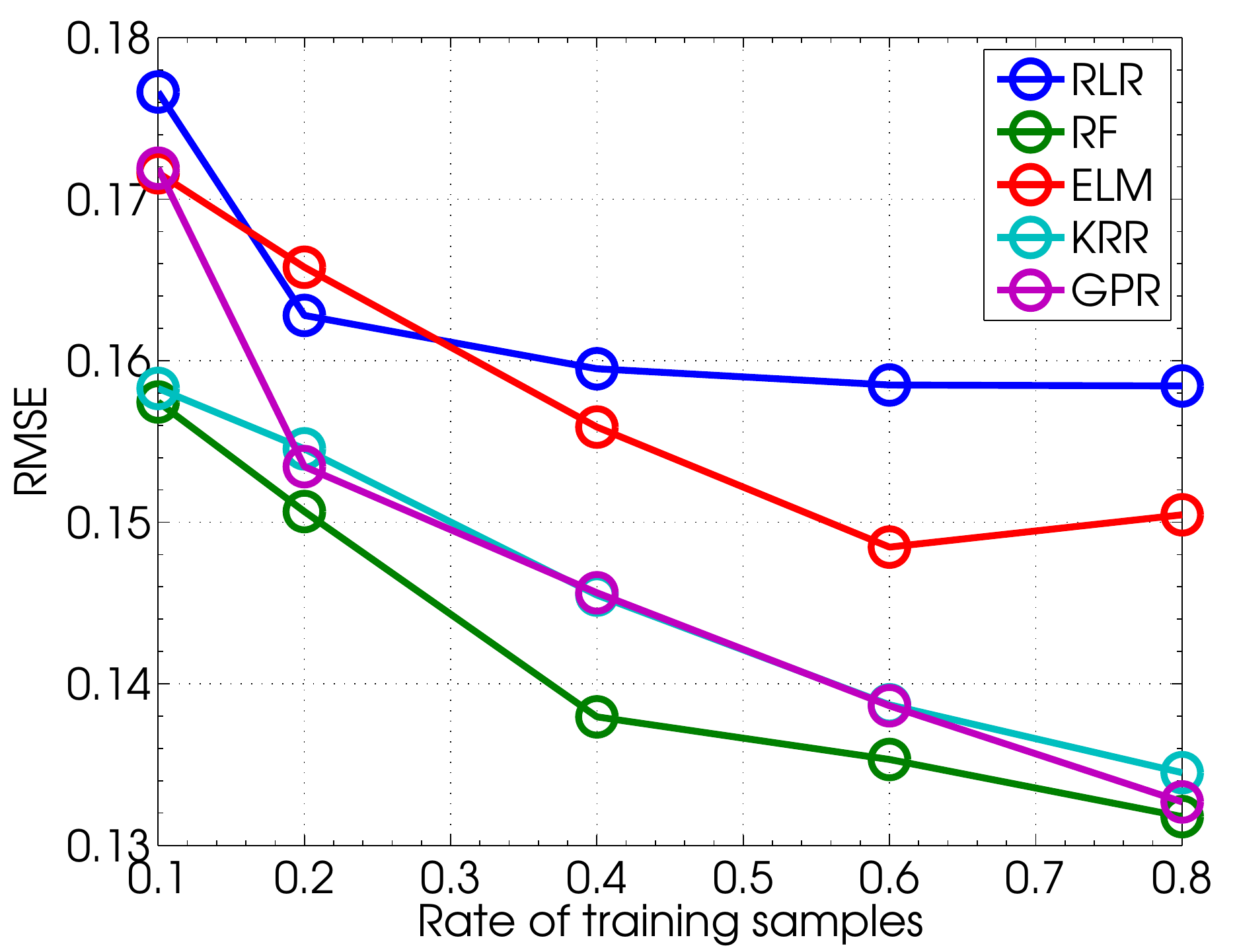} \\
\rotatebox{90}{\bf LNPR}   & \includegraphics[width=3.7cm]{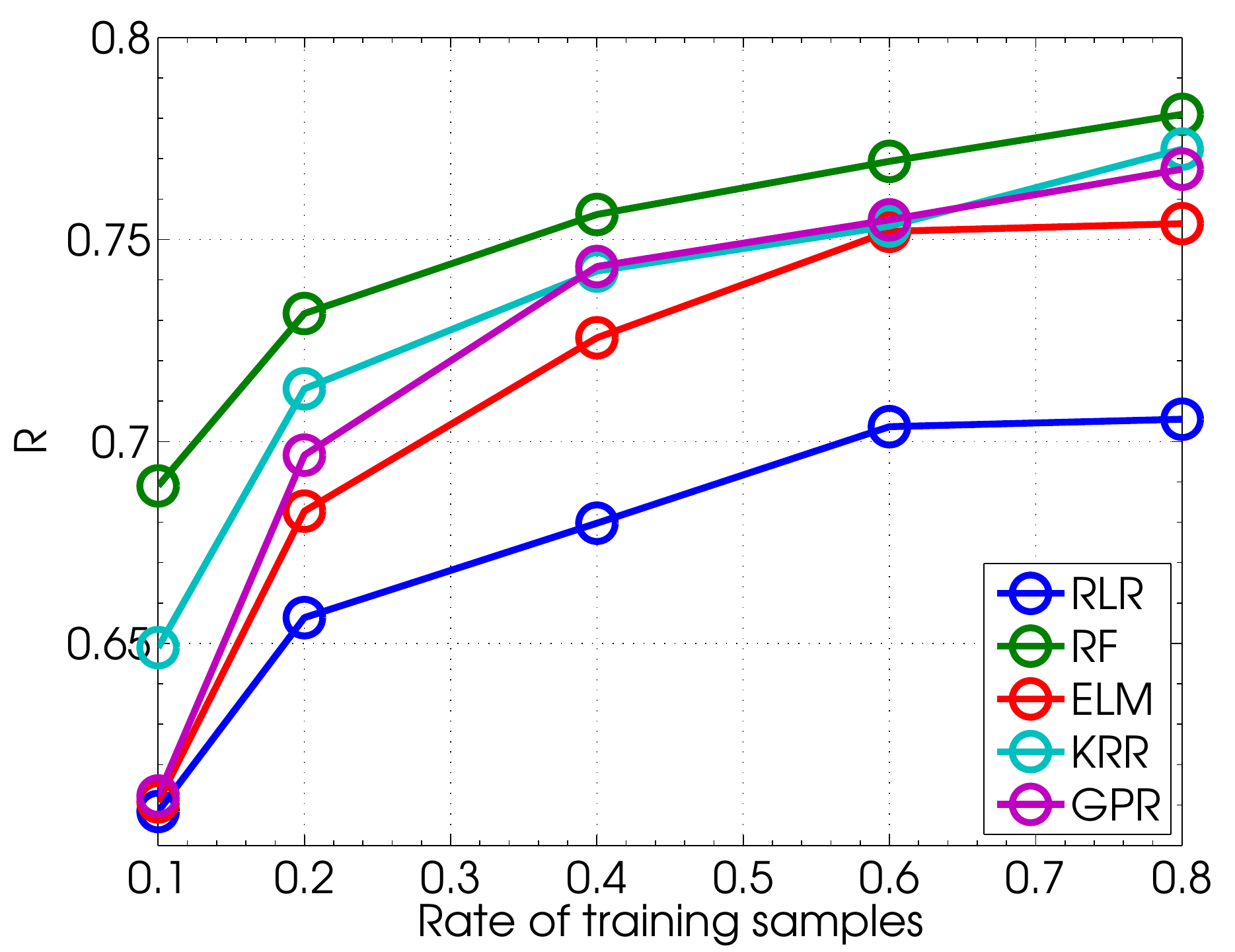} & \includegraphics[width=3.7cm]{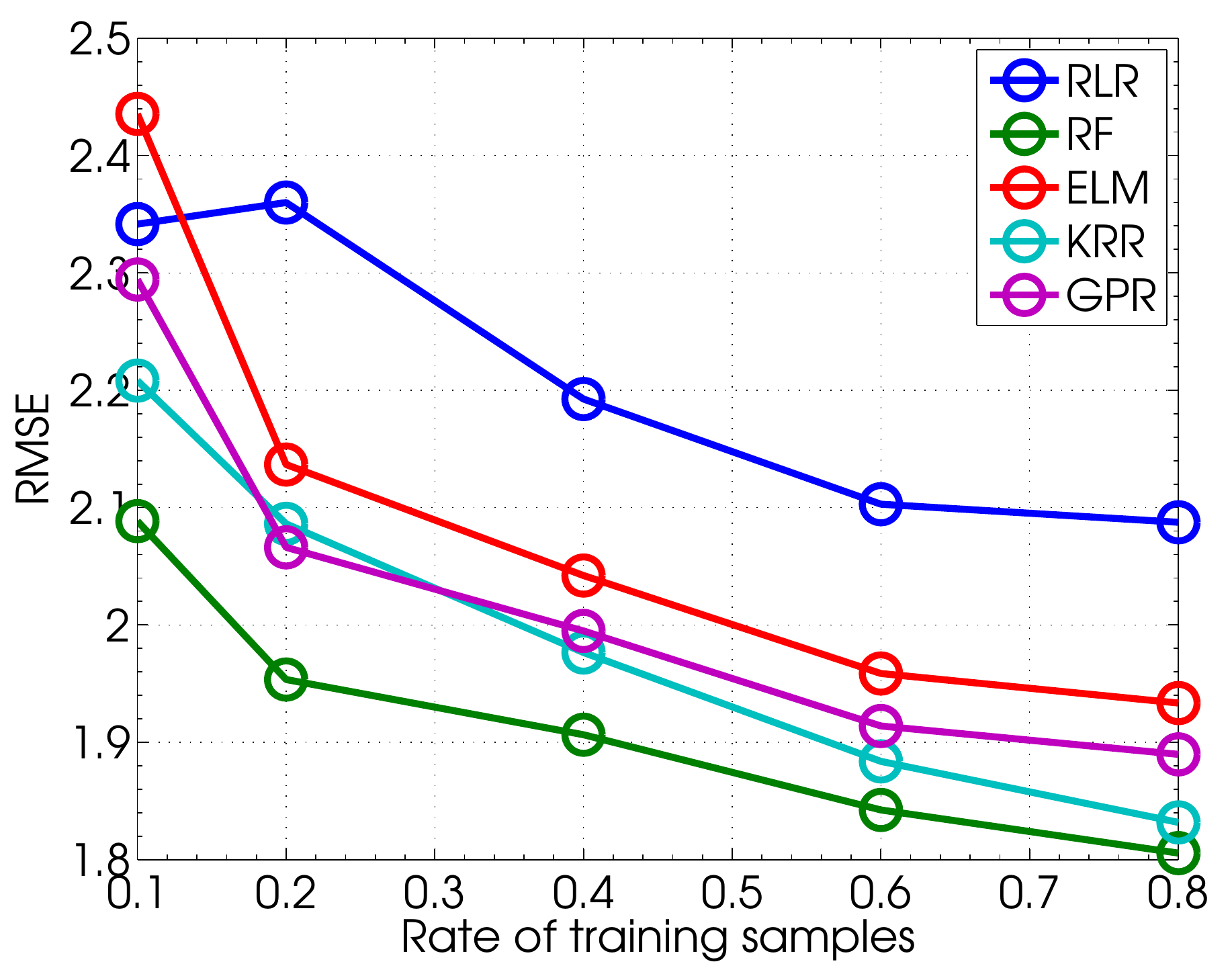} \\
\rotatebox{90}{\bf LDMC}  & \includegraphics[width=3.7cm]{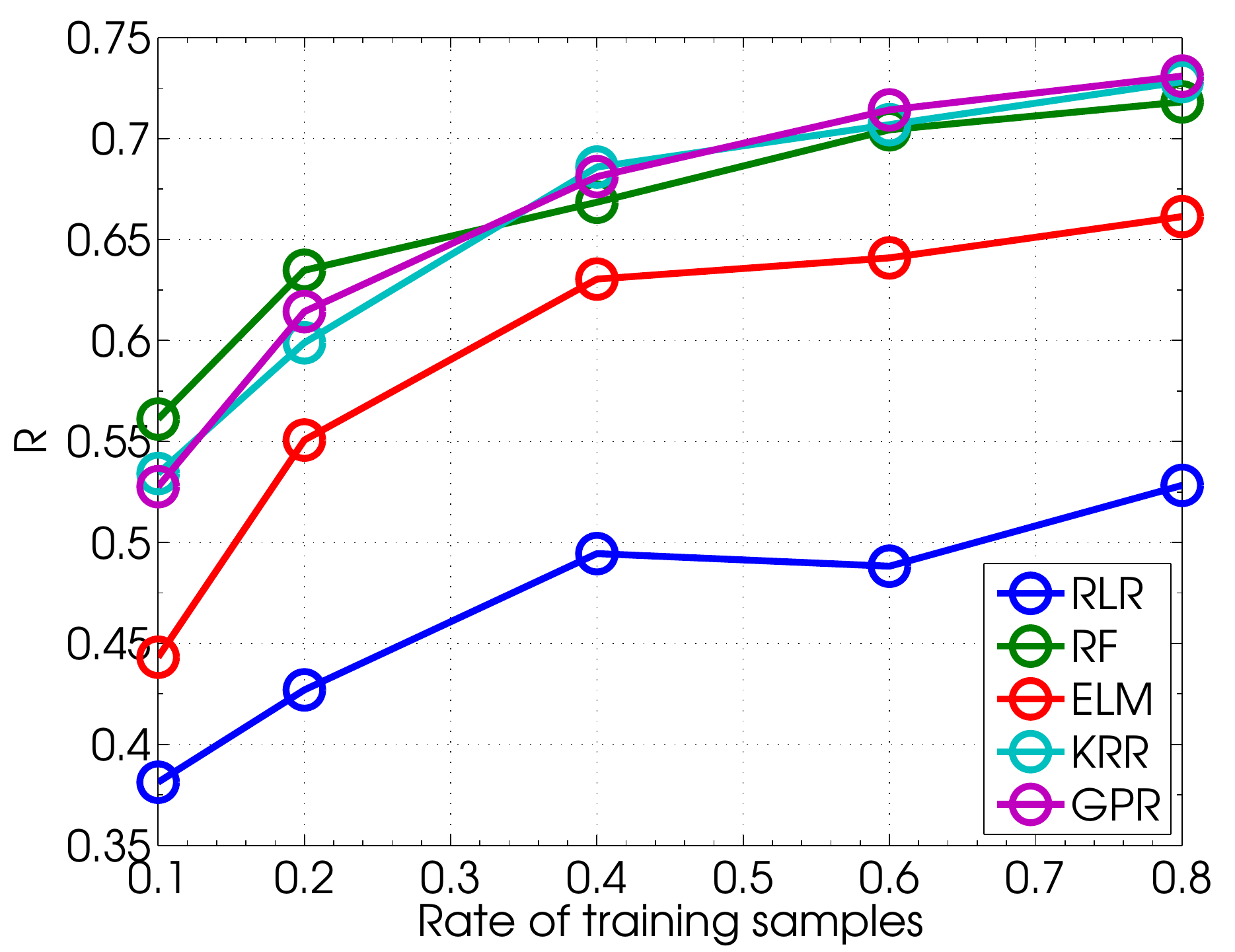} & \includegraphics[width=3.7cm]{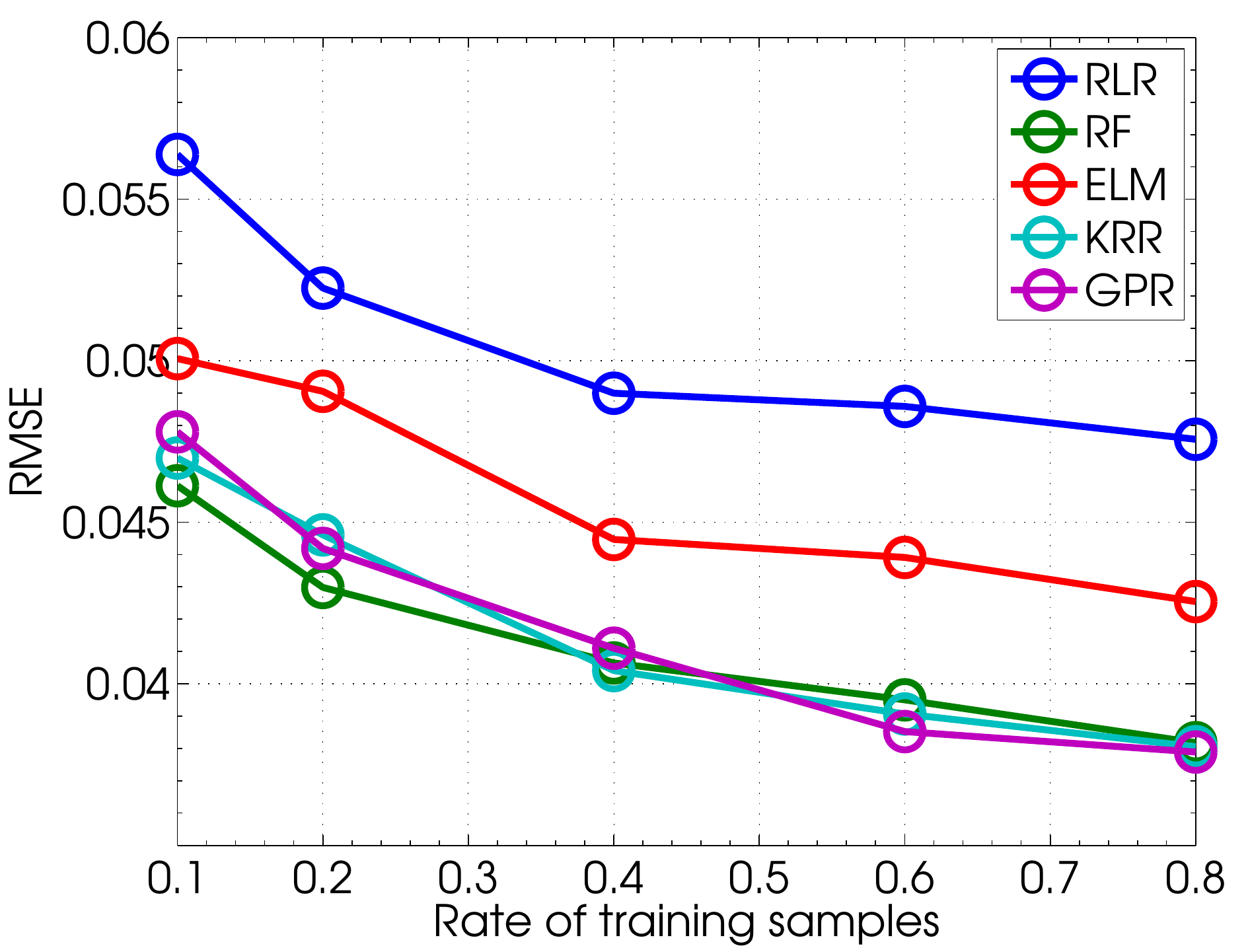} \\
%\rotatebox{90}{\bf FMNC}  & \includegraphics[width=3.7cm]{images/freshmass_R.pdf} & \includegraphics[width=3.7cm]{images/freshmass_RMSE.pdf} \\
\end{tabular}
\end{center}
\caption{Results in the test set for all methods and leaf traits as a function of the rate of training data used.}\label{fig:rates}
\end{figure}

\newpage

\section{Sensitivity analysis in the gap filling of the TRY database}
\label{ap:senGF}
\setcounter{figure}{0}
\setcounter{table}{0}
\setcounter{equation}{0}
Table~\ref{tab:gapfillingfeatures} shows the relative importance of the predictors considered for each gap filled trait. The predictor importance has been computed for every tree composing the random forests by summing changes in MSE due to splits on every predictor and dividing the sum by the number of branch nodes. As our trees are grown with surrogate splits, this sum is taken over all splits at each branch node including surrogate splits.

\begin{table}[!h]
\centering
\caption{Ranking of the five most relevant variables for the gap-filling. Bio1 to Bio11 are temperature related climatological variables while Bio 12 to Bio19 are precipitation related climatological variables. For a more detailed climatic variables description see \ref{ap:climatedata}.}
\label{tab:gapfillingfeatures}
\begin{tabular}{|c|c|c|c|c|c|}
\hline
\hline
{\textbf{Ranking}}            & {\textbf{SLA}} & {\textbf{LDMC}} & {\textbf{LNC}} & {\textbf{LPC}} & \textbf{LNPR} \\ \hline
{\textbf{1}}               & Species                   & Species                              & Genus                             & LNPR                              & Genus          \\ \cline{1-1}
{\textbf{2}}              & Genus                     & Genus                               & LPC                             & Species                             & Family          \\ \cline{1-1}
{\textbf{3}}             & LDMC                      & Family                               & Species                              & BIO16                              & Species          \\ \cline{1-1}
{\textbf{4}}          & BIO14                       & Growth form                           & LNPR                            & BIO3                             & BIO16          \\ \cline{1-1}
{\textbf{5}}                & BIO19                  & BIO11                                & Family                              & BIO1                              & BIO3          \\
\hline
\hline
\end{tabular}
\end{table}

Results show that the taxonomy (categorical information) plays a crucial role in the gap filling of all traits, being the species names and the genus the most influential ones. Those results match conclusions of previous works and confirm the effectiveness of the taxonomic hierarchy information in order to predict trait values \citep{cordlandwehr2013plant}. Results also confirm previous studies that showed close relationships between vegetation composition and environmental conditions. For example, SLA predictions are firstly controlled by climatological information related with availability of water for the vegetation. \cite{marron2003impact} pointed that species with typically low SLA values are more conservative with the acquired resources, due to their LDMC, high concentrations of cell walls and secondary metabolites, and high leaf and root longevity. For the case of LDMC, a variable strongly related to SLA, only mean temperature information is included amongst the most 5 significant variables. This observation is also in agreement with results in \citep{schrodt2015bhpmf,albert2010intraspecific}, which showed that temperature-related climatological variables are an important source of information to describe functional variation within species along environmental gradients. Leaf N and P concentrations are controlled by mean temperature in different ways. On the one hand, concentrations of N and P in plant tissues increase to offset the decreases in plant metabolic rate as the ambient temperature decreases \citep{reich2004global}. On the other hand, leaf N concentrations may increase in arid regions because plants may contain higher N concentrations to better adapt to more limited environments \citep{wright2003least}. Lastly, leaf P concentrations are influenced by precipitation because an increase in soil water availability facilitate decomposition processes of litter and amplify P availability especially in arid regions \citep{yang2016variations}.

\section{Sensitivity analysis in the trait prediction models}
\label{ap:sentraitmaps}
\setcounter{figure}{0}
\setcounter{table}{0}
\setcounter{equation}{0}

We assessed the relative importance of the climatic and remote sensing data that contributed to the globally mapped canopy traits. Table~\ref{tab:traitmapfeatures} shows the relative importance of the seven most important predictors considered by the RF model. The predictor importance has been computed for every tree composing the random forests by summing changes in MSE due to splits on every predictor and dividing the sum by the number of branch nodes. As our trees are grown with surrogate splits, this sum is taken over all splits at each branch node including surrogate splits.

\begin{table}[!h]
\centering
\caption{Ranking of the seven most relevant variables for the mapping of the considered traits. Bio1 to Bio11 are temperature related climatological variables while Bio 12 to Bio19 are precipitation related climatological variables, B1med to B7med are the yearly median values of the different MODIS bands and VI*max, VI*min, VI*sum, VI*std are maximum, minimum, annual sum and standard deviation of considered vegetation index (VI*). For a more detailed climatic and remote sensing variables description see \ref{ap:climatedata} and Section \ref{ap:CWMs calc}.}
\label{tab:traitmapfeatures}
\begin{tabular}{|c|c|c|c|c|c|}
\hline
\hline
{\textbf{Ranking}}            & {\textbf{SLA}} & {\textbf{LDMC}} & {\textbf{LNC}} & {\textbf{LPC}} & \textbf{LNPR} \\ \hline
{\textbf{1}}               & EVImax                  & BIO14          & B5med            & BIO14                & EVIsum         \\ \cline{1-1}
{\textbf{2}}              & B6med                     & BIO5        & EVImax           & BIO16                    & BIO16          \\ \cline{1-1}
{\textbf{3}}             & EVIstd                     & Elevation     & B6med                & EVIstd                & EVImin          \\ \cline{1-1}
{\textbf{4}}          & B5med                       & EVIsum       & B2med              & BIO3                    & BIO1          \\ \cline{1-1}
{\textbf{5}}                & BIO15                 & B6med         & EVIstd                  & BIO10                & BIO14          \\ \cline{1-1}
{\textbf{6}}                & B2med                  & BIO17        & BIO3                 & BIO17                     & BIO3          \\ \cline{1-1}
{\textbf{5}}                & Elevation                 & BIO12      & NDWImax                & NDWImax               & BIO5          \\
\hline
\hline
\end{tabular}
\end{table}

\section{Traits values and variation between and within PFTs}
\label{ap:TRYtypicalvalues}
\setcounter{figure}{0}
\setcounter{table}{0}
\setcounter{equation}{0}
We have computed mean values and the standard deviation of in-situ leaf level trait measurements for the different PFTs. In order to associate leaf level measurements with PFTs we have used the ancillary data provided by the TRY database which relates species names of each trait measurement with information describing the conventional PFTs definitions.  Table \ref{tab:typvaluesTRY} shows the computed statistics.

\begin{table}[t!]
\small
\caption{In-situ trait leaf level mean values and standard deviation for each considered PFT. }
\label{tab:typvaluesTRY}
\begin{center}
\renewcommand{\arraystretch}{0.8}
\begin{tabular}{|p{3cm}|p{3cm}|p{3cm}|}
\hline
\hline
{\bf PFT}	 & {\bf Mean}	 & {\bf Std}  \\
\hline
\hline
\multicolumn{3}{|l|}{Specific Leaf Area (SLA)} \\
\hline
ENF	 & 4.82	 & 2.67	   \\
EBF	 & 12.07   & 5.15	   \\
DNF	 & 9.19	 & 	2.85   \\
DBF	 & 18.57	 & 8.38	   \\
SHL	 & 14.54	 & 7.46	   \\
GRL	 & 20.97	 & 8.58	   \\
\hline
\multicolumn{3}{|l|}{Leaf Nitrogen Concentration (LNC)} \\
\hline
ENF	 & 12.18	 & 2.47	   \\
EBF	 & 18.82   & 5.85	   \\
DNF	 & 20.01	 & 	3.24   \\
DBF	 & 22.78	 & 4.82   \\
SHL	 & 17.56	 & 7.15	   \\
GRL	 & 20.52	 & 7.81	   \\
\hline
\multicolumn{3}{|l|}{Leaf Phosphorus Concentration (LPC)} \\
\hline
ENF	 & 1.15	 & 0.36	   \\
EBF	 & 0.70   & 0.30	   \\
DNF	 & 1.86	 & 	0.40   \\
DBF	 & 1.40	 & 0.52   \\
SHL	 & 1.11	 & 0.53	   \\
GRL	 & 1.28	 & 0.58   \\
\hline
\multicolumn{3}{|l|}{Leaf Nitrogen-Phosphorus Ratio (LNPR)} \\
\hline
ENF	 & 9.48	 & 2.03	   \\
EBF	 & 19.10   & 4.48	   \\
DNF	 & 10.18	 & 	2.77   \\
DBF	 & 14.20	 & 4.34  \\
SHL	 & 15.29	 & 5.30	   \\
GRL	 & 11.88	 & 5.60   \\
\hline
\multicolumn{3}{|l|}{Leaf Dry Matter Content (LDMC)} \\
\hline
ENF	 & 0.31	 & 0.10	   \\
EBF	 & 0.34   & 0.06	   \\
DNF	 & 0.28	 & 0.06   \\
DBF	 & 0.33	 & 0.06  \\
SHL	 & 0.29	 & 0.06	   \\
GRL	 & 0.26	 & 0.07   \\
\hline
\end{tabular}
\end{center}
\end{table}